\documentclass[prd,aps,twocolumn,nofootinbib,showpacs,superscriptaddress]{revtex4-1}
\usepackage{amsfonts}
\usepackage{amsmath}
\usepackage{amssymb}
\usepackage{bm}
\usepackage{dcolumn}
\usepackage[dvips]{graphicx}
\usepackage{graphics}
\usepackage[latin1]{inputenc}
\usepackage{latexsym}
\usepackage{rotating}
\usepackage[colorlinks=true]{hyperref}
\usepackage{xspace} 
\usepackage[usenames]{color}
\usepackage{mathrsfs}
\usepackage{multirow}
\usepackage{pifont}
\usepackage{enumitem}

\definecolor {darkgreen}{rgb}{0.2,0.7,0.2}
\definecolor{purple}{rgb}{0.5,0,0.5}

\newcommand\be{\begin{equation}}
\newcommand\ba{\begin{eqnarray}}
\newcommand\ee{\end{equation}}
\newcommand\ea{\end{eqnarray}}

\newcommand\bw{\begin{widetext}}
\newcommand\ew{\end{widetext}}

\newcommand{\nn}{\nonumber}

\newcommand{\ppE}{{\mbox{\tiny ppE}}}
\newcommand{\dCS}{{\mbox{\tiny dCS}}}
\newcommand{\MD}{{\mbox{\tiny MDR}}}
\newcommand{\MG}{{\mbox{\tiny MG}}}
\newcommand{\EDGB}{{\mbox{\tiny EdGB}}}

\newcommand{\ED}{{\mbox{\tiny ED}}}

\newcommand{\EI}{{\mbox{\tiny EI}}}
\newcommand{\I}{{\mbox{\tiny I}}}
\newcommand{\RD}{{\mbox{\tiny RD}}}

\newcommand{\Int}{{\mbox{\tiny Int}}}
\newcommand{\MR}{{\mbox{\tiny MR}}}
\newcommand{\BD}{{\mbox{\tiny BD}}}

\newcommand{\QK}{{\mbox{\tiny QK}}}
\newcommand{\KG}{{\mbox{\tiny KG}}}
\newcommand{\EA}{{\mbox{\tiny EA}}}

\newcommand{\Kerr}{{\mbox{\tiny Kerr}}}

\newcommand{\ISCO}{{\mbox{\tiny ISCO}}}
\newcommand{\GR}{{\mbox{\tiny GR}}}

\newcommand{\ST}{{\mbox{\tiny ST}}}

\newcommand{\hor}{{\mbox{\tiny H}}}

\newcommand{\NS}{{\mbox{\tiny NS}}}
\newcommand{\BS}{{\mbox{\tiny BS}}}
\newcommand{\BH}{{\mbox{\tiny BH}}}

\newcommand{\mrm}{\mathrm}

\begin{document}

\title{Theoretical Physics Implications of the Binary Black-Hole Mergers \\ GW150914 and GW151226}

\author{Nicol\'as Yunes}
\affiliation{eXtreme Gravity Institute, Department of Physics, Montana State University, Bozeman, MT 59717, USA.}

\author{Kent Yagi}
\affiliation{Department of Physics, Princeton University, Princeton, New Jersey 08544, USA.}

\author{Frans Pretorius}
\affiliation{Department of Physics, Princeton University, Princeton, New Jersey 08544, USA.}

\date{\today}

\begin{abstract} 

The gravitational wave observations GW150914 and GW151226 by Advanced LIGO provide the first opportunity to learn about physics in the extreme gravity environment of coalescing binary black holes. 
The LIGO Scientific Collaboration and the Virgo Collaboration have verified that this observation is consistent with Einstein's theory of General Relativity, constraining the presence of certain parametric anomalies in the signal. 
This paper expands their analysis to a larger class of anomalies, highlighting the inferences that can be drawn on non-standard theoretical physics mechanisms that could otherwise have affected the observed signals.
We find that these gravitational wave events constrain a plethora of mechanisms associated with the generation and propagation of gravitational waves, including the activation of scalar fields, gravitational leakage into large extra dimensions, the variability of Newton's constant, the speed of gravity, a modified dispersion relation, gravitational Lorentz violation and the strong equivalence principle. Though other observations limit many of these mechanisms already, GW150914 and GW151226 are unique in that they are direct probes of dynamical strong-field gravity and of gravitational wave propagation.
We also show that GW150914 constrains inferred properties of exotic compact object alternatives to Kerr black holes. We argue, however, that the true potential for GW150914 to both rule out exotic objects and constrain physics beyond
General Relativity is severely limited by the lack of understanding of the coalescence regime in almost all relevant modified gravity theories. This event thus significantly raises the bar that these theories have to pass, both in terms of having a sound theoretical underpinning, and reaching the minimal level of being able to solve the equations of motion for binary merger events.
We conclude with a discussion of the additional inferences that can be drawn if the lower-confidence observation of an electromagnetic counterpart to GW150914 holds true, or such a coincidence is observed with future events; this would provide dramatic constraints on the speed of gravity and gravitational Lorentz violation. 

\end{abstract}


\maketitle

\section{Introduction}  

The Laser Interferometer Gravitational-Wave Observatory (LIGO) Scientific Collaboration and the Virgo Collaboration (LVC) recently announced the first direct detection of gravitational waves (GWs)~\cite{Abbott:2016blz,Abbott:2016nmj}. With signal-to-noise ratios (SNR) of 24 and 13 respectively for GW150914~\cite{Abbott:2016blz} and GW151226~\cite{Abbott:2016nmj}, and associated statistical $\sigma > 5$ for both, there is little doubt that these are true GW observations. The details of the signals indicate the GWs were produced during the late quasi-circular inspiral, merger and ringdown of binary black hole (BH) systems. The loudness of GW150914 is due to a combination of Advanced LIGO's (aLIGO's)~\cite{TheLIGOScientific:2014jea} remarkable sensitivity (dimensionless strains of $h \sim 10^{-21}$), the source's proximity to Earth ($420^{+150}_{-180}$ Mpc~\cite{TheLIGOScientific:2016wfe,TheLIGOScientific:2016pea}) and how massive the binary was (source-frame component masses $(m_{1}, m_{2}) = (36^{+5}_{-4},29^{+4}_{-4}) M_{\odot}$~\cite{TheLIGOScientific:2016wfe,TheLIGOScientific:2016pea}), the latter property fortuitously leading to the intrinsically loudest part of the signal lying in aLIGO's most sensitive frequency band. GW151226 occurred at a similar distance ($440^{+180}_{-190}$ Mpc~\cite{Abbott:2016nmj,TheLIGOScientific:2016pea})  though with lower source-frame masses of $(m_{1}, m_{2}) = (14^{+8}_{-4},8^{+2}_{-2}) M_{\odot}$~\cite{Abbott:2016nmj,TheLIGOScientific:2016pea}, resulting in a weaker overall signal but with many more GW cycles in band compared to GW150914. These events are thus ideal to learn about theoretical physics in extreme gravity. 

The social scientist and epistemologist Karl Popper argued that scientists can never truly ``prove'' that a theory is correct, but rather all we can do is disprove, or more accurately constrain, alternative hypothesis~\cite{popper}. The theory that remains and cannot be disproven by observations becomes the \emph{status quo}. Indeed, this was the case for Newtonian gravity before the 1900s, and it is the case today for Einstein's theory of General Relativity (GR). The latter has been subjected to a battery of tests through Solar System~\cite{Will:2014kxa}, binary pulsar~\cite{stairs,Wex:2014nva} and cosmological observations~\cite{Jain:2010ka,Clifton:2011jh,Joyce:2014kja,Koyama:2015vza,Salvatelli:2016mgy}, with no signs of failure\footnote{Some have argued dark matter~\cite{Milgrom:1983ca,Milgrom:2008rv,Famaey:2011kh} or dark energy~\cite{Jain:2010ka,Koyama:2015vza,Salvatelli:2016mgy} could be explained by modified gravity theories, though the observational evidence does not favor this over a simple cosmological constant or as-of-yet undiscovered dark matter particles.}. These tests, however, cannot effectively probe the \emph{extreme gravity regime}: where the gravitational field is strong and dynamical, where the curvature of spacetime is large, and where characteristic velocities are comparable to the speed of light~\cite{Yunes:2013dva}. 

The events GW150914 and GW151226 allow for just that. The putative BHs that generated these GWs are intrinsically strong-field sources, they reached speeds $\sim0.5$ times the speed of light prior to merger, and the GW luminosities peaked at $\sim 10^{56}$ergs/s, within 3 orders of magnitude of the Planck luminosity. Consequently, the gravitational fields were not only immense, but they changed violently and rapidly during the less than 1 second observable durations of these events. 

The LVC began to test extreme gravity with GW150914 by first showing that the residual, i.e.~the signal after subtracting the best-fit GR model, is consistent with noise~\cite{TheLIGOScientific:2016src}. Moreover, the collaboration searched for the presence of certain anomalies, i.e.~features in the signal that deviate from the GR prediction, using a parameterized model and found no evidence for any. GW151226 was also shown to pass the latter test, but its SNR was too low to give a meaningful result from the residual test~\cite{Abbott:2016nmj,TheLIGOScientific:2016pea}.

Having established GR as the status quo in the extreme gravity regime, we here follow Popper and study the theoretical physics implications of these detections. More specifically, we examine what the verification of GR and absence of anomalies in the data imply for theoretical physics (see~\cite{Will:2014kxa,Yunes:2013dva,Gair:2012nm,Berti:2015itd,Yagi:2016jml} for reviews on modified theories of gravity and testing them in the extreme gravity regime with GWs). For example, GW150914 and GW151226 constrain new radiative channels, such as dipole scalar field emission or GW polarizations beyond the plus and cross polarizations predicted by GR, BH mass leakage into extra dimensions, and temporal variability of Newton's gravitational constant during the coalescence. These implications affect the viability of physical mechanisms that play an important role in quantum gravitational phenomenology and high-energy model-building~\cite{Giddings:2016tla}.

Before summarizing our results, let us first discuss what is perhaps one of the most important consequence of these detections for testing GR, in particular with GW150914: our ability to use this exquisite piece of data to probe extreme gravity is today limited by our woeful lack of understanding of how gravity can differ from GR in this regime. Most\footnote{The only exception is a particular class of scalar-tensor theories, when one or both of the compact objects are neutron stars (NSs)~\cite{ST1,Palenzuela:2013hsa,Shibata:2013pra}, or when both are BHs but embedded in a prescribed scalar field background~\cite{Healy:2011ef,Berti:2013gfa}.} of the existing studies of compact binary coalescences that are alternatives to binary BH mergers in GR are limited to two regimes: 
\begin{itemize}[itemsep=0.1pt, topsep=5pt, partopsep=0pt]
\item[(i)] the early inspiral where post-Newtonian (PN) expansions\footnote{This approximation solves the field equations using an expansion in small velocities (relative to the speed of light) and weak gravitational fields. A term proportional to $(v/c)^{2N}$ relative to its leading-order expression is said to be of $N$PN order.} (to some order) have been computed, or 
\item[(ii)] isolated stationary compact object alternatives to BHs in GR, where their quasi-normal mode (QNM) structure is putatively relevant to the late-time dynamics of the post-merger remnant.
\end{itemize}
Prior to GW150914 the mainstream consensus was that the binary BH mergers aLIGO would likely hear would be lower mass~\cite{Mandel:2009nx,abadie}; hence, a significant portion of the earlier inspiral would contribute to the SNR where the perturbative calculations in (i) have more discriminating power, and the plunge/merger regime is less crucial\footnote{This is even more so for binary NS mergers, another primary target for aLIGO~\cite{abadie}, which are expected to have lower masses than BHs and thus merge at even higher frequencies.}. Similarly, the calculations in (ii) are adequate for electromagnetic wave tests~\cite{psaltis-review,Bambi:2015kza,Johannsen:2015mdd,Johannsen:2016uoh} with e.~g.~the Event Horizon Telescope~\cite{2009astro2010S..68D} and with future space-based measurements of the ringdown phase of supermassive BH mergers~\cite{2006PhRvD..73f4030B,Berti:2007zu}. GW150914 has now presented us with data we did not anticipate (at least not immediately), where most of the SNR is coming from a regime where the applicability of calculations based solely on (i) and (ii) to describe GWs are questionable at best.

One striking feature of the GW150914 signal in particular serves to highlight all of this: after reaching peak amplitude, the GW emission drops to below the noise threshold within the light-crossing time of the length-scale implied by the total mass of the binary. This is of course entirely as expected in vacuum GR, though the physics within the brief transition is extremely rich:
\begin{itemize}[itemsep=0.1pt, topsep=5pt, partopsep=0pt]
\item[(a)] Cosmic censorship~\cite{Penrose_CC} is respected, and hence, the no-bifurcation theorem requires the horizons merge into a single structure~\cite{1973lsss.book.....H}.
\item[(b)] The BH uniqueness (``no-hair'') results~\cite{israel,1971PhRvL..26..331C,hawking-uniqueness,1975PhRvL..34..905R} together with the apparent stability of the Kerr family of solutions~\cite{1972PhRvD...5.2419P} implies the end-state must be a Kerr BH\footnote{The non-linear stability of Kerr has not yet been proven in a strict mathematical sense (see e.g.~\cite{Dafermos:2016uzj} and the discussion therein).}.
\item[(c)] Numerical solutions show that the time it takes from formation of a common dynamical horizon to when the spacetime settles down to a linearly perturbed Kerr solution is remarkably short~\cite{2007PhRvD..75l4018B}. 
\end{itemize}
The fact that it has taken over half a century of dedicated research by the GR community to allow us to make this short itemized summary of the physics of a BH merger is a testament to how non-trivial this feature of the GW150914. On the flip side, it also highlights how poor our understanding is with regards to the nature of {\em conceivable} theories of gravity in this regime.

Another very important consequence of the use of LVC observations to test GR is that ``exotic'' compact object alternatives to BHs within GR, such as boson stars, gravastars, or traversable wormholes, can no longer claim viability based only on (1) the demonstration of the existence of stationary solutions, and (2) consistency with the properties of X-ray binary systems harboring one of these putative objects. In fact, certain exotic objects do not even have theories that describe how they form, let alone their dynamics in the highly non-linear, violent regime of a collision (of those listed above, boson stars are the only exception), which is why the LVC did not attempt to constrain such exotica~\cite{TheLIGOScientific:2016src}. To be consistent with the LVC observations then, one must assume not only that such theories exist, but also that when solved, the non-linear, {\em matter} oscillations inevitably excited during the collision will damp on the remnant's light-crossing time. 

Does such a damping naturally occur during the collision of non-vacuum compact objects in GR? Certainly not, as shown in the merger of binary NSs. Here we know that the remnant either promptly collapses to a BH, or a hypermassive NS is formed with a highly non-spherical, time-dependent structure that emits strong GWs for a long time relative to light-crossing (see~\cite{Lehner:2016lxy,Radice:2016dwd,Hotokezaka:2016bzh,Dietrich:2015pxa, Foucart:2015gaa,East:2015vix,Clark:2015zxa} for some recent studies). Therefore, if NS mergers offer any guidance as to what one might expect in the collision of exotica, it is that, immediately following merger, the stationary, isolated compact object solution does {\em not} provide a good starting point to understand what this phase of the signal looks like.  \emph{Thus, the LVC observations have significantly raised the bar that exotic matter alternatives to BHs within GR must pass to still be considered viable: their merger dynamics must be well-understood \emph{and} shown to be consistent with the signals.}

With these observations in mind, the main goal of this paper is to study what GW150914 and GW151226 imply about the theoretical nature of extreme gravity. Even though earlier experiments and observations in the weak field have placed bounds on mechanisms we will discuss, and in some cases these will be stronger than what we extract from the GW events at present, the latter bounds are for the very first time coming directly from the extreme (dynamical \emph{and} strong field) gravity regime. Moreover, many existing bounds are plagued by systematics associated with models of non-gravitational physics required to interpret observations. For GW150914 and GW151226 the errors in the analytic and numerical GR waveforms used to interpret the aLIGO signal  (the GR ``mismodeling error''~\cite{Cutler:2007mi}) are a small part of the error budget~\cite{TheLIGOScientific:2016wfe}. In fact, we will show in this paper that mismodeling error does not affect the bounds on non-GR effects derived here for events with SNR comparable to (or less than) that of GW150914 and GW151226.

As this paper is quite long, a roadmap is in order to guide a wide audience with different interests. In the remainder of the Introduction we summarize all key results, breaking them down into 4 categories: implications on emission mechanisms, implications on propagation effects, implications on the nature of the compact objects involved in the merger, and more speculative conclusions associated with electromagnetic counterparts. The rest of the paper is then split following similar categories: 
\begin{itemize}
\item[(I)] a review of the extreme gravity properties of the GW150914 and GW151226 signals and waveform modeling (Sec.~\ref{sec:testsofgravwithBHS}), 
\item[(II)] GW150914 and GW151226 constraints on mechanisms that affect the generation and propagation of GWs (Sec.~\ref{sec:parametrizedtests}),
\item[(III)] generic properties of the remnant and inferences on the existence of exotic alternatives to BHs as inferred from the GW150914 signal (Sec.~\ref{sec:generictests}),
\item[(IV)] inferences that can be drawn from the more speculative coincidence of GW150914 with a short gamma-ray burst (GRB)~\cite{Connaughton:2016umz} (Sec.~\ref{sec:spec-imps}).
\end{itemize}
Readers familiar with (I) may wish to skip Sec.~\ref{sec:testsofgravwithBHS}, but those who are not familiar with waveform modeling in non-GR theories may find Secs.~\ref{sec:ppE} and~\ref{subsec:par-def-to-physics} useful.~The most important parts of (II) are Secs.~\ref{sec:generation2},~\ref{sec:gen-constr-in-generation} and~\ref{sec:mod-disp-rel}. The first two deal with constraints on generation effects, with the first mapping them to bounds on specific modified theories and the second relating them to model-independent bounds on modifications to the binding energy and energy flux. Section~\ref{sec:gen-constr-in-generation} also discusses constraints on effects that suddenly activate or deactivate during the late inspiral. Section~\ref{sec:mod-disp-rel} concludes the discussion of constraints by focusing on propagation effects and mapping these to bounds on specific modified theories. These results are shown to be robust to mismodeling bias in Appendix~\ref{app:BvsD}, which studies constraints with two different phenomenological inspiral-merger-ringdown GW models, and in Appendix~\ref{app:effect-of-hi-PN}, which deals with the effect of higher PN order terms. 

\subsection{Summary of Key Results}
{
\newcommand{\minitab}[2][l]{\begin{tabular}{#1}#2\end{tabular}}
\renewcommand{\arraystretch}{1.4}
\begingroup
\squeezetable
\begin{table*}[htb]
\begin{centering}
\begin{tabular}{c|c|c|c|c|c|c|c|c}
\hline
\hline
\noalign{\smallskip}
 \multirow{2}{*}{Theoretical Mechanism} &  \multirow{2}{*}{GR Pillar} & \multirow{2}{*}{PN} & \multicolumn{2}{c|}{$|\beta|$} &\multicolumn{4}{c}{Example Theory Constraints}\\ 
 & & & {\bf{GW150914}} & {\bf{GW151226}}  & Repr. Parameters & {\bf{GW150914}} & {\bf{GW151226}} & Current Bounds \\
\hline \hline
 \multirow{2}{*}{Scalar Field Activation}  & \multirow{2}{*}{SEP} & \multirow{2}{*}{$-1$}  & \multirow{2}{*}{$\mathbf{1.6 \times 10^{-4}}$}& \multirow{2}{*}{$\mathbf{4.4 \times 10^{-5}}$} & $\sqrt{|\alpha_\EDGB|}$ [km] & 
 --- & --- &  $10^7$~\cite{Amendola:2007ni}, 2~\cite{Kanti:1995vq,kent-LMXB,Pani:2009wy}\\
   &  & &   & & $|\dot{\phi}|$ [1/sec] & 
 --- &  --- & $10^{-6}$~\cite{Horbatsch:2011ye} \\ 
 \hline
 Scalar Field Activation & SEP, PI & $+2$ & $\mathbf{1.3\times 10^1}$  & $\mathbf{4.1}$ & $\sqrt{|\alpha_\dCS|}$ [km] & --- &  --- &  $10^8$~\cite{alihaimoud-chen,kent-CSBH}\\ 
\hline 
\multirow{2}{*}{Vector Field Activation}  & \multirow{2}{*}{SEP, LI} & \multirow{2}{*}{$0$}& \multirow{2}{*}{$\mathbf{7.2 \times 10^{-3}}$} & \multirow{2}{*}{$\mathbf{3.4 \times 10^{-3}}$} & $(c_+,c_-)$ & $\mathbf{(0.9,2.1)}$ & $\mathbf{(0.8,1.1)}$ & $(0.03,0.003)$~\cite{Yagi:2013qpa,Yagi:2013ava}\\ 
  & & &  & & $(\beta_\KG,\lambda_\KG)$ & $\mathbf{(0.42,-)}$ & $\mathbf{(0.40,-)}$ & $(0.005,0.1)$~\cite{Yagi:2013qpa,Yagi:2013ava}\\ 
\hline 
Extra Dimensions  & 4D & \multirow{1}{*}{$-4$}& \multirow{1}{*}{$\mathbf{9.1 \times 10^{-9}}$}& \multirow{1}{*}{$\mathbf{9.1 \times 10^{-11}}$} & $\ell$ [$\mu$m] & $\mathbf{5.4 \times 10^{10}}$ & $\mathbf{2.0 \times 10^{9}}$ & 10--10$^3$~\cite{Johannsen:2008tm,johannsen2,Adelberger:2006dh,psaltis-RS,gnedin}\\
\hline
Time-Varying $G$  & SEP & \multirow{1}{*}{$-4$} &  \multirow{1}{*}{$\mathbf{9.1 \times 10^{-9}}$} &\multirow{1}{*}{$\mathbf{9.1 \times 10^{-11}}$} & $|\dot G|$ [10$^{-12}$/yr] & $\mathbf{5.4 \times 10^{18}}$ & $\mathbf{1.7 \times 10^{17}}$ & 0.1--1~\cite{Manchester:2015mda,2011Icar..211..401K,Hofmann,Copi:2003xd,Bambi:2005fi}\\ 
\hline \hline
\multirow{1}{*}{Massive graviton}  & \multirow{1}{*}{$m_g=0$} & \multirow{1}{*}{$+1$} &\multirow{1}{*}{$\mathbf{1.3 \times 10^{-1}}$} &  \multirow{1}{*}{$\mathbf{8.9 \times 10^{-2}}$}  & \multirow{1}{*}{$m_g$ [eV]} & $\mathbf{10^{-22}}$~\cite{TheLIGOScientific:2016src} & $\mathbf{10^{-22}}$~\cite{TheLIGOScientific:2016pea} & \multirow{1}{*}{$10^{-29}$--$10^{-18}$~\cite{talmadge,sutton,goldhaber,Hare:1973px,Brito:2013wya}} \\
\hline
Mod. Disp. Rel.  & \multirow{2}{*}{LI} &  \multirow{2}{*}{$+4.75$} &   \multirow{2}{*}{$\mathbf{1.1 \times 10^{2}}$} &   \multirow{2}{*}{$\mathbf{2.6 \times 10^{2}}$}  & $E_*^{-1}$ [eV$^{-1}$] (time) &  \multirow{1}{*}{$\mathbf{5.8 \times 10^{-27}}$} &  \multirow{1}{*}{$\mathbf{3.3 \times 10^{-26}}$} & ---\\ 
(\emph{Multifractional})  &  &   &   &    &$E_*^{-1}$ [eV$^{-1}$] (space) & $\mathbf{1.0 \times 10^{-26}}$ & $\mathbf{5.7 \times 10^{-26}}$ & $3.9 \times 10^{-53}$~\cite{Kiyota:2015dla}\\ 
\hline
Mod. Disp. Rel. & \multirow{2}{*}{LI} &  \multirow{2}{*}{$+5.5$} &   \multirow{2}{*}{$\mathbf{1.4 \times 10^{2}}$}  &   \multirow{2}{*}{$\mathbf{4.3 \times 10^{2}}$}  & $\eta_{\rm dsrt}/L_{\mrm{Pl}} >0$ &  \multirow{2}{*}{$\mathbf{1.3 \times 10^{22}}$}  &  \multirow{2}{*}{$\mathbf{3.8 \times 10^{22}}$} & ---\\ 
(\emph{Modified Special Rel.})  &  &   &    &  &$\eta_{\rm dsrt}/L_{\mrm{Pl}} <0$  &  &  & $2.1 \times 10^{-7}$~\cite{Kiyota:2015dla}\\ 
\hline
Mod. Disp. Rel.    & \multirow{2}{*}{4D} &   \multirow{2}{*}{$+7$} &  \multirow{2}{*}{$\mathbf{5.3 \times 10^{2}}$}  &  \multirow{2}{*}{$\mathbf{2.4 \times 10^{3}}$} & $\alpha_{\rm edt}/L_{\mrm{Pl}}^2 > 0$  &  \multirow{2}{*}{$\mathbf{5.5 \times 10^{62}}$} &  \multirow{2}{*}{$\mathbf{2.5 \times 10^{63}}$} &  $2.7 \times 10^{2}$~\cite{Kiyota:2015dla}\\ 
(\emph{Extra Dim.}) &  &    &   &  & $\alpha_{\rm edt}/L_{\mrm{Pl}}^2 <0$  & & & --- \\ 
\hline
   & \multirow{6}{*}{LI} &   \multirow{2}{*}{$+4$} &  \multirow{2}{*}{---}  &  \multirow{2}{*}{---} & $\mathring{k}_{(I)}^{(4)} > 0$  &  \multirow{1}{*}{---} &  \multirow{1}{*}{---} & $6.1 \times 10^{-17}$~\cite{Kostelecky:2010ze,Kiyota:2015dla}\\ 
 &  &    &   &  & $\mathring{k}_{(I)}^{(4)} <0$ & $\mathbf{0.64}$ & $\mathbf{19}$ & --- \\ 
 \multirow{1}{*}{Mod. Disp. Rel.}  &  &   \multirow{2}{*}{$+5.5$} &  \multirow{2}{*}{$\mathbf{1.4 \times 10^{2}}$}  &  \multirow{2}{*}{$\mathbf{4.3 \times 10^{2}}$} & $\mathring{k}_{(V)}^{(5)} > 0$ [cm] &  \multirow{2}{*}{$\mathbf{1.7 \times 10^{-12}}$~\cite{Kostelecky:2016kfm}} &  \multirow{2}{*}{$\mathbf{3.1 \times 10^{-11}}$} & $1.7 \times 10^{-40}$~\cite{Kostelecky:2010ze,Kiyota:2015dla}\\ 
\multirow{1}{*}{(\emph{Standard Model Ext.})}  &  &    &   &  & $\mathring{k}_{(V)}^{(5)} <0$ [cm] & & & --- \\ 
 & &   \multirow{2}{*}{$+7$} &  \multirow{2}{*}{$\mathbf{5.3 \times 10^{2}}$}  &  \multirow{2}{*}{$\mathbf{2.4 \times 10^{3}}$} & $\mathring{k}_{(I)}^{(6)} > 0$ [cm$^2$] &  \multirow{2}{*}{$\mathbf{7.2 \times 10^{-4}}$} &  \multirow{2}{*}{$\mathbf{3.3 \times 10^{-3}}$} & $3.5 \times 10^{-64}$~\cite{Kostelecky:2010ze,Kiyota:2015dla}\\ 
 &  &    &   &  & $\mathring{k}_{(I)}^{(6)} <0$ [cm$^2$] & & & --- \\ 
\hline
Mod. Disp. Rel.    & \multirow{2}{*}{LI} &   \multirow{2}{*}{$+7$} &  \multirow{2}{*}{$\mathbf{5.3 \times 10^{2}}$}  &  \multirow{2}{*}{$\mathbf{2.4 \times 10^{3}}$} & \multirow{2}{*}{$\kappa_{\mrm{hl}}^4 \mu_{\mrm{hl}}^2$ [1/eV$^2$]} &  \multirow{2}{*}{$\mathbf{1.5 \times 10^{6}}$} &  \multirow{2}{*}{$\mathbf{6.9 \times 10^{6}}$} & \multirow{2}{*}{---}\\ 
(\emph{Ho\v rava-Lifshitz}) &  &    &   &  & & & &\\ 
\hline
Mod. Disp. Rel.    & \multirow{2}{*}{LI} & \multirow{2}{*}{$+4$} & \multirow{2}{*}{---}   & \multirow{2}{*}{---}   & \multirow{2}{*}{$c_+$} & \multirow{2}{*}{$\mathbf{0.7}$~\cite{Blas:2016qmn}} & \multirow{2}{*}{$\mathbf{0.998}$} & \multirow{2}{*}{$0.03$~\cite{Yagi:2013qpa,Yagi:2013ava}}\\ 
(\emph{Lorentz Violation}) & & & & & & & \\   
\noalign{\smallskip}
\hline
\hline
\end{tabular}
\end{centering}
\caption{
Theoretical mechanisms (first column) that arise in modified theories of gravity and how they violate fundamental pillars of GR (second column). The numbers in boldface show the approximate, 90\%-confidence upper bounds placed by GW150914 and GW151226 on ppE parameters (fourth and fifth columns) and on parameters (seventh and eighth columns) representing specific theoretical mechanisms realized in a set of gravitational theories (sixth column) that enter at different PN order (third column) relative to GR; prior constraints on these example theories are shown in the last column. The top section of the table shows constraints on modifications to GW generation, while the bottom corresponds to constraints on GW propagation. Constraints on scalar field activation, which violates the strong equivalence principle (SEP) or gravitational parity invariance (PI), are exemplified by realizations in Einstein-dilaton Gauss-Bonnet (EdGB) gravity, dynamical Chern-Simons (dCS) gravity and scalar-tensor theories, controlled by the coupling constants $\alpha_\EDGB$, $\alpha_\dCS$ and the scalar field growth rate $\dot \phi$ respectively. The GW events cannot constrain these theories from the leading PN order correction to the waveform phase within the small coupling approximation, which assumes that the deformation away from GR is small. Constraints on the activation of vector fields, which violates LI and SEP by breaking LPI and LLI, are exemplified in Einstein-\AE ther (EA) theory with dimensionless coupling constants $(c_+,c_-)$ and khronometric theory with $(\beta_\KG, \lambda_\KG)$. A constraint on BH mass leakage into extra dimensions is exemplified by a realization of a RS-II braneworld model, where $\ell$ is the size of the large extra dimension. Constraints on the time variation of the gravitational constant, which also violates SEP by breaking LPI, are characterized by limits on $\dot{G}$.  Constraints on massive gravity are exemplified by kinematical constructions that modify the GW dispersion relation, whose magnitude is controlled by the graviton mass $m_g$. We also present constraints on the modified dispersion relation of the graviton in five different well-motivated cases, with some of them normalized by the Planck length $L_\mrm{Pl}$. For comparison, we present the constraint on violations of gravitational LI from the arrival time delay of GWs between Hanford and Livingston detectors (last row, seventh and eighth columns). 
 }
\label{tab:summary2}
\end{table*}
\endgroup
}

\vspace{0.2cm}
\fbox{\parbox{8cm}{
{\bf GW150914 and GW151226 constrain a plethora of emission mechanisms beyond GR radiation reaction.}
}}
\vspace{0.2cm}

The first half of Table~\ref{tab:summary2} (up to the double line) presents a summary of the GW emission or generation mechanisms that can be constrained. In particular, the GW events constrain the presence of (i) dipole radiation in the signal due to e.g.~the activation or growth of a scalar field, (ii) BH mass leakage due to large extra dimensions, (iii) a time-varying gravitational constant due to e.g.~the existence of a time-varying scalar field, and (iv) Lorentz-violating effects in the production of GWs. We also derived bounds on the sudden activation of a scalar field, as predicted e.g.~through dynamical scalarization in certain scalar-tensor theories (for non-vacuum spacetimes)~\cite{ST1,Palenzuela:2013hsa,Shibata:2013pra,Taniguchi:2014fqa,sennett,Sampson:2014qqa}, which can be constrained most strongly when the sudden scalarization occurs in band. Taking a more agnostic approach to particular causes of deviations, GW150914 and GW151226 place constraints on generic deviations from the GR prediction of the evolution of the binding energy and radiated flux during a binary merger. These constraints can be used to bound generic scalar hair around BHs~\cite{Stein:2013wza}. 

We arrive at these conclusions through a Fisher parameter estimation study, which we show agrees with the Bayesian analysis of~\cite{TheLIGOScientific:2016pea,TheLIGOScientific:2016src} to 30--50\%, wherever we can compare our results (i.e.~for corrections at positive PN orders). We use the same (parametrically-deformed) inspiral-merger-ringdown waveform model (so-called gIMR) employed by the LVC~\cite{TheLIGOScientific:2016pea,TheLIGOScientific:2016src}, but without precession. As there seems to be some confusion in the literature about the scope of these models, we mathematically show in Sec.~\ref{sec:ppE} that gIMR is a subset of the parameterized post-Einsteinian (ppE) framework of~\cite{PPE} when the baseline GR waveform is taken to be the phenomenological waveform model of~\cite{Ajith:2007qp,Ajith:2009bn,Santamaria:2010yb,Husa:2015iqa,Khan:2015jqa}. 

We also illustrate that ignorance of the higher-order PN corrections to the \emph{inspiral} waveform in modified gravity does not necessarily weaken the constraints and inferences on the leading-order physics obtained with a model that only uses the corresponding leading-order PN deformation\footnote{The inclusion of modifications in the merger phase would also not weaken the constraints presented here; more likely they would improve them somewhat.}. Specifically, using BD theory~\cite{Fierz:1956zz,Jordan:1959eg,Brans:1961sx,zaglauer,Yunes:2011aa,Mirshekari:2013vb} as the test-case, we estimate that a higher PN order modified waveform model would only correct the constraints by $\mathcal{O}(10\%)$ at most\footnote{Such a correction would be important if one is attempting to characterize a measured anomaly, but is less of an issue when constraining its existence, as explained e.g.~in~\cite{Sampson:2013jpa}.}. On the other hand, unlike in BD theory, if the system parameters conspire to suppress the coefficients of the leading PN order term, the next-to-leading order term will become dominant and the bounds on non-GR effects modeled with the leading PN order correction become \emph{conservative}, i.e.~including higher PN order corrections would make the bounds stronger. Indeed, such a suppression of the leading PN effect is precisely why GW150914 and GW151226 cannot place meaningful constraints on EdGB gravity. We present this material in Appendix~\ref{app:effect-of-hi-PN}.  

The fact that the gIMR model is a subset of the ppE framework allows us to use the many years of work on ppE and modified gravity theory to draw theoretical physics implications from the absence of ppE-like anomalies in the GW150914 and GW151226 data. In particular, each ppE exponent (or equivalently, the relative PN order shown in the third column of Table~\ref{tab:summary2}) describing how the frequency response of a chirping binary is altered can be related to a set of physical mechanisms that are responsible for the effect. This allows us to test some of the fundamental pillars of GR, which typically are related to tests of the SEP~\cite{Will:2014kxa}: 
\begin{itemize}[itemsep=0.1pt, topsep=5pt, partopsep=0pt]
\item[(A)] The trajectories of freely-falling test bodies, including self-gravitating ones, are independent of their internal structure and composition (the weak equivalence principle (WEP) extended to self-gravitating bodies).
\item[(B)] Results of any local experiment, including gravitational experiments, are independent of when and where they are performed (local position invariance (LPI)), and of the velocity of the experimental apparatus (local Lorentz invariance (LLI)).
\end{itemize}
Dipole gravitational radiation due to the activation of additional dynamical fields can violate item (A). Time variation of Newton's gravitational constant would violate LPI. The presence of Lorentz-violating effects due to, for example the presence of dynamical vector fields, breaks both LPI and LLI. GW150914 and GW151226 are therefore much more than a probe of the structure of spacetime of a binary BH merger; it also allows for the verification of some of the most important pillars of Einstein's theory.  

\vspace{0.6cm}
\fbox{\parbox{8cm}{
{\bf GW150914 and GW151226 constrain a number of theoretical mechanisms that modify GW propagation.}
}}
\vspace{0.2cm}

The second half of Table~\ref{tab:summary2} (below the double line) presents a summary of the propagation mechanisms that can be constrained. The LVC observations not only constrains the mass of the graviton~\cite{TheLIGOScientific:2016pea,TheLIGOScientific:2016src}, but they more generically constrain the dispersion relation of GWs, both super- and subluminal GW propagation, and the presence of Lorentz violation in their propagation.

As with effects active during the generation of GWs, we arrive at these conclusions with a Fisher analysis, which we have also checked is consistent with the Bayesian study of~\cite{TheLIGOScientific:2016pea,TheLIGOScientific:2016src} wherever possible. For example, we have verified that the Fisher constraint on the graviton mass $m_g$ with a simple massive graviton dispersion relation is consistent with the Bayesian bound of~\cite{TheLIGOScientific:2016pea,TheLIGOScientific:2016src}, both of which are a few times more stringent than the current Solar System bound~\cite{talmadge}. All of the inferences on the propagation of GWs from GW150914 and GW151226 come from information on the phasing of the GW, a much more powerful tool than information derived solely from the difference in GW time of arrival between the Hanford and Livingston detectors. In particular, the bound presented here and in~\cite{TheLIGOScientific:2016pea,TheLIGOScientific:2016src} is twenty orders of magnitude stronger than that based only on a time delay argument~\cite{Blas:2016qmn}, except when the graviton propagation speed acquires a frequency-independent correction (since then the GW phase modification becomes degenerate with the time of coalescence).

However, unlike in the GW generation case, the constraints on GW propagation mechanisms are often significantly stronger than other current constraints from binary pulsar and Solar System observations. In particular, the GW constraints on ($\mathbb{A}>0$) superluminal propagation and on sub- or superluminal GW propagation entering at low PN order [see Fig.~\ref{fig:A-constraint}] are the best found to date. The GW constraints on the mass of the graviton are also the best to date, except for constraints coming from observations of galaxy clusters~\cite{goldhaber}. Although theories that predict modifications to GW propagation also typically modify the GW generation mechanism, we show here that the former typically dominate the latter. This is because modifications to GW propagation accumulate over the propagation time (i.e.~the distance), while modifications to GW generation accumulate while the system is generating GWs in band. The latter will never be comparable to the former for aLIGO sources, unless the binary happens to coalesce in the Solar System. 

\vspace{0.6cm}
\fbox{\parbox{8cm}{
{\bf{GW150914 allows for inferences to be made regarding the validity of the Kerr hypothesis, and likewise it constrains properties of exotic compact object alternatives to Kerr BHs.}}
}}
\vspace{0.2cm}

GW150914 was a \emph{golden} event~\cite{Hughes:2004vw}, which allows a measurement of the amount of energy and angular momentum carried away by GWs during coalescence~\cite{TheLIGOScientific:2016src}. This information can, in turn, be used to infer properties of the geometry of the compact objects, such as the location of the innermost-stable circular orbit (ISCO) of the remnant (for a ``test particle'' moving in such a spacetime). Such an inference is made by using the relation between the spin of the final BH, the spin of the individual BHs before merger and the location of the ISCO of the remnant in an effective-one-body treatment as a null test, which was established in GR using numerical relativity simulations~\cite{Buonanno:2007sv}. Inferences about the location of the ISCO can then be used to constrain string-inspired BH solutions~\cite{Ayzenberg:2016ynm,kent-CSBH} and parametrically deformed Kerr metrics~\cite{glampedakis}. The observed energy loss could also be used to limit the amount of exotic ``hair'' (e.g.~\cite{Herdeiro:2014goa,Herdeiro:2016tmi}) the BHs in this event have. If a sizable fraction of the initial mass of each BH is attributable to such hair, presumably a correspondingly large fraction could be radiated during merger (exactly how much would need to be calculated via numerical simulations). Although the SNR of GW150914 is not high enough to allow for interesting constraints on the above spacetimes, future louder signals would allow for tighter bounds via this analysis. 

As discussed earlier in the Introduction, the dramatic drop in the observed signal within $\sim 4$ milliseconds after reaching peak amplitude is consistent with the rapid hair-loss experienced by the Kerr remnant in GR. On the flip side, this observation places a severe constraint on the properties of hypothesized exotic matter alternatives to Kerr BHs. Assuming the collision excites matter oscillations in the remnant that emit observable GWs, we can place bounds on an effective viscosity that the exotic matter must have to be consistent with the observed signal. Alternatively, lack of observed damped normal-mode oscillations of such exotic matter can be used to place restrictions on the initial amplitude and damping time-scales of these putative modes.

\vspace{0.6cm}
\fbox{\parbox{8cm}{
{\bf{If the Fermi Gamma-ray Burst Monitor (GBM) signal is an actual counterpart to GW150914, this observation places more stringent constraints on GW propagation mechanisms than GW150914 alone.}}
}}
\vspace{0.2cm}

If the GBM signal~\cite{Connaughton:2016umz}  was a short GRB counterpart to GW150914, then the speed of GWs could be constrained in a model-independent fashion. The strength of this constraint depends on the intrinsic time delay between the gamma-ray and GW emission~\cite{Nishizawa:2014zna}, which is currently uncertain due to ignorance of the gamma-ray emission mechanism. If one assumes that the Fermi event was a prompt emission counterpart to GW150914 and GWs do not propagate subluminally, the speed of GWs can be constrained to be equal to the speed of light to one part in $10^{17}$~\cite{Ellis:2016rrr,Li:2016iww,Collett:2016dey}. This, in turn, would impose dramatic constraints on gravitational Lorentz violation~\cite{Jacobson:2000xp,Jacobson:2008aj,Blas:2009qj,Blas:2010hb}, restricting the latter ten orders of magnitude more stringently than current binary pulsar bounds~\cite{Yagi:2013qpa,Yagi:2013ava}, as predicted in~\cite{Hansen:2014ewa}. However these conclusions are premature at this stage, given the low-confidence of the GBM event.

The remainder of this paper presents the details of the calculations that led to the above conclusions. All throughout, we follow the conventions of Misner, Thorne and Wheeler~\cite{MTW}, and unless otherwise stated use geometric units where $G = 1 = c$. In particular, note that we do not employ Planck units, and thus $\hbar \neq 1$. Conversion between geometric units and SI units can be achieved by noting that $1 M_{\odot} = 1.476 \; {\rm{km}} = 4.925 \times 10^{-6} \; {\rm{Hz}}$.

\begin{figure*}[htb]
\begin{center}
\includegraphics[width=\columnwidth,clip=true]{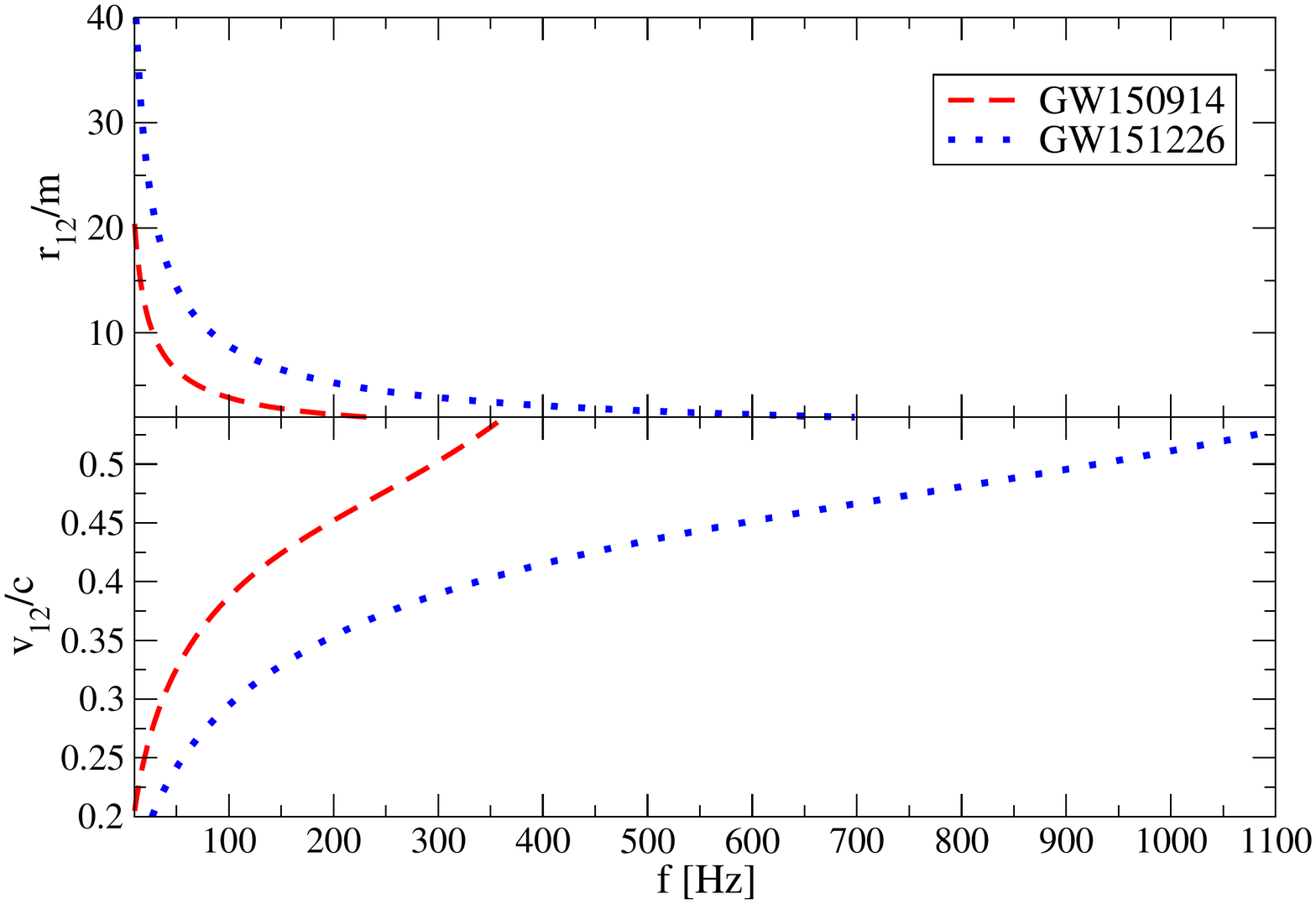}  \quad 
\includegraphics[width=\columnwidth,clip=true]{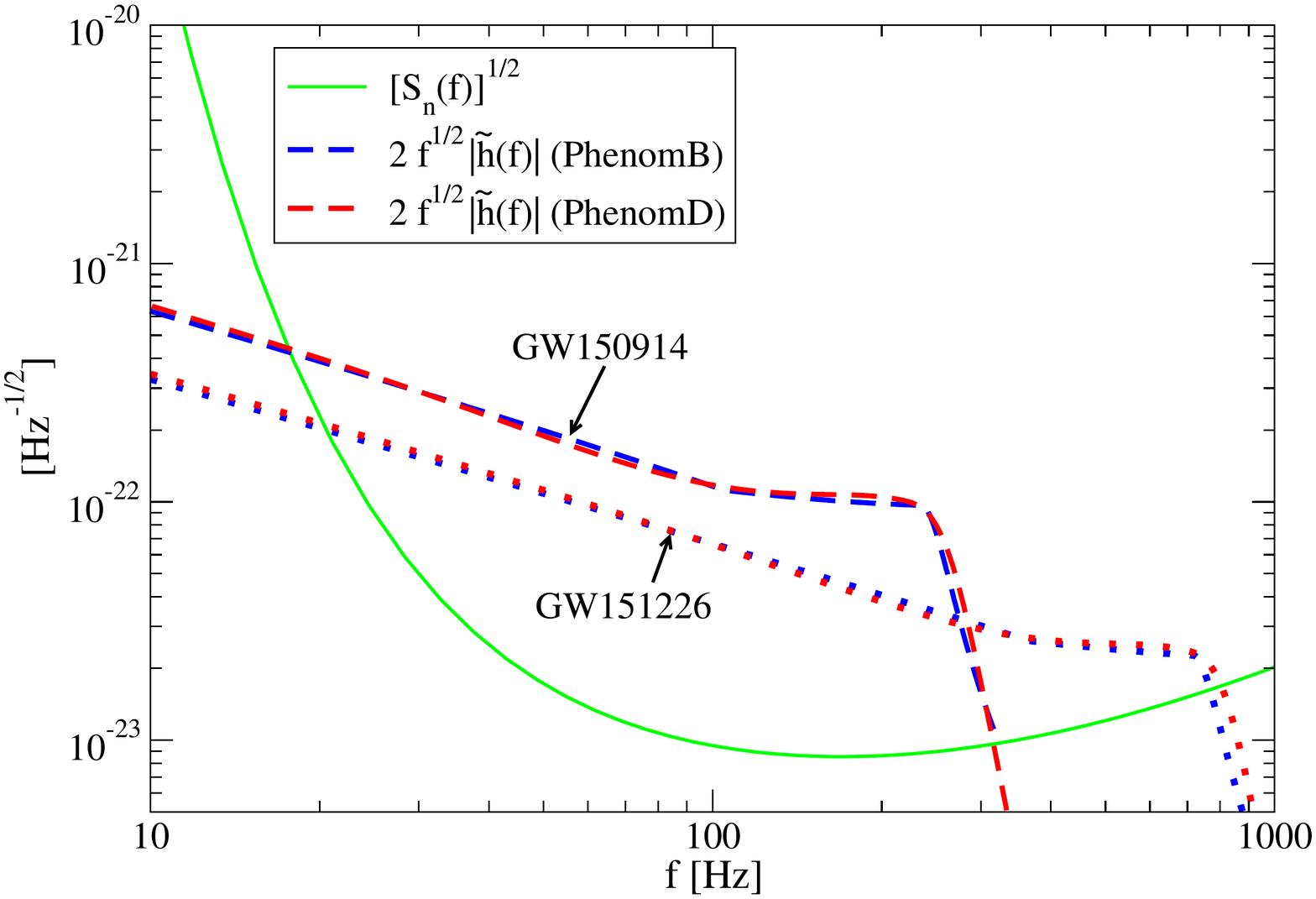}  
\caption{\label{fig:randvoff-hoff} (Color online) (Left) Third-order PN estimates of the orbital separation (top) and velocity (bottom) as a function of the GW frequency (see also Fig.~2 of~\cite{Abbott:2016blz}). (Right) An estimate of the square root of the spectral noise density curve of aLIGO when GW150914 was detected (as interpolated from the data made publicly available by the LVC~\cite{noise-data} as described in Appendix~\ref{app:noise-fit}), and two models (PhenomB~\cite{Ajith:2009bn} and PhenomD~\cite{Husa:2015iqa,Khan:2015jqa}) of the amplitude of the GW Fourier spectrum of GW150914 (GW151226) multiplied by twice the square root of the frequency, and scaled to SNR 24 (13). 
} 
\end{center}
\end{figure*}

\section{BH Coalescence as a Probe of Extreme Gravity} 
\label{sec:testsofgravwithBHS}

This section begins by describing the different phases of the GW events in detail, and how they can probe extreme gravity. We then describe the GW models used in GR to describe the phases of coalescence, as well as the parametric models that capture deformations from GR. When discussing the latter, we show that the parametrically-deformed model used by the LVC in~\cite{TheLIGOScientific:2016src} is an implementation of the ppE framework~\cite{PPE} for a particular GR model. This implies that one can work in the ppE framework to interpret constraints on departures from GR as constraints on different physical mechanisms, whose mappings are summarized at the end of this section. 

\subsection{Description of Coalescence}  
\label{sec:phase-definition}

The coalescence of a comparable mass binary system can be roughly divided into 3 phases~\cite{Flanagan:1997sx,2007arXiv0710.1338P,Centrella:2010zf}: 
\begin{itemize}[itemsep=0.1pt, topsep=5pt, partopsep=0pt]
\item{\bf{Inspiral.}} The compact objects are well-separated with respect to the total mass ($r_{12}/m \gg 1$), the characteristic orbital velocity is much smaller than the speed of light ($v/c \ll 1$), and the inspiral rate is slow relative to the timescale of the orbit.
\item{\bf{Plunge and Merger.}} The compact objects are so close to each other that GW emission has reached a level at which the inspiral timescale is comparable to the orbital timescale. The evolution of the orbit then transitions from an inspiral to a plunge at velocities approaching the speed of light, and the two objects coalesce.
\item{\bf{Ringdown.}} The highly-distorted remnant formed after merger oscillates, radiating away any deformations and relaxes to a stationary state.  
\end{itemize}
Even though this classification is clean in concept, in reality the transition from one phase to another is not abrupt, and there is no stark demarcation in the waveform when one ends and the next begins. However, keeping this picture in mind is helpful to better understand how the two aLIGO detections, and GW150914 in particular, informs us about extreme gravity.

\begin{figure}[htb]
\begin{center}
\includegraphics[width=\columnwidth,clip=true]{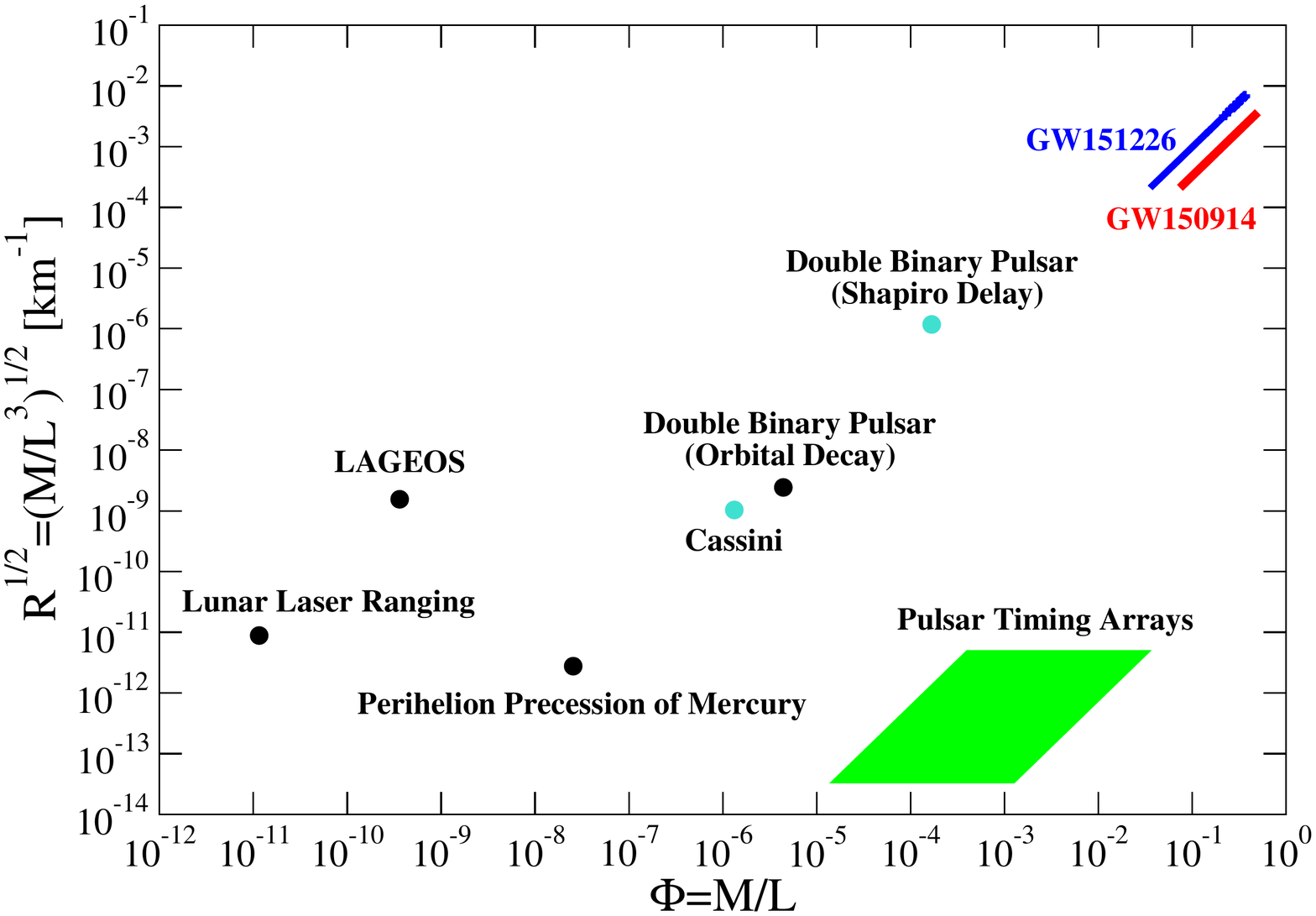}
\caption{\label{fig:phase-diagram-new} 
(Color online) Schematic diagram of the curvature-potential phase space sampled by various experiments that test GR. The vertical axis shows the inverse of the characteristic curvature length scale, while the horizontal axis shows the characteristic gravitational potential, based on Table~\ref{table:mass-length}. GW150914 and GW151226 sample a regime where the curvature and the potential are both simultaneously large and dynamical, indicated here by the finite range the curves sweep in the figure. The finite area of pulsar timing arrays is due to the range in the GW frequency and the total mass of supermassive BH binaries that such arrays may detect in the future. Figure~\ref{fig:phase-diagram-new-time-scale} is a companion plot that illustrates the dynamical aspects of gravity probed by these experiments; the lighter (blue) dots here are to indicate that the Shapiro time delay from binary pulsars and the Cassini satellite do not give information on the dynamical regime. 
}
\end{center}
\end{figure}

\begin{figure*}[htb]
\begin{center}
\includegraphics[width=\columnwidth,clip=true]{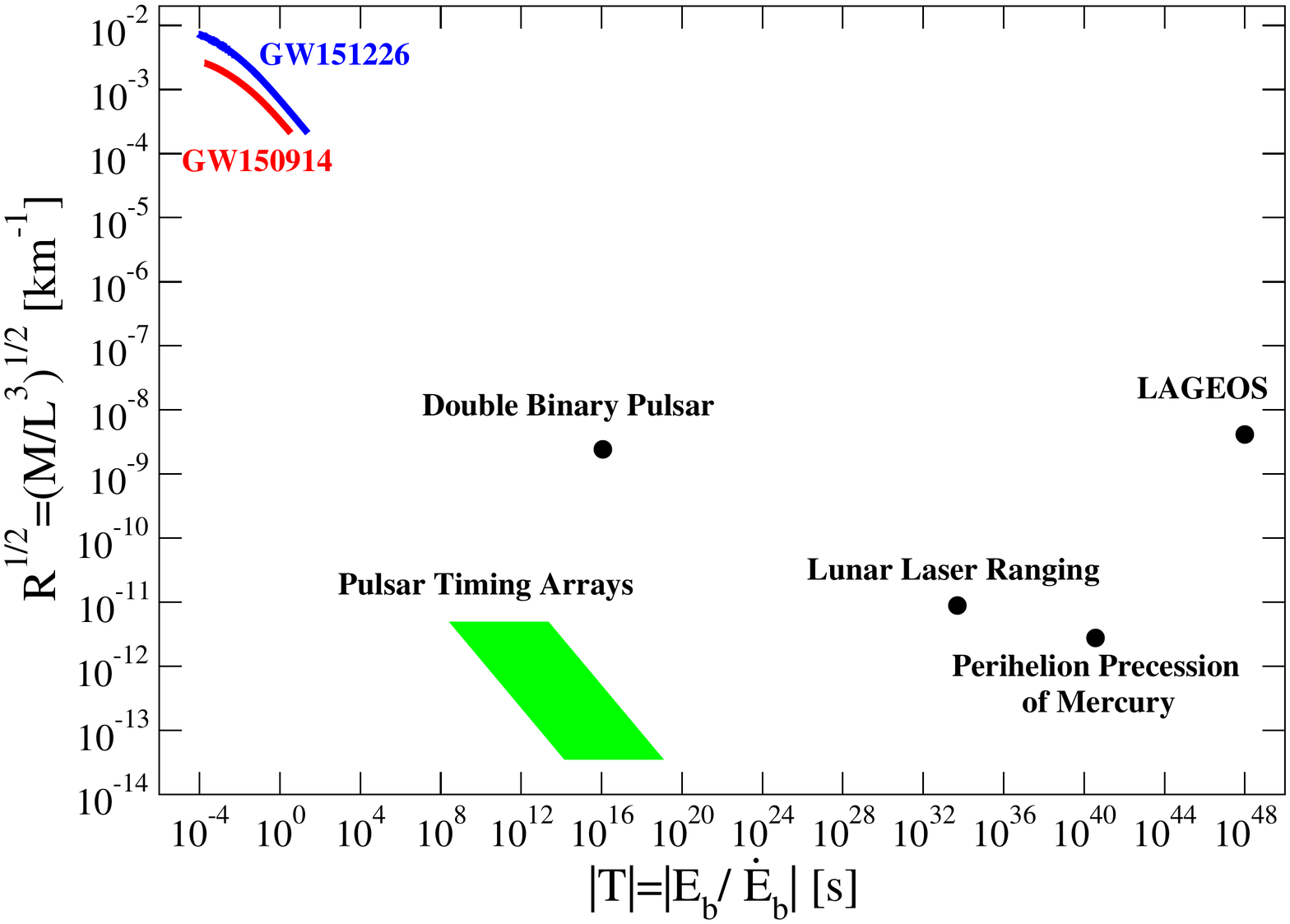} \quad 
\includegraphics[width=\columnwidth,clip=true]{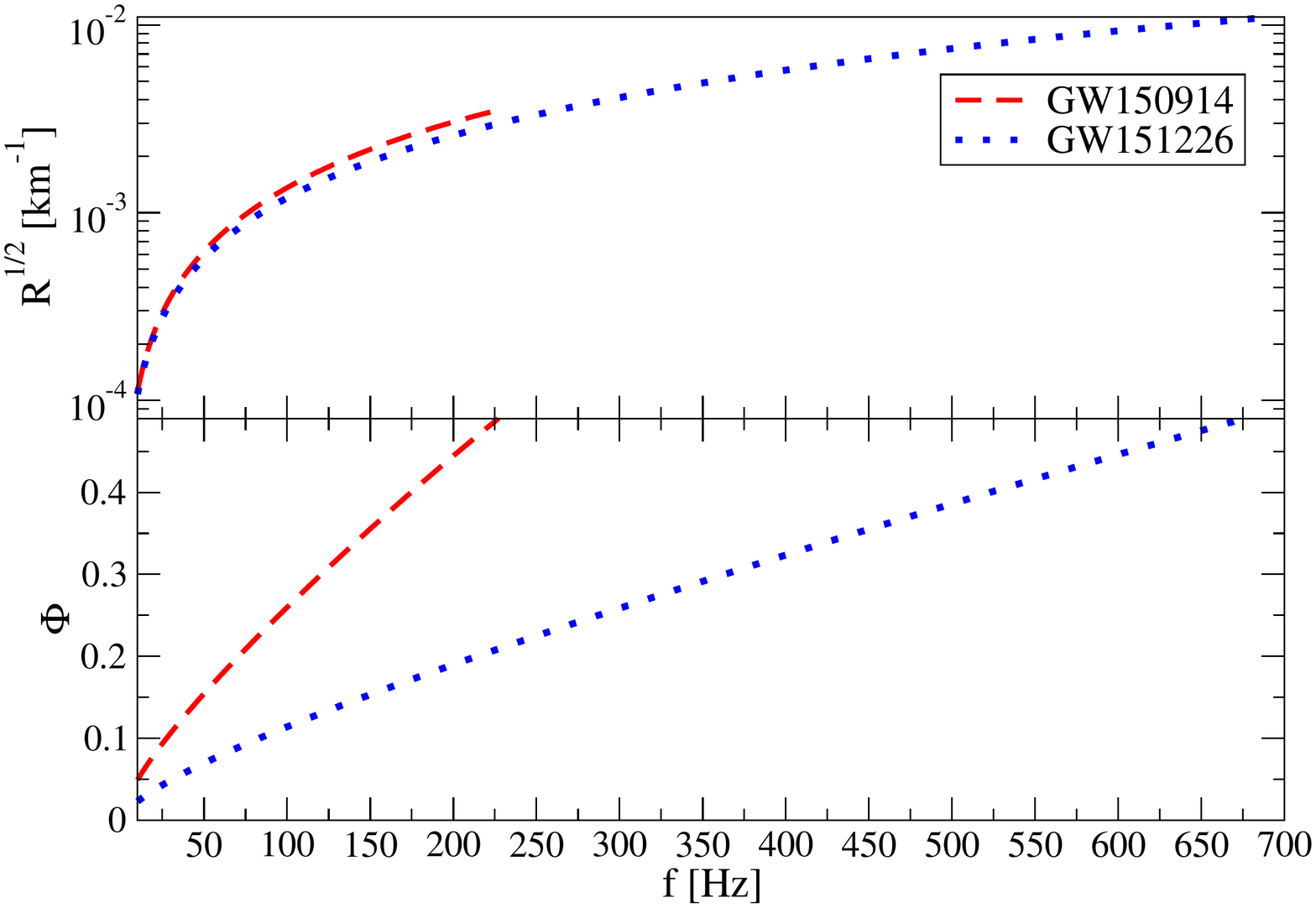}
\caption{\label{fig:phase-diagram-new-time-scale} (Color online) (Left) Schematic diagram of the curvature-radiation reaction time-scale phase space sampled by relevant experiments shown in Fig.~\ref{fig:phase-diagram-new}. As is evident, GW150914 and GW151226 sample a regime of dynamic gravity where the radiation-reaction timescale is the shortest by many orders of magnitude. (Right) Characteristic curvature and strength of the Newtonian gravitational potential as a function of GW frequency. 
}
\end{center}
\end{figure*}

Both events entered the aLIGO sensitivity band when the compact objects were already quite close. The left panel of Fig.~\ref{fig:randvoff-hoff} shows the orbital velocity as a function of GW frequency, estimated here via $v_{12} = (\pi m f)^{1/3}$, with $m$ the total mass and $f$ the GW frequency. For example, at 10 Hz, the two compact objects that produced GW150194 were already traveling at $v_{12}/c \sim 0.2$ and with orbital separation\footnote{The mapping between orbital separation and frequency $r_{12} = r_{12}(f)$ is gauge dependent, but for estimation purposes we use the 3PN accurate relation reviewed in~\cite{blanchet-review}.} of about $r_{12} \sim 20 m$, or approximately $1960$ km. When the frequency reached $132$ Hz, the binary's orbital velocity was roughly $v_{12}/c \sim 0.4$ and the orbital separation approached $r_{12} \sim 3 m \sim 300$ km; the latter is close to the light-ring of a test particle in the Schwarzschild spacetime of a BH with mass equal to the binary's total mass. Beyond $132$ Hz, the binary rapidly plunged, merging at approximately $230$ Hz as the  orbital separation decreased to $r_{12} \sim 2 m$, where two Schwarzschild BHs would have ``touched''. This frequency also roughly coincides with the start of the ringdown, as the Fourier amplitude of the GW150194 signal shows a break at around this frequency (see the right panel of Fig.~\ref{fig:randvoff-hoff}). Due to its smaller component masses, the signal for GW151226 entered the aLIGO band at 10Hz with $v_{12}/c \sim 0.15$ and a corresponding separation of $r_{12} \sim 40 m = 1280 {\rm{km}}$. The separation reached $r_{12} \sim 3 m \sim 100$km at $f=410$Hz, with merger happening roughly at roughly 800Hz.

The binary BH coalescences that generated these two GW events are solidly in the extreme gravity regime. Further ways to quantify this are to compute the characteristic curvature ${\cal{R}} = M/{\cal{L}}^{3}$ and the characteristic gravitational potential $\Phi = M/{\cal{L}}$, where M and ${\cal{L}}$ are the characteristic mass and size of the system. Following~\cite{psaltis-review,Baker:2014zba}, Fig.~\ref{fig:phase-diagram-new} shows these quantities evaluated from $f=20$Hz to merger for events GW150914 and GW151226, taking ${\cal{L}} = r_{12}$ and $M=m$. For comparison, we also show the curvature and the gravitational potential for the LAGEOS~\cite{LAGEOS} and Cassini~\cite{Bertotti:2003rm} satellites, the Earth-Moon System used in lunar laser ranging~\cite{Williams:2004qba}, the Mercury-Sun system used in perihelion precession observations, pulsar timing observations~\cite{Hobbs:2009yy}, and the double binary pulsar~\cite{burgay,lyne,kramer-double-pulsar}; the mass and length scale of each system is summarized in Table~\ref{table:mass-length}. 

Both GW150914 and GW15226 land in the far top right corner of the phase space of Fig.~\ref{fig:phase-diagram-new}, precisely where gravity is strong. What is not clearly depicted in this figure is how \emph{dynamical} the gravitational field for each observation is. Some of this can be inferred from the fact that the GW events in Fig.~\ref{fig:phase-diagram-new} are shown as lines instead of points. A better illustration of the time variation of spacetime is shown in the left panel of Fig.~\ref{fig:phase-diagram-new-time-scale}, which is similar to Fig.~\ref{fig:phase-diagram-new}, except that the abscissa is now the radiation-reaction time-scale sampled by each observation. We model this via $|T| = |E_{b}/\dot{E}_{b}|$, where $E_{b}$ is the characteristic gravitational binding energy and $\dot{E}_{b}$ is the rate of change of this energy, which for a binary system we approximate as the GW energy flux at spatial infinity, i.e.~$|T| = (5/64) (m/\eta) v_{12}^{-8}$, where $\eta = m_{1} m_{2}/m^{2}$ is the symmetric mass ratio. For a binary system, this quantity is exactly the same as $\Phi/\dot{\Phi}$ and (up to factors of order unity) ${\cal{R}}/\dot{{\cal{R}}}$. Thus, $T$ is a measure of how long it takes the system, and in particular the gravitational field and the curvature, to change appreciably.  GW150914 and GW151226 land in the top left region of the left panel of Fig.~\ref{fig:phase-diagram-new-time-scale}, at least four orders of magnitude away from the double binary pulsar. The Shapiro time delay and Cassini observation do not appear in this figure, as they do not sample the dynamical sector of GR. For the GW events, how rapidly the curvature and the potential sweep through the detector's sensitivity band is shown on the right panel of Fig.~\ref{fig:phase-diagram-new-time-scale}. 
{
\newcommand{\minitab}[2][l]{\begin{tabular}{#1}#2\end{tabular}}
\renewcommand{\arraystretch}{1.2}
\begingroup
\squeezetable
\begin{table}[tb]
\begin{centering}
\begin{tabular}{c|cc}
\hline
\hline
\noalign{\smallskip}
 & $M$  & ${\cal{L}}$ \\ \hline
GW150914~\cite{Abbott:2016blz,TheLIGOScientific:2016wfe,TheLIGOScientific:2016pea} & $65.3 M_\odot$  & $190$--$1300$km \\ 
GW151226~\cite{Abbott:2016nmj,TheLIGOScientific:2016pea} & $21.7 M_\odot$  & $64$--$900$km \\ 
Pulsar Timing Arrays~\cite{Lommen:2015gbz} & $10^6$--$10^9M_\odot$  & $10^{9.6}$--$10^{12}$km \\ 
Bin. Pulsar (Shapiro Delay)~\cite{wex-private} & $1.34M_\odot$  & $1.04 \times 10^4$km \\ 
Bin. Pulsar (Orbital Decay)~\cite{kramer-double-pulsar} & $2.59 M_\odot$  & $8.7 \times 10^5$km \\ 
LAGEOS~\cite{LAGEOS} & $1 M_\oplus$  & $1.9 R_\oplus$ \\ 
Lunar Laser Ranging~\cite{Merkowitz:2010kka} & $1 M_\oplus$  & $3.8 \times 10^5$km \\ 
Cassini~\cite{Bertotti:2003rm} & $1 M_\odot$  & $1.6 R_\odot$ \\ 
Pericenter Precession of Mercury~\cite{TEGP,Will:2014kxa} & $1 M_\odot$  & $5.8 \times 10^7$km \\ 
\noalign{\smallskip}
\hline
\hline
\end{tabular}
\end{centering}
\caption{
The characteristic mass and length scale chosen to compute the characteristic curvature and potential in Fig.~\ref{fig:phase-diagram-new}. For GW150914, GW151226 and pulsar timing arrays, we extract the length scale from the observed frequency via ${\cal{L}} = [M/(\pi f)^2]^{1/3}$, where for the former two we choose $f=20$Hz up to contact, while for the latter we choose $f = 3 \times 10^{-9}$--$5 \times 10^{-7}$Hz. The length scale for the binary pulsar Shapiro delay corresponds to the sum of the minimum impact parameter with an inclination of 89.3$^{\circ}$ ($\sim 9800$km) and the effect of lensing ($\sim 600$km) of the double binary pulsar PSR J0737-3039~\cite{wex-private}.
}
\label{table:mass-length}
\end{table}
\endgroup
}

\subsection{GW Model in GR and outside GR}  
\subsubsection{IMRPhenom Model in GR}
\label{sec:IMRPhenom-GR}

The LVC employed two main waveform models (both calculated within GR) to reconstruct the signal~\cite{TheLIGOScientific:2016wfe,TheLIGOScientific:2016qqj}. One of these, the so-called \emph{IMRPhenom} model~\cite{Ajith:2007qp,Ajith:2009bn,Santamaria:2010yb,Husa:2015iqa,Khan:2015jqa}, was heavily used to validate GR in~\cite{TheLIGOScientific:2016pea,TheLIGOScientific:2016src}. In particular, the LVC employed the most recent IMRPhenom model (PhenomPv2), which is a modified version of PhenomD~\cite{Husa:2015iqa,Khan:2015jqa} that includes precession by rotating a spin-aligned waveform to a precessing frame~\cite{Schmidt:2012rh}. In this paper, we will use the PhenomD model and ignore precession effects\footnote{The LVC was unable to precisely extract the individual spin components of each BH binary prior to coalescence for either event, nor the spin parameter combination that characterizes the amount of precession~\cite{TheLIGOScientific:2016wfe,Abbott:2016izl,Abbott:2016nmj,TheLIGOScientific:2016pea}.}. The differences in the constraints on non-GR effects obtained with an older version of the IMRPhenom model (PhenomB~\cite{Ajith:2009bn}) and PhenomD waveforms are discussed in Appendix~\ref{app:BvsD}. This appendix also provides a rough estimate of the impact of mismodeling error in tests of GR, showing that this error does not affect tests for the modified gravity effects considered here using events GW150914 and GW151226.

Let us then briefly summarize the PhenomD model of~\cite{Khan:2015jqa}. This phenomenological approach models the Fourier transform of the response function as a piecewise function with 3 distinct pieces or phases, where each phase $i$ takes the following form: 
\be
\label{eq:waveform}
\tilde{h}_{i}(f) = A_{i}(f) e^{i \Phi_{i}(f)}\,.
\ee
The three phases that are distinguished are the inspiral, an intermediate phase and the merger-ringdown phase. In the inspiral phase, the waveform is modeled as follows. The amplitude is treated in PN theory, including terms up to 3PN order that are known analytically, and higher-order functionals (up to 4.5PN order) fitted to numerical simulations. In particular, the early-inspiral part of the phase is simply given by
\be
\label{eq:early-insp-IMR}
\Phi_{\EI}(f) = 2 \pi f t_{c} - \phi_{c} - \frac{\pi}{4} + \frac{3}{128 \eta} \left(\pi m f\right)^{-5/3} \sum_{i=0}^{7} \phi_{i} \left(\pi m f\right)^{i/3}\,. 
\ee
The early inspiral parameters $(t_{c},\phi_{c})$ correspond to a constant time and phase offset, and the $\phi_{i}$ coefficients are functions of the component masses and the component spins (see e.g.~Appendix A in~\cite{Khan:2015jqa}). 

\subsubsection{Parametrically Deformed Models}
\label{sec:ppE}
 
The IMRPhenom model is constructed within GR, and thus, it must be generalized in order to account for modified gravity effects. The LVC decided to introduce a generalized IMRPhenom (gIMR) model through the substitution rule
\be
\label{eq:pmapping}
\vec{p} \rightarrow \vec{p} \left(1 + \delta p\right)\,, 
\ee
where $\vec{p}$ denotes parameters for either the early inspiral ($\phi_{i}$), the late inspiral ($\sigma_{i}$), the intermediate phase ($\beta_{i}$) or the merger-ringdown ($\alpha_{i}$). For example, when the modification is introduced in the early inspiral, then $\phi_{i} \to \phi_{i}(1 + \delta \phi_{i})$ and the gIMR model is schematically
\be
\label{eq:gIMR}
\tilde{h}_{\rm gIMR}(f) =
\begin{cases} 
      A_{\I}(f) e^{i \Phi_{\I}(f)} e^{i \delta\Phi_{\I,\rm gIMR}}  & f\leq f_{\Int}\,, \\
      A_{\Int}(f) e^{i \Phi_{\Int}(f)} & f_{\Int}\leq f\leq f_{\MR}\,, \\
      A_{\MR}(f) e^{i \Phi_{\MR}(f)} & f_{\MR}\leq f\,, 
   \end{cases}
\ee
where we have defined
\be
\label{eq:deltaPhi-gIMR}
\delta \Phi_{\I,\rm gIMR} = \frac{3}{128 \eta} \sum_{i=0}^{7} \phi_{i} \; \delta \phi_{i} \; (\pi m f)^{(i-5)/3}\,.
\ee
and where $f_{\Int}$ and $f_{\MR}$ are the frequencies where the waveforms transition from the inspiral to the intermediate phase, and from the intermediate to the merger-ringdown phase respectively\footnote{The PhenomB waveform model~\cite{Ajith:2009bn} can be expressed as a piecewise function as in Eq.~(\ref{eq:gIMR}). Such a waveform was upgraded to PhenomC in~\cite{Santamaria:2010yb} and then to Phenom D in~\cite{Husa:2015iqa,Khan:2015jqa}. The gIMR model is based on PhenomD, which cannot technically be written as such a 3-part  piecewise function because the transition frequencies for the amplitudes and phases are not exactly the same. Nonetheless, the arguments we present next continue to hold if one considers each phase of the coalescence (inspiral, intermediate, merger-ringdown) separately.}. The phases (including the above correction $\delta \Phi_{\I,\rm gIMR}$ in the inspiral) are forced to be continuous and smooth at the transitions. The term proportional to $\delta \phi_1$ is absent if we follow the definition in Eq.~\eqref{eq:pmapping} as $\phi_1=0$ in GR; rather, $\delta \phi_{1}$ is taken to be the absolute correction at 0.5PN order, which corresponds to setting $\phi_1=1$ in Eq.~\eqref{eq:deltaPhi-gIMR}. In principle, there could also be GR modifications that are proportional to $\ln f$ in Eq.~\eqref{eq:deltaPhi-gIMR}, but in practice, there are no known theories that predict such a behavior; nonetheless, the arguments we present below continue to hold for such high-PN order modifications if the ppE framework is also extended to higher-PN order~\cite{Sampson:2013lpa}.  

As we now show mathematically, the gIMR model is an implementation of the ppE framework~\cite{PPE} applied to the IMRPhenom waveform. This may seem obvious, though there has been some debate in the literature as to the overlap of the various methods of deforming GR waveforms; we thus thought it would be instructive to describe this in more detail here. The ppE framework was devised to capture how theoretical mechanisms that deviate from GR impact the waveform. The general idea was to introduce amplitude and phase deformations to the best GR waveform model available. At the time~\cite{PPE} was written, the IMRPhenom model had not yet been developed, so~\cite{PPE} used its predecessor~\cite{Ajith:2007kx} to model the GR waveform. Thus, one of the first (and simplest because of the use of a single ppE phase deformation) IMR ppE model proposed in~\cite{PPE} (see Eq.~(1) in that paper) was 
\be
\label{eq:IMR-ppE}
\tilde{h}_{\ppE}(f) =
\begin{cases} 
      A_{\I}(f) e^{i \Phi_{\I}(f)} e^{i \delta \Phi_{\I,\ppE}} & f\leq f_{\Int}\,, \\
      A_{\Int}(f) e^{i \Phi_{\Int}(f)} & f_{\Int}\leq f\leq f_{\MR}\,, \\
      A_{\MR}(f) e^{i \Phi_{\MR}(f)} & f_{\MR}\leq f\,, 
   \end{cases}
\ee
where back then the amplitudes did not include PN corrections and the ringdown was modeled with a Lorentzian, following the PhenomB model (i.e.~the predecessor of the PhenomD model). Neglecting negative PN terms which were included in~\cite{PPE}, the ppE inspiral phase deformation takes the form (see Eq.~$(45)$ of~\cite{PPE}) 
\begin{align}\label{eq:deltaPhi-IMR-ppE}
\delta \Phi_{\I,\ppE}(f) &=  \frac{3}{128} \left(\pi {\cal{M}} f\right)^{-5/3} \sum_{i=0}^{7} \phi_{i}^{\ppE} \left(\pi {\cal{M}} f\right)^{i/3}\,,
\end{align}
where the coefficients $\phi_{i}^{\ppE}$ are ppE parameters and ${\cal{M}} = \eta^{3/5} m$ is the chirp mass. We recognize immediately that the gIMR phase deformation in Eqs.~\eqref{eq:gIMR} and ~\eqref{eq:deltaPhi-gIMR} is mathematically identical to the ppE phase deformation in Eqs~\eqref{eq:IMR-ppE} and~\eqref{eq:deltaPhi-IMR-ppE} with the mapping $\phi_{i}^{\ppE} = \phi_{i} \delta \phi_{i} \eta^{-i/5}$.

One can of course introduce ppE parameters at every PN order, thus greatly enlarging the parameter space, but a much more informative test is to consider one deformation at a time. Indeed, \cite{Sampson:2013lpa} first showed and~\cite{TheLIGOScientific:2016src} verified that using many GR deformations in the phase greatly dilutes the amount of information that can be extracted from the signal, without a noticeable gain in the ability to detect an anomaly. Furthermore, there is no reason to expect an alternative theory will follow the GR PN sequence of rational exponents, and limiting to these thus weakens the scope of the test. Considering then a single deformation at a time, the inspiral ppE phase takes the form (see Eq.~$(1)$ in~\cite{PPE})
\be
\label{eq:ppEphase}
\delta \Phi_{\I,\ppE}(f) = \beta \left(\pi {\cal{M}} f\right)^{b/3}\,,
\ee
where $\beta$ is called the amplitude coefficient and $b$ is the exponent coefficient. The former controls the magnitude of the deformation to GR, while the latter controls the type of physical mechanism that is responsible for the modification. If one considers the gIMR model with only a single PN coefficient modification that enters at $(n/2)$-PN order with $n \in \mathbb{N}$ (as also done in~\cite{TheLIGOScientific:2016pea,TheLIGOScientific:2016src}), then the mapping between gIMR and ppE is simply 
\be
b = n-5\,,
\ee
and
\begin{align}
\label{eq:map1}
\beta &= \frac{3}{128} \phi_{n} \; \delta \phi_{n} \; \eta^{-n/5}\,, \quad {\rm{if}} \quad \phi_{n} \neq 0\,,
\end{align}
or  
\begin{align}
\label{eq:map2}
\beta &= \frac{3}{128} \delta \phi_{n} \; \eta^{-n/5}\,, \quad {\rm{if}} \quad  \phi_{n} = 0\,,
\end{align}
(no summed implied over $n$ in these equations). Evaluating the first few terms, for example,
\begin{align}
\label{eq:beta-i0}
\beta &= \frac{3}{128}\delta \phi_{0}\,, \qquad {\rm{at}} \; 0{\rm{PN}} \; {\rm{order}}\,,
\\
\beta &= \frac{3}{128} \delta \phi_{1} \; \eta^{-1/5}\,, \qquad {\rm{at}} \; 0.5{\rm{PN}} \; {\rm{order}}\,,
\\
\beta &=\frac{3}{128} \phi_{2} \; \delta \phi_{2} \; \eta^{-2/5}\,, \qquad {\rm{at}} \; 1{\rm{PN}} \; {\rm{order}}\,,
\end{align}
where we used $\phi_0=1$ in Eq.~\eqref{eq:beta-i0}, $\phi_{2} =  {3715}/{756} + ({55}/{9})  \eta$, and the other inspiral phase coefficients of GR can be found in Appendix A of~\cite{Khan:2015jqa}. 

The gIMR waveforms are then a restricted subset of the ppE waveforms presented in~\cite{PPE}. We say subset because the gIMR waveforms only consider positive PN order deformations to the waveform phase and no deformations to the amplitude. The ppE framework allows for both of these, and the negative PN order deformations are particularly important when extracting information about certain physical effects, as discussed in the introduction. 

\subsubsection{From Parametric Deformations to Theoretical Physics}
\label{subsec:par-def-to-physics}

Now that we have established that the gIMR model is a subset of the ppE framework applied to the IMRPhenom waveform family, we can use all of the machinery of the ppE formalism to connect GR deformations to specific theoretical mechanisms. The latter can be classified into \emph{generation mechanisms} and \emph{propagation mechanisms}. The mapping between these mechanisms and ppE parameters $\beta$ (or $\delta \phi_{i}$) has been developed over the past several years in~\cite{PPE,Alexander:2007kv,Yunes:2009bv,Yunes:2010yf,cornish-PPE,Yagi:2011xp,yagi:brane,yunesstein,vigelandnico,Chatziioannou:2012rf,mirshekari,kent-CSBH,Yagi:2012vf,Sampson:2013wia,Sampson:2013jpa,Hansen:2014ewa,Sampson:2014qqa,Yagi:2015oca} and it is summarized in the review paper~\cite{Yunes:2013dva}. For completeness, we present this mapping here, correcting a few typos that appeared in the literature and updating the mapping with recent results.

\begin{center}
{\small\emph{$3.1$~Generation Mechanisms}}
\end{center}

Generation mechanisms refer to those that are active close to the binary system, where the GWs are being generated. Typically, one refers to this region as the near-zone in the PN formalism, and the inner-zone in BH perturbation theory. One can think of generation mechanisms as modification to the Poisson equation in the weak-field limit of the Einstein equations, i.e.~modifications to the structure of the fields in terms of the source multipole moments. Such modifications then propagate into corrections to the binding energy and angular momentum of a binary, and thus to the equations of motion. Generation mechanisms also affect the equations of motion via modifications to the energy and angular momentum flux. The character of the modification depends sensitively on the particular mechanism that activates. For example, when a scalar field activates in the near zone, it typically leads to dipolar energy loss from the binary system and a faster rate of orbital decay than predicted in GR. 

{
\newcommand{\minitab}[2][l]{\begin{tabular}{#1}#2\end{tabular}}
\renewcommand{\arraystretch}{1.2}
\begingroup
\squeezetable
\begin{table*}[tb]
\begin{centering}
\begin{tabular}{c|c|c|c|c|c}
\hline
\hline
\noalign{\smallskip}
Theoretical Effect & Theoretical Mechanism  & Theories & ppE $b$ & Order & Mapping \\
\hline
\multirow{2}{*}{Scalar Dipolar Radiation} & Scalar Monopole Activation  & EdGB~\cite{Metsaev:1987zx,Maeda:2009uy,yunesstein,Yagi:2011xp} & $-7$ &  $-1$PN & $\beta_{\EDGB}$~\cite{Yagi:2011xp}\\
 & BH Hair Growth & Scalar-Tensor Theories~\cite{Jacobson:1999vr,Horbatsch:2011ye} & $-7$ &  $-1$PN & $\beta_{\ST}$~\cite{Jacobson:1999vr,Horbatsch:2011ye}\\ \hline
\multirow{2}{*}{Anomalous Acceleration} & Extra Dim. Mass Leakage  & RS-II Braneworld~\cite{Randall:1999ee,Randall:1999vf} & $-13$ &  $-4$PN & $\beta_{\ED}$~\cite{yagi:brane}\\
 & Time-Variation of $G$ & Phenomenological~\cite{1937Natur.139..323D,Yunes:2009bv} & $-13$ &  $-4$PN & $\beta_{\dot{G}}$~\cite{Yunes:2009bv}\\ \hline
Scalar Quadrupolar Radiation & Scalar Dipole Activation & \multirow{3}{*}{dCS~\cite{CSreview,Yagi:2011xp}} & \multirow{3}{*}{$-1$} &\multirow{3}{*}{$+2$PN} & \multirow{3}{*}{$\beta_{\dCS}$~\cite{Yagi:2012vf}} \\
Scalar Dipole Force & due to  &  &  &   &  \\
Quadrupole Moment Deformation & Grav. Parity Violation  &  &  &  & \\ \hline
\multirow{2}{*}{Scalar/Vector Dipolar Radiation} & Vector Field Activation & \multirow{3}{*}{EA~\cite{Jacobson:2000xp,Jacobson:2008aj}, Khronometric~\cite{Blas:2009qj,Blas:2010hb}}  & \multirow{2}{*}{$-7$} & \multirow{2}{*}{$-1$PN} & \multirow{2}{*}{$\beta_{\AE}^{(-1)}$, $\beta_{\KG}^{(-1)}$~\cite{Hansen:2014ewa}}\\
\multirow{2}{*}{Modified Quadrupolar Radiation} & due to &  & \multirow{2}{*}{$-5$}  & \multirow{2}{*}{$0$PN} &  \multirow{2}{*}{$\beta_{\AE}^{(0)}$, $\beta_{\KG}^{(0)}$~\cite{Hansen:2014ewa}} \\
 & Lorentz Violation &  &  &  & \\ \hline
\multirow{7}{*}{Modified Dispersion Relation} & \multirow{7}{*}{GW Propagation} & Massive Gravity~\cite{Will:1997bb,Rubakov:2008nh,Hinterbichler:2011tt,deRham:2014zqa} & $-3$ &  $+1$PN & \\
 &  & Double Special Relativity~\cite{AmelinoCamelia:2000ge,Magueijo:2001cr,AmelinoCamelia:2002wr,AmelinoCamelia:2010pd} & $+6$ &  $+5.5$PN & \\
 &  & Extra Dim.~\cite{Sefiedgar:2010we}, Horava-Lifshitz~\cite{Horava:2008ih,Horava:2009uw,Vacaru:2010rd} & \multirow{1}{*}{$+9$} &  \multirow{1}{*}{$+7$PN} & \\
 &  & gravitational SME ($d=4$)~\cite{Kostelecky:2016kfm} & $+3$ & $+4$PN &    $\beta_{\MD}$\\
 &  & gravitational SME ($d=5$)~\cite{Kostelecky:2016kfm} & $+6$ & $+5.5$PN & \cite{Will:1997bb,mirshekari}\\
 &  & gravitational SME ($d=6$)~\cite{Kostelecky:2016kfm} & $+9$ & $+7$PN & \\
 &  & Multifractional Spacetime~\cite{Calcagni:2009kc,Calcagni:2011kn,Calcagni:2011sz}& 3--6 &  $4$--$5.5$PN &\\
\noalign{\smallskip}
\hline
\hline
\end{tabular}
\end{centering}
\caption{%
Theoretical effects introduced into the GW observable due to various theoretical mechanisms, together with example theories where such mechanisms arise. For each effect and mechanism, we specify the ppE exponent which would signal its appearance, the relative PN order at which these effects first enter the Fourier GW phase, and the mapping to the ppE magnitude coefficient $\beta$.
}
\label{tab:summary}
\end{table*}
\endgroup
}
Table~\ref{tab:summary} summarizes the theoretical effects and mechanisms in the generation of GWs that can be constrained from GWs emitted in the coalescence of a BH binary. By ``theoretical effect'' we mean the type of modification that is induced on the GW observable, while ``theoretical mechanism'' refers to the process that produces the aforementioned modification. The table also provides examples of theories where these effects and mechanisms arise, together with the relative PN order at which they first enter the Fourier GW phase and the mapping between the ppE coefficient $\beta$ (or alternatively $\delta \phi_i$) and the system parameters and coupling constants that control the magnitude of the modification. We have not included in the table any effect or mechanism that can only be constrained with binary systems when at least one component is required to be a NS (for those mappings refer to the review paper~\cite{Yunes:2013dva}). 

Example theories in Table~\ref{tab:summary} that modify GW generation mechanisms are as follows: 

\begin{itemize}[itemsep=0.1pt, topsep=5pt, partopsep=0pt]

\item \emph{Einstein-dilaton Gauss-Bonnet (EdGB) Gravity}: 
BHs have scalar monopole charge (a measure of the dependence of the BH mass on the scalar field) as sourced by the Kretchmann curvature. Such charges induce scalar dipole radiation, which then speeds up the rate at which the binary inspirals. The magnitude of this modification is proportional to the dimensionless EdGB coupling parameter $\zeta_{\EDGB} \equiv 16 \pi \alpha_\EDGB^2/m^4$. The mapping between $\beta$ and the system and coupling parameters are given by~\cite{Yagi:2011xp}
\begin{align}
\label{eq:beta-EdGB}
\beta_{\EDGB} &= - \frac{5}{7168} \zeta_{\EDGB} \frac{\left(m_1^2 s_2^\EDGB - m_2^2 s_1^\EDGB \right)^2}{m^4 \eta^{18/5}}\,.
\end{align}
Here, $s_A^{\EDGB}$ are the spin-dependent factors of the BH scalar charges in EdGB gravity, which are given by $s_A^\EDGB \equiv 2(\sqrt{1-\chi_A^2} - 1 + \chi_A^2)/\chi_A^2$~\cite{barausse-prep}, with $\chi_A$ the magnitude of the spin angular momentum of the $A$th body normalized by its mass squared. 

\item \emph{dynamical Chern-Simons (dCS) Gravity}: 
Similar to EdGB gravity, BHs have scalar dipole charge sourced by the Pontryagin invariant that induce scalar quadrupolar radiation. The magnitude of the correction to the rate at which the binary inspirals is proportional to the dimensionless dCS coupling parameter $\zeta_{\dCS} \equiv 16 \pi \alpha_\dCS^2/m^4$. The mapping for $\beta$ is given by~\cite{Yagi:2012vf}
\begin{align}
\label{eq:beta-dCS}
\beta_{\dCS} &= \frac{1549225}{11812864} \frac{\zeta_{\dCS}}{\eta^{14/5}} \left[\left(1 - \frac{231808}{61969} \eta\right) \chi_{s}^{2} 
\right. \nn \\
&\left.
+ \left(1 - \frac{16068}{61969} \eta\right) \chi_{a}^{2} - 2 \delta_{m} \chi_{s} \chi_{a}\right]\,,
\end{align}
where we introduced the symmetric and antisymmetric dimensionless spin combinations\footnote{$\chi_s$ is different from $\chi_\mrm{eff} \equiv  (a_{\|1} +  a_{\|2})/m$ in~\cite{TheLIGOScientific:2016wfe}.} $\chi_{s,a} = (a_{\|1}/m_{1} \pm a_{\|2}/m_{2})/2$ with $a_{\|A}$ representing the projection of the (dimensional) spin vector $\vec a_A$ onto the unit orbital angular momentum vector, and the dimensionless mass difference $\delta_{m} = (m_{1}-m_{2})/m$. 

\item \emph{Scalar-Tensor (ST) Theories}: 
A BH can acquire a scalar charge if the scalar field is evolving in time with a growth rate $\dot{\phi}$ due to e.g.~a cosmological background~\cite{Jacobson:1999vr,Horbatsch:2011ye,Berti:2013gfa}. The mapping for $\beta$ is given by~\cite{Jacobson:1999vr,Horbatsch:2011ye}
\be
\label{eq:beta-ST}
\beta_{\ST} =- \frac{5}{1792} \dot{\phi}^2 \eta^{2/5} \left(m_1 s_1^\ST - m_2 s_2^\ST \right)^2\,,
\ee
where $s_A^\ST \equiv  [1+(1-\chi_A^2)^{1/2}]/2$ are the spin-dependent factor of BH scalar charges in scalar-tensor theories with BH hair growth~\cite{Horbatsch:2011ye}.

\item \emph{RS-II Braneworld Scenario}: 
The leakage of gravitons into extra dimensions induces an anomalous acceleration that is proportional to the rate of leakage into the bulk $dm/dt$, which in turn is proportional to the square of the ratio of the size of the extra dimension to the total mass. The mapping for $\beta$ is given by~\cite{yagi:brane}\footnote{This corrects a small typo (the numerical prefactor and the dependence on $\eta$) in the review paper~\cite{Yunes:2013dva}, and recasts the constraint in terms of the rate of change of the total mass.}
\be
\beta_{\ED}  = \frac{25}{851968} \left(\frac{dm}{dt}\right)  \frac{3 - 26 \eta + 34 \eta^{2}}{\eta^{2/5} (1-2\eta)}\,.
\ee
The denominator never vanishes since $\eta < 1/4$.
 
\item \emph{Phenomenological Varying-$G$ Theories}: 
Phenomenological models where Newton's gravitational constant $G$ has a small time-variation also induce an anomalous acceleration, and thus modifications to the Fourier waveform phase scale with $\dot{G}$. The mapping for $\beta$ is given by~\cite{Yunes:2009bv}
\be
\beta_{\dot{G}} = - \frac{25}{65536} \dot{G} {\cal{M}}\,.
\ee
In fact, one can view the RS-II braneworld scenario as a particular realization of these phenomenological varying-$G$ theories. 
 
\item \emph{Einstein-\AE ther (EA) and Khronometric Theory}: 
EA theory generically predicts gravitational Lorentz violation, where the magnitude of the latter is controlled by four dimensionless coupling parameters; two of these are $(c_+,c_-)$, and the remaining two, including $c_{14}$, can be expressed in terms of $(c_+,c_-)$ when imposing Solar System constraints. The mapping for $\beta$ entering at 0PN order is given by~\cite{Hansen:2014ewa}
\begin{align}
\label{eq:beta-EA}
\beta_{\AE}^{(0PN)} &= - \frac{3}{128} \left[ \left( 1-\frac{c_{14}}{2} \right) \left(\frac{1}{w_2^{\AE}} \right.\right. \nn \\ &\left. \left. +\frac{2c_{14}c_+^2}{(c_++c_--c_-c_+)^2 w_1^{\AE}}
	+\frac{3c_{14}}{2w_0^{\AE} (2-c_{14})} \right) - 1 \right]\,.
\end{align}
Here, $w_0^{\AE}$, $w_1^{\AE}$ and $w_2^{\AE}$ are the propagation speeds of the spin-0, spin-1 and spin-2 modes, which depend on $(c_+,c_-)$. 

A similar mapping can be found for khronometric theory, which contains three dimensionless coupling constants; two of these are $(\beta_\KG, \lambda_\KG)$ and the remaining one $\alpha_\KG$ is expressed in terms of $\beta_\KG$ when saturating Solar System bounds. The mapping for the ppE parameter $\beta$ entering at 0PN order is given by~\cite{Hansen:2014ewa}\footnote{In Eq.~(91) in~\cite{Hansen:2014ewa}, the overall factor in the first term should be $1-\alpha_\KG/2$ instead of $1-2/\beta_\KG$. We also imposed $\alpha_\KG = 2 \beta_\KG$ to satisfy Solar System bounds.}
\begin{align}
\label{eq:beta-kh}
\beta_{\KG}^{(0PN)} &= - \frac{3}{128} \left[ \left( 1 - \beta_\KG \right) \left(\frac{1}{w_2^{\KG}} 
\right.\right. \nn \\  
&+ \left.\left.
	\frac{3 \beta_\KG}{2 w_0^{\KG} (1- \beta_\KG)} \right) - 1 \right]\,.
\end{align}
Similar to the EA case, $w_0^{\KG}$ and $w_2^{\KG}$ are the propagation speeds of the spin-0 and spin-2 modes that depend on $(\beta_\KG, \lambda_\KG)$.

Both $\beta_{\AE}^{(0PN)}$ and $\beta_{\KG}^{(0PN)}$ should also contain terms that depend on the BH scalar charges, which are currently unknown; we neglect such terms in this paper because Ref.~\cite{Hansen:2014ewa} showed that they are subdominant for NS binaries. We do not present the ppE parameter $\beta_{\AE}^{(-1PN)}$ and $\beta_{\KG}^{(-1PN)}$ that enter at $-1$PN order, since they are proportional to the square of the difference in the individual BH scalar charges.  See Sec.~\ref{sec:counterpart} for a more detailed explanation of EA and khronometric theory.  

\end{itemize}

\begin{center}
{\small\emph{$3.2$~Propagation Mechanisms}}
\end{center}

Propagation mechanisms refer to those that activate while the wave is propagating away from the source in a regime at least several GW wavelengths away from the binary's center of mass (in the so-called far-zone, radiation zone or wave-zone). One can think of propagation mechanisms as modifications to the plane-wave propagator in field theory, i.e.~modifications to the inverse of the wave operator. This can introduce modifications to the amplitude of the waves, such as amplitude mixing generated by gravitational parity violation~\cite{Yunes:2010yf}, or they can introduce modifications to the phase of the waves, due to real corrections to the wave's dispersion relation~\cite{mirshekari}. Modifications to the dispersion relation are typically also associated with either modifications to the Lorentz group or to its action in real or momentum space. Thus, such modifications are associated with gravitational Lorentz-violating effects, which are typically found in quantum-gravitational models, such as loop quantum gravity~\cite{2008PhRvD..77b3508B} and string theory~\cite{2005hep.th....8119C,2010GReGr..42....1S}.  

Table~\ref{tab:summary} also summarizes the theoretical mechanisms that can be constrained in the propagation of GWs using GWs from compact binaries. As in the generation case, the table provides the mapping between the ppE coefficient $\beta$ and the system parameters and coupling constants of particular theories with modified dispersion relation. For a generic modified dispersion relation of the form 
\begin{align}
\label{dispersion}
E^2 &=  \left(p c\right)^2 + \mathbb{A} \left(p c\right)^{\alpha}\,,
\end{align}
the modification to the Fourier GW phase takes on the ppE form of Eq.~\eqref{eq:ppEphase} with
\begin{align}
\label{zeta}
\beta_{\MD} &= \frac{\pi^{2-\alpha}}{(1-\alpha)} \frac{D_\alpha}{\lambda_{\mathbb{A}}^{2-\alpha}}  \frac{\mathcal{M}^{1-\alpha}}{(1+z)^{1-\alpha}}\,,
\\
b &= 3 \alpha - 3\,.
\end{align}
In these equations, $E$ and $p$ are the energy and momentum of the graviton respectively, while $\mathbb{A}$ is the strength of the dispersion modification (that depends on the coupling constants of the particular theory) and $\lambda_{\mathbb{A}} \equiv h \, \mathbb{A}^{{1}/{(\alpha-2)}}$ is similar to a Compton wavelength. For events at small redshift, the distance $D_\alpha$ is given by 
\be
D_{\alpha} = \frac{z}{H_{0} \sqrt{\Omega_{M} + \Omega_{\Lambda}}} \left[1 - \frac{z}{4} \left(\frac{3 \Omega_{M}}{\Omega_{M} + \Omega_{\Lambda}} + 2 \alpha \right) + {\cal{O}}(z^{2})\right]\,,
\ee
with $z$ the redshift and $H_{0}$ the current value of the Hubble constant, while $\Omega_M$ and $\Omega_{\Lambda}$ are the energy density of matter and dark energy respectively. This modification to the Fourier phase is independent of the generation mechanism, and in particular, independent of the particular waveform model used for the inspiral, merger and ringdown.  Thus, one can add this modification onto any waveform model directly.

A GW that obeys a modified dispersion relation will also travel at a speed different from that of light. Using Eq.~\eqref{dispersion} we can calculate the group velocity $v_{g}$ to find 
\be
\label{eq:prop-speed}
\frac{v_{g}}{c}  =  \frac{1}{c} \frac{d\omega}{dk}= 1 + \frac{(\alpha -1)}{2} \mathbb{A} \, E^{\alpha-2}\,,
\ee
to leading order in $\mathbb{A} E^{\alpha-2} \ll 1$. For $\alpha > 1$ and $\mathbb{A}<0$, GWs travel slower than the speed of light. When this is the case, high energy massive particles may travel faster than GWs and emit gravitational Cherenkov radiation~\cite{Caves:1980jn,Moore:2001bv}. The fact that observed high energy cosmic ray particles have traveled extragalactic distances without losing energy to this type of radiation places a stringent constraint on the magnitude of $\mathbb{A}$ in the $\mathbb{A}<0$ sector, in particular when $\alpha \geq 2$~\cite{Kiyota:2015dla}. 

The dimensionless constant $\alpha$ controls the type of dispersion modification. In the limit where $E$ and $p$ are large compared to $(\mathbb{A} p^\alpha)^{1/2}$, but small compared to the Planck energy $E_{p}$, this generic parameterization can capture the following theories or phenomenological models: 
\begin{itemize}[itemsep=0.1pt, topsep=5pt, partopsep=0pt]
\item {\emph{Double Special Relativity}}~\cite{AmelinoCamelia:2000ge,Magueijo:2001cr,AmelinoCamelia:2002wr,AmelinoCamelia:2010pd}: $ \mathbb{A}  = \eta_{\rm dsrt}$ and $\alpha = 3$,  where $\eta_{\rm dsrt}$ is a parameter that characterizes an observer-independent length scale, commonly taken to be the Planck length,
\item {\emph{Extra-Dimensional Theories}}~\cite{Sefiedgar:2010we}: $ \mathbb{A}  = - \alpha_{\rm edt}$ and $\alpha = 4$,  where $\alpha_{\rm edt}$ is a constant that characterizes the square of the Planck length in extra dimensional theories,
\item {\emph{Ho\v{r}ava-Lifshitz Gravity}}~\cite{Horava:2008ih,Horava:2009uw,Vacaru:2010rd,Blas:2011zd}: $ \mathbb{A}  = \kappa^{4}_{\rm hl}  \mu^{2}_{\rm hl}/16$ and $\alpha = 4$, where $\kappa_{\rm hl}$ (related to the bare gravitational constant) and $\mu_{\rm hl}$ (related to the deformation in the ``detailed balance'' condition imposed to reduce the number of coupling constants) are constants of the theory,
\item {\emph{Massive Graviton}}~\cite{Will:1997bb,Rubakov:2008nh,Hinterbichler:2011tt,deRham:2014zqa}: $\mathbb{A} = \left(m_{g}c^2\right)^{2}$ and $\alpha = 0$, where $m_{g}$ is the mass of the graviton,
\item {\emph{Multifractional Spacetime Theory}}~\cite{Calcagni:2009kc,Calcagni:2011kn,Calcagni:2011sz,Calcagni:2016zqv} $\mathbb{A}  = 2 E_*^{2-\alpha}/(3-\alpha)$ (timelike fractal spacetime)  or $\mathbb{A}  = - 2 \cdot 3^{1-\alpha/2} E_*^{2-\alpha}/(3-\alpha)$ (spacelike fractal spacetime) with $\alpha = 2-3$ ($\alpha=2.5$ being a typical choice), where $E_*$ is the characteristic length scale above which spacetime is discrete,
\item {\emph{Gravitational Standard Model Extension (SME)}}~\cite{Kostelecky:2016kfm}: $\mathbb{A} =- 2 \mathring{k}_{(I)}^{(d)}$ for even $d \geq 4$ and $\mathbb{A} = \pm 2 \mathring{k}_{(V)}^{(d)}$ for odd $d \geq 5$ with $\alpha = d-2$ in the rotation-invariant limit to linear order in $\mathring{k}_{(V)}^{(d)}$, where $\mathring{k}_{(I)}^{(d)}$ and $\mathring{k}_{(V)}^{(d)}$ are constant coefficients that control the Lorentz-violation operators. The modified dispersion relation without rotation invariance is given by Eq.~(5) in~\cite{Kostelecky:2016kfm}.
\end{itemize} 
The first two theories in the above list also typically predict a constant (massive graviton) term, but we have left this term out of the list above. The modifications to the dispersion relation need not automatically be Planck suppressed~\cite{2004PhRvL..93s1301C,2006hep.th....3002C}. This is because Planck suppression typically arises because of Lorentz-invariance; in theories that lack this symmetry, modifications to the dispersion relation can be dramatically enhanced upon regularization and renormalization~\cite{Gambini:2011nx}.

Theories that predict modified dispersion relations for the graviton are also likely to modify GWs in the generation phase, so which one dominates? Let us argue that the former typically dominates the latter due to the accumulation of the modified gravity effect with distance, using massive gravity as an example. Reference~\cite{sutton} showed that the fractional correction to the radiated GW energy flux from a binary in massive gravity is given by $\sim 1/(\lambda_g^2 f^2)$. Therefore, the ppE parameter due to GW generation is roughly given by $\beta_\MG^{(\mathrm{gen})} \sim (3/128) \pi^2 \mathcal{M}^2/\lambda_g^2$ with $b=-11$, namely a $-3$PN correction. Comparing this with the ppE parameter for the modified propagation, $\beta_\MG^{(\mathrm{prop})} \sim \pi^2 D_0 \mathcal{M}/\lambda_g^2$ with $b=-3$ [see Eq.~\eqref{zeta} with $\alpha=0$], one finds
\ba
\frac{\beta_\MG^{(\mathrm{prop})} \left(\pi \mathcal{M} f\right)^{-1}}{\beta_\MG^{(\mathrm{gen})}\left(\pi \mathcal{M} f\right)^{-11/3}} &\sim& 10^{18} \left( \frac{\mathcal{M}}{28 M_\odot} \right)^{5/3} \nn \\
&\times& \left( \frac{D_0}{380\mathrm{Mpc}} \right) \left( \frac{f}{100\mathrm{Hz}} \right)^{8/3}\,.
\ea
This clearly shows that the propagation effect dominates the generation effect, even when the former is of higher PN order relative to the latter. Therefore, when we consider modifications to the propagation of GWs we neglect related modifications to GW generation.

\section{Theoretical Implications of events GW150914 and GW151226} 
\label{sec:parametrizedtests}

This section discusses the theoretical implications of events GW150914 and GW151226. We classify theoretical implications into those that affect the generation of GWs and those that affect the propagation of GWs, and determine the precise implications the LVC observations have for each class. For both events, we relate Bayesian as well as Fisher estimates on parameter constraints to different physical mechanisms. 

When we carry out Fisher analyses~\cite{cutlerflanagan} in Secs.~\ref{sec:generation} and~\ref{sec:mod-disp-rel}, the parameter vector that determines our IMRPhenom waveform is $\bm{\theta} = (\ln \mathcal{M}_z, \ln \eta, \chi_s, \chi_a, t_c, \phi_c, \ln A_0, \beta)$, where $A_0$ is an overall amplitude factor proportional to $\mathcal{M}_z^{5/6}/D_L$, with $\mathcal{M}_z$ and $D_L$ the redshifted chirp mass and luminosity distance respectively. The non-zero parameter values for the injections are given in Table~\ref{table:injection} ($t_c=\phi_c=\beta=0$), except for $A_0$ that is determined from the total (network) SNR\footnote{The GW amplitude depends not only on the masses, distance and redshift, but also on the sky location and the inclination of the source, which are poorly constrained~\cite{TheLIGOScientific:2016wfe}. Moreover, we use a fit for the noise curve, which naturally has some (small) error, discussed in Appendix~\ref{app:noise-fit}. In order to minimize the effect of such uncertainties, we choose the amplitude $A_{0}$ to give SNRs of 24 and 13 for GW150914 and GW151226 respectively.}. The injected $(\chi_s,\chi_a)$ for GW151226 corresponds to $(\chi_1,\chi_2) = (0.49,-0.32)$ and an effective dimensionless spin of $0.21$, which is consistent with the measured values as reported by the LVC. We start the integration of the Fisher matrix $\Gamma_{ij}$ at 20Hz to be consistent with the aLIGO sensitivity curve during the O1 observation period, and we use a fit to the spectral noise sensitivity curve during O1 (see Appendix~\ref{app:noise-fit}). We follow~\cite{cutlerflanagan,Poisson:1995ef,Berti:2004bd} and impose a physical Gaussian prior on $\chi_s$ and $\chi_a$ that ensures that $|\chi_s| \leq 1$ and $|\chi_a| \leq 1$; this is done by multiplying the likelihood function (in the Fisher approximation) by a Gaussian function of $\chi_s$ and $\chi_a$ with a standard deviation of unity. 

{
\newcommand{\minitab}[2][l]{\begin{tabular}{#1}#2\end{tabular}}
\renewcommand{\arraystretch}{1.2}
\begingroup
\begin{table}[htb]
\begin{centering}
\begin{tabular}{c|cc}
\hline
\hline
\noalign{\smallskip}
 & GW150914  & GW151226 \\ \hline
$(m_1,m_2)$ & $(35.7,29.1) M_\odot$  & $(14.2,7.5) M_\odot$ \\ 
$(\chi_s,\chi_a)$ & $(0,0)$  & $(0.085,0.41)$ \\ 
$z$ & $0.088$  & $0.09$ \\ 
SNR & 24  & 13 \\ 
\noalign{\smallskip}
\hline
\hline
\end{tabular}
\end{centering}
\caption{
Injected parameters for Fisher analyses with GW150914 and GW151226. These parameters are consistent with the aLIGO measurement~\cite{Abbott:2016blz,TheLIGOScientific:2016wfe,Abbott:2016nmj,TheLIGOScientific:2016pea}. 
}
\label{table:injection}
\end{table}
\endgroup
}

\subsection{Implications on the Generation of GWs} 
\label{sec:generation}
\subsubsection{Constraining Generation Mechanisms} 
\label{sec:generation2}

As described in Sec.~\ref{subsec:par-def-to-physics}, constraints on a plethora of mechanisms that may be active in the generation of GWs can be captured within the ppE formalism. Therefore, constraints on the ppE amplitude coefficients $\beta$ (or $\delta \phi_{i}$) as a function of PN order provide  constraints on physical mechanisms as well. This is one of the benefits and power of the ppE framework.

\begin{figure}[t]
\begin{center}
\includegraphics[width=\columnwidth,clip=true]{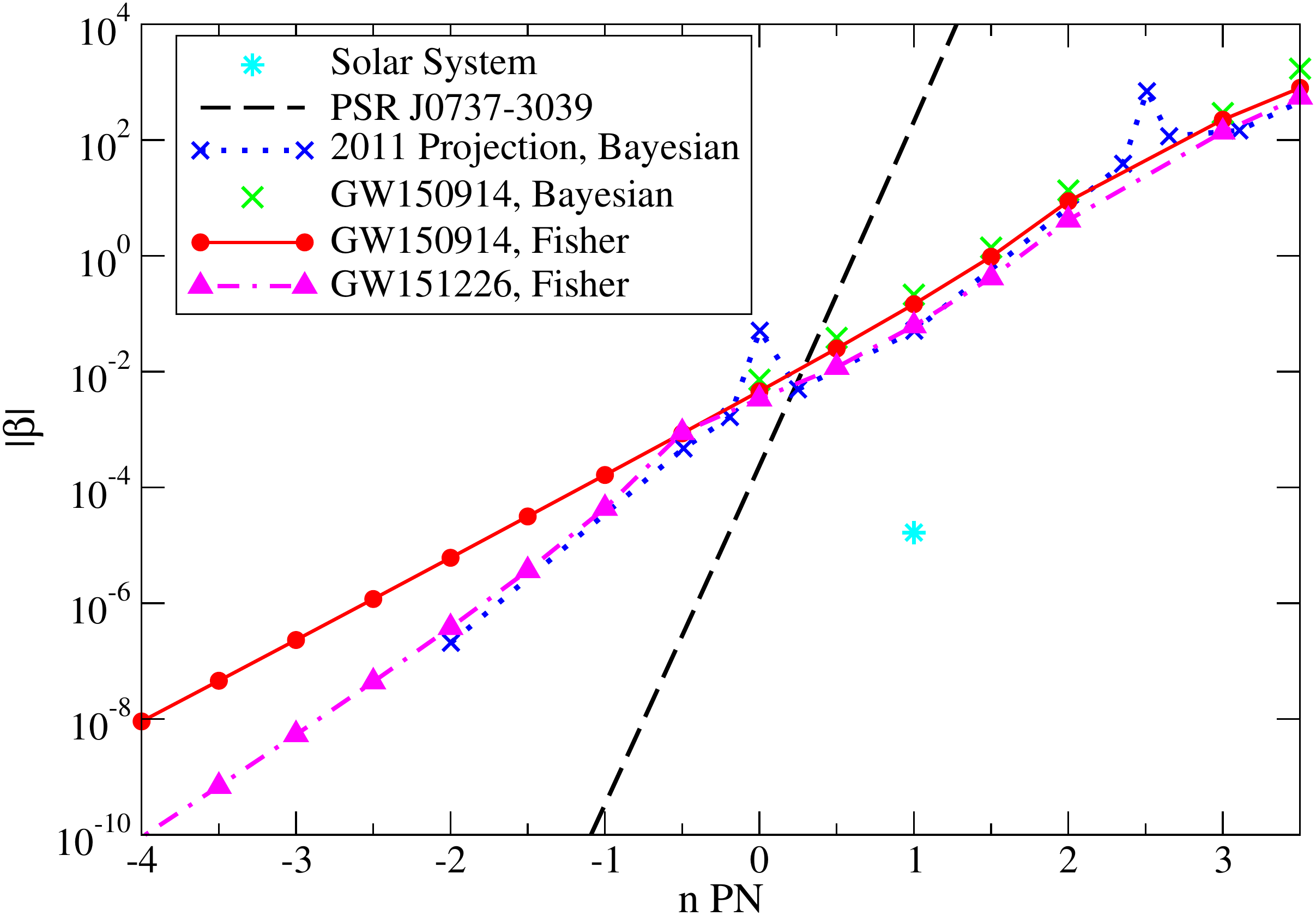} 
\caption{\label{fig:beta-cons}~(Color online)~90\%-confidence constraints on the ppE parameter $|\beta|$ at $n$th PN order. The green crosses represent the bounds reported in~\cite{TheLIGOScientific:2016pea,TheLIGOScientific:2016src} through a Bayesian analysis of event GW150914, mapped to constraints on $\beta$. The red (magenta) dots and line represent bounds from GW150914 (GW151226) estimated with a Fisher analysis, using the IMRPhenom waveform (without spin precession) and a fit to the aLIGO spectral noise density. The constraints obtained with a Fisher analysis agree very well with the Bayesian constraint reported in~\cite{TheLIGOScientific:2016pea,TheLIGOScientific:2016src}. The blue dotted line shows projected constraints predicted in 2011 by~\cite{cornish-PPE} for a system similar to GW151226. The dashed black line is a rough estimate on the constraints that the double binary pulsar PSR J0737-3039~\cite{burgay,lyne,kramer-double-pulsar} can place on the ppE $\beta$ parameter~\cite{Yunes:2010qb}, while the cyan star refers to the bound on $\beta$ at 1PN from the perihelion precession of Mercury~\cite{Sampson:2013wia}. Binary pulsar observations can constrain negative PN order deviations better than aLIGO, while aLIGO does better than binary pulsar observations at higher PN order, as first calculated in~\cite{Yunes:2010qb}. However, note also that binary pulsar and Solar System bounds cannot be directly compared to GW ones as the binary pulsar (Solar System) one corresponds to the extreme case of no conservative (no dissipative) corrections. Moreover, stronger constraints on $\beta$ for these latter tests do not necessarily mean stronger constraints on modifications to GR for BH mergers, as $\beta$ depends not only on theoretical coupling parameters but also on system parameters, and in certain theories (like EdGB gravity), non-GR corrections are suppressed in stars compared to BHs.} 
\end{center}
\end{figure}

The LVC in~\cite{TheLIGOScientific:2016pea,TheLIGOScientific:2016src} performed a Bayesian analysis of the constraints that event GW150914 places on the ppE coefficients $\delta \phi_{i}$ (at $90\%$ credible level), using the IMRPhenom model with precessing spins. Figure~\ref{fig:beta-cons} plots these constraints mapped to constraints on $\beta$ as a function of relative PN order in the Fourier GW phase (green crosses). For example, a constraint on $\beta$ at 0PN order means a constraint on a ppE term in the Fourier GW phase that is proportional to $(\pi {\cal{M}} f)^{-5/3}$.

Another way to estimate constraints on $\beta$ is to carry out a Fisher analysis~\cite{Finn:1992xs,Chernoff:1993th}. We have done this using IMRPhenom waveforms without precessing spins (as these have a minimal effect on the waveforms~\cite{TheLIGOScientific:2016wfe,Abbott:2016izl})\footnote{Given that aLIGO could not measure precessional effects with GW150914 or GW151226, we expect the inclusion of precessional effects in the tests of GR that we carried out here to be negligible. However, if an event is detected in the near future with large precession, then the inclusion of such effects could allow us to break degeneracies in parameter estimation, in particular among spins and masses (see e.g.~\cite{Chatziioannou:2014coa}), improving constraints on non-GR effects~\cite{stavridis,Yagi:2009zm,Yagi:2009zz,Klein:2009gza}.} and a spectral noise density constructed by fitting the aLIGO data for the initial 16 days of coincident observations (see Appendix~\ref{app:noise-fit}). We only include the ppE correction in the inspiral phase (i.e. for $f< f_\Int$ with $f_\Int = 52$Hz (154Hz) for GW150914 (GW151226)) due to the lack of merger simulations in non-GR theories, which should allow us to find conservative bounds on $\beta$. The results are plotted in Fig.~\ref{fig:beta-cons} as a function of PN order. The GW150914 Fisher estimate is a good approximation to the more complete Bayesian analysis of~\cite{TheLIGOScientific:2016wfe}, overestimating the constraint by roughly 15--50\%. We do not show constraints at 2.5PN order as such a correction is degenerate with a constant phase shift. The Fisher analysis here includes negative PN order effects, since many theoretical implications require such constraints; the Bayesian analysis of~\cite{TheLIGOScientific:2016pea,TheLIGOScientific:2016src} does not report on these negative PN order constraints, which is why they are not shown in Fig.~\ref{fig:beta-cons}. Based on the fact that the phase of the IMRPhenom waveform agrees with that of the numerical relativity waveform within $\sim 0.015$ rad or better for any frequency~\cite{Khan:2015jqa}, we roughly estimate a systematic error on $\beta$ due to the waveform mismodeling within GR to be 3--4 orders of magnitude smaller than the bounds in Fig.~\ref{fig:beta-cons} (see Appendix~\ref{app:BvsD} for more details).

Figure~\ref{fig:beta-cons} shows that GW151226 places stronger constraints on $\beta$ than GW150914~\cite{Abbott:2016nmj,TheLIGOScientific:2016pea} especially at negative PN orders. This is because GW151226 consists of a BH binary with lower total mass than GW150914, and thus, (i) the velocity of the binary constituents at a fixed frequency (e.g. $f \sim 50$Hz) is smaller and (ii) the observed frequency range is larger than for GW150914. The first fact makes the negative-PN-order, ppE correction terms in the phase and the total number of GW cycles in band larger than for GW150914. This, together with the second point above, make $\beta$ less degenerate with other binary parameters, leading to stronger constraints. Regarding corrections at high positive PN orders, point (i) results in a deterioration of the constraints, while point (ii) strengthens them compared to GW150914~\cite{Abbott:2016nmj,TheLIGOScientific:2016pea}. Taken together then these opposing effects lead to similar bounds at positive PN orders for GW150914 and GW151226. We also calculated the bounds on $\beta$ by combining those of GW150914 and GW151226 using Eq.~(4.12d) in~\cite{sutton} and found that such a combined bound is almost indistinguishable from that of GW151226 alone (the improvement reaches at most $\sim 30\%$ at $n \sim 0$PN). This finding is consistent with a similar analysis performed by the LVC~\cite{TheLIGOScientific:2016pea}.

Our analysis and the study of the LVC in~\cite{TheLIGOScientific:2016pea,TheLIGOScientific:2016src} differ in many ways, and yet, the two yield similar constraints on $\beta$. The main differences between these studies are that the former (latter) uses 
\begin{itemize}[itemsep=0.1pt, topsep=5pt, partopsep=0pt]
\item[(i)] a Fisher (Bayesian) analysis,
\item[(ii)] non-precessing (precessing) waveform templates, 
\item[(iii)] a fit for the noise curve (the real data),
\item[(iv)] a simulated waveform injection compatible with the real signal (the real signal), and 
\item[(v)] includes only statistical (both statistical and systematic) errors. 
\end{itemize}
Probably, differences (i)--(iii) do not have a large impact on the $\beta$ constraints for the following reasons. The difference in statistical errors between Fisher and Bayesian studies scales as $\mathcal{O}(1/\mrm{SNR}^2)$~\cite{cutlerflanagan,Vallisneri:2007ev}, which is only $\sim \mathcal{O}(0.2\%)$ ($\mathcal{O}(0.6\%)$) given the SNR of GW150914 (GW151226). Precession for both events was too small to be measurable by the LVC~\cite{TheLIGOScientific:2016wfe,Abbott:2016izl,Abbott:2016nmj,TheLIGOScientific:2016pea}. The real noise spectral density contains many spikes, but these are very thin, and thus, for the same SNR, they affect constraints on $\beta$ by only a few percent (see Appendix~\ref{app:noise-fit})\footnote{See the related work by~\cite{Berry:2014jja}, which shows that the effect of non-Gaussianity in the noise on parameter estimation is negligible.}. We do not include any specific noise realization in our Fisher analysis, since (i) such a noise realization only shifts the posterior distribution without affecting its spread~\cite{Sampson:2013lpa}, and (ii) the uncertainties in parameters averaged over different noise realizations are the same as those with zero noise injection~\cite{Nissanke:2009kt}. On the other hand, differences (iv) and (v) are probably more important. For example, in our Fisher analysis we set the spin magnitudes of the injection to zero, but the posteriors found by the LVC~\cite{TheLIGOScientific:2016wfe} are quite wide, and a different choice of spin magnitude can affect our Fisher estimates by a factor of $\sim 2$. Even using the Bayesian analysis of~\cite{TheLIGOScientific:2016wfe}, the mapping between $\delta \phi_i$ and $\beta$ [see Eqs.~\eqref{eq:map1} and~\eqref{eq:map2}] depends on the posterior distribution of other parameters, and different choices can also affect constraints on $\beta$ at high PN order by a factor of $\sim 2$. As another example, consider the systematic errors on the GW150914 measurement of $\delta \phi_{i}$ (or $\beta$) reported in~\cite{TheLIGOScientific:2016pea,TheLIGOScientific:2016src}, i.e. the distance from the peak of the posterior to zero; these systematic errors are comparable to the statistical error, the former of which is not included in the Fisher bound. In spite of these differences, the constraints on $\beta$ that we found through our analysis are very close to those reported by the LVC, whenever we can compare them directly. 

Interestingly, Ref.~\cite{cornish-PPE} had already estimated the accuracy to which the ppE parameter $\beta$ could be constrained, using a detailed Bayesian analysis and inspiral-only waveforms. For comparison, we have taken the results of~\cite{cornish-PPE} directly and plotted them in Fig.~\ref{fig:beta-cons} (dotted blue line) without modification, i.e.~for a system with an SNR of 20, (source) component masses $(m_{1},m_{2}) = (12,6) M_{\odot}$ ($\eta = 0.222$) and luminosity distance $D_{L} = 462 \; {\rm{Mpc}}$. It is quite a coincidence that these parameters are so close to those of the detected events. The constraints predicted in 2011 are in good agreement with the actual constraints placed by aLIGO, in particular with GW151226. The spikes in the constraints of~\cite{cornish-PPE} arise due to degeneracies with the chirp mass and the phase of coalescence, the former of which is partially broken when one incorporates the merger-ringdown phase, as done in~\cite{TheLIGOScientific:2016pea,TheLIGOScientific:2016src}. 

For comparison, Fig.~\ref{fig:beta-cons} also includes an estimate of the bounds that the double binary pulsar PSR J0737-3039~\cite{Yunes:2010qb,Sampson:2013wia} (black dashed line) and the perihelion precession of Mercury~\cite{Sampson:2013wia} (cyan star)\footnote{We updated bounds on parameterized post-Newtonian (ppN) parameters from~\cite{Sampson:2013wia} using~\cite{Will:2014kxa}.} can place on the ppE $\beta$ parameter. The latter places a stronger bound on $\beta$ at 1PN order than the GW events, though it can only probe corrections in the conservative sector, such as those to the binding energy of the source. 
The binary pulsar estimates are rough, since they do not take into account possible covariances between $\beta$ and other binary pulsar parameters, and only dissipative corrections (namely those in the energy flux) are included. Nonetheless, they are enough to illustrate that binary pulsars can do a much better job at constraining ``negative PN'' order effects, while aLIGO (and in particular event GW150914) can beat binary pulsar constraints above Newtonian ($0$PN) order, as first suggested in~\cite{Yunes:2010qb}. 

One must keep in mind, however, that a \emph{direct} comparison of bounds on $\beta$ from binary pulsar observations and Solar System experiments to those obtained with GWs is misleading, since the former cannot probe gravity when compact objects merge (in particular, when the compact objects are BHs). Moreover, a stronger constraint on $\beta$ from one class of observation compared to another does not necessarily translate to a comparable improvement in bounds on non-GR theories, as the latter depends on how $\beta$ is related to theoretical coupling constants and binary parameters. Similarly, ppN bounds from the Sun-Mercury system cannot directly be used to bound the $\beta$ ppE coefficient, as these parameters may depend on both theoretical and system parameters. For example, if such parameters are proportional to the mass ratio, Solar System experiments may not be as sensitive as GW observations. Therefore, the binary pulsar and Solar System bounds on $\beta$ in Fig.~\ref{fig:beta-cons} should be considered as a mere reference and should not be compared directly to those from GW150914 and GW151226.
 
Let us now map the constraints on the ppE parameter $\beta$ to specific theoretical mechanisms, some of which we already summarized in Table~\ref{tab:summary2} (see also Table ~\ref{tab:summary}). Constraints listed in the top part of this table are obtained from modifications to GW generation. For reference, we also present current constraints on example alternative theory parameters obtained from other observations, such as table-top experiments and observations with low-mass X-ray binaries. We find the following theoretical implications of the two detected events on GW generation: 

\begin{itemize}[itemsep=0.1pt, topsep=5pt, partopsep=0pt]

\item \emph{EdGB gravity}:
GW150914 cannot place constraints on EdGB gravity due to degeneracies between the spin magnitudes, the component masses and the EdGB coupling constant in the leading-order (dipole) EdGB waveform correction. If one were to assume \emph{a priori} that the spins of the binary constituents of GW150914 are zero and the components masses are given by their posterior peaks, then one would be able to constrain $\sqrt{|\alpha_\EDGB|} \lesssim 22$km, which is consistent with the prediction of~\cite{Yagi:2011xp,Yagi:2015oca}. Without this \emph{a priori} information and using spin combinations within the 90\% posterior distribution measured by aLIGO~\cite{TheLIGOScientific:2016wfe}, the constraint on the EdGB coupling constant weakens dramatically (see Appendix~\ref{app:EdGB}). Saturating this weakened constraint leads to a large value of $\zeta_{\EDGB}$, which violates the small-coupling approximation used to derive the waveform correction in the first place [Eq.~\eqref{eq:beta-EdGB}].  

Repeating the analysis with GW151226, one finds the bound $\sqrt{|\alpha_\EDGB|} \lesssim 5.1$km, which gives $\zeta_\EDGB = 0.76$ and satisfies the small coupling approximation. However, if one further varies the mass ratio, which is less strongly constrained for GW151226 than GW150914, one finds that a set of masses and spins can shut off the scalar dipole radiation. This leads to a very weak constraint that violates the small coupling approximation. 

These GW events may still be able to place meaningful constraints on EdGB gravity from higher PN order corrections and from the waveform structure during merger and ringdown, but these have not yet been calculated. 

\item \emph{dCS gravity}:
As in the EdGB case, GW150914 and GW151226 cannot place meaningful constraints on dCS gravity because of degeneracies between the spin magnitudes, the component masses and the dCS coupling constant in the leading-order dCS waveform correction. If one were to assume \emph{a priori} that the spins of the BHs in GW150914 are zero, which is consistent with the 90\% posterior distribution measured by aLIGO~\cite{TheLIGOScientific:2016wfe}, then the leading-order dCS modification would vanish exactly, and the next-to-leading order correction would enter at very high PN order~\cite{pani-DCS-EMRI,Yagi:2011xp}. This would lead to an extremely weak constraint on the dCS coupling constant that would violate the small coupling approximation adopted to derive Eq.~\eqref{eq:beta-dCS}.  GW151226 is inconsistent with both of the BHs being nonspinning, but the resulting bound on dCS of $\zeta_\dCS \lesssim 10^3$ violates the small coupling approximation.

\item \emph{Scalar-tensor theories with BH scalar growth due to the excitation of a time-dependent scalar field~\cite{Jacobson:1999vr,Horbatsch:2011ye}}: GW150914 and GW151226 cannot place constraints on $\dot{\phi}$ because of degeneracies between this quantity and the component masses and the spin magnitudes in the waveform correction. Choosing spin magnitudes that lead to the weakest (most conservative) constraint with masses fixed to the injected ones, one finds the bound $\dot{\phi} < {\cal{O}}(10^{4}/{\rm{sec}})$ for GW150914. Saturating this constraint, however, leads to a dimensionless expansion parameter that violates the small coupling approximation $m_{A} \dot{\phi} \ll 1$, which was used to construct the waveform deformation~\cite{Jacobson:1999vr,Horbatsch:2011ye}.  With GW151226, the most conservative bound is $\dot{\phi} \lesssim 5\times 10^{3}/{\rm{sec}}$. Although such a bound satisfies the approximation, if one further varies the mass ratio, one can find a set of masses and spins such that scalar dipole radiation is highly suppressed. This leads to a very weak constraint on $\dot{\phi}$ that again violates the small coupling approximation. 

\item \emph{EA and khronometric theory}:
GW150914 and GW151226 can place constraints on EA and khronometric theory, although these are weaker than the current binary pulsar bound. Since the bounds on EA and khronometric theory in Table~\ref{tab:summary2} are derived from the bounds on $\beta_{\AE}^{(0PN)}$ and $\beta_{\KG}^{(0PN)}$, they correspond to assuming that scalar and vector dipole radiation are suppressed \emph{a priori}. This can be justified for NSs, but not yet for BHs, as their scalar charges have not been calculated. If one includes both $\beta_{\AE}^{(-1PN)}$ and $\beta_{\AE}^{(0PN)}$ or $\beta_{\KG}^{(-1PN)}$ and $\beta_{\KG}^{(0PN)}$ in the parameter set, as done e.g.~in the projected constraints of~\cite{Hansen:2014ewa}, the bounds on $\beta_{\AE}^{(0PN)}$ and $\beta_{\KG}^{(0PN)}$ become weaker due to degeneracies. Nonetheless, although the GW constraints are weaker than current bounds, they arise entirely from interactions that take place in BH spacetimes with extreme gravity, where one could have expected such effects to be enhanced.

\item \emph{Extra dimensions and temporal variation of $G$}: 
GW150914 and GW151226 can place constraints on the size of large extra dimensions\footnote{The aLIGO constraint on the size of the extra dimension in the RS-II braneworld model is based on the conjecture that classical BHs evaporate~\cite{emparan-conj,tanaka-conj}. Given that static, brane-localized BH solutions have now been constructed~\cite{Figueras:2011gd,Abdolrahimi:2012qi,Wang:2016nqi}, it is not clear whether classical BHs actually do evaporate. If they do not, then BH observations cannot be used to constrain the size of extra dimensions in the way discussed here. An alternative approach is to use the correction to the binding energy discussed in~\cite{Garriga:1999yh,Inoue:2003di}.} and any time-variation in $G$, but these are worse than other current constraints, such as those imposed with binary pulsars. This is because these effects enter at $-4$PN order, which implies that binary pulsar observations (as shown in Fig.~\ref{fig:beta-cons}) and (low-mass) BH X-ray binaries lead to much stronger limits. Constraints that could be placed by space-borne detectors, such as eLISA~\cite{Seoane:2013qna} and DECIGO~\cite{setoDECIGO}, could be many orders of magnitude stronger than aLIGO (and competitive with current bounds)~\cite{Yunes:2009bv,yagi:brane}. Once more, nonetheless, the GW constraints are unique in that they use data from the extreme gravity regime. 

\end{itemize}
 
\subsubsection{Generic Constraints on the Generation of GWs}  
\label{sec:gen-constr-in-generation}

\begin{figure}[t]
\begin{center}
\includegraphics[width=\columnwidth,clip=true]{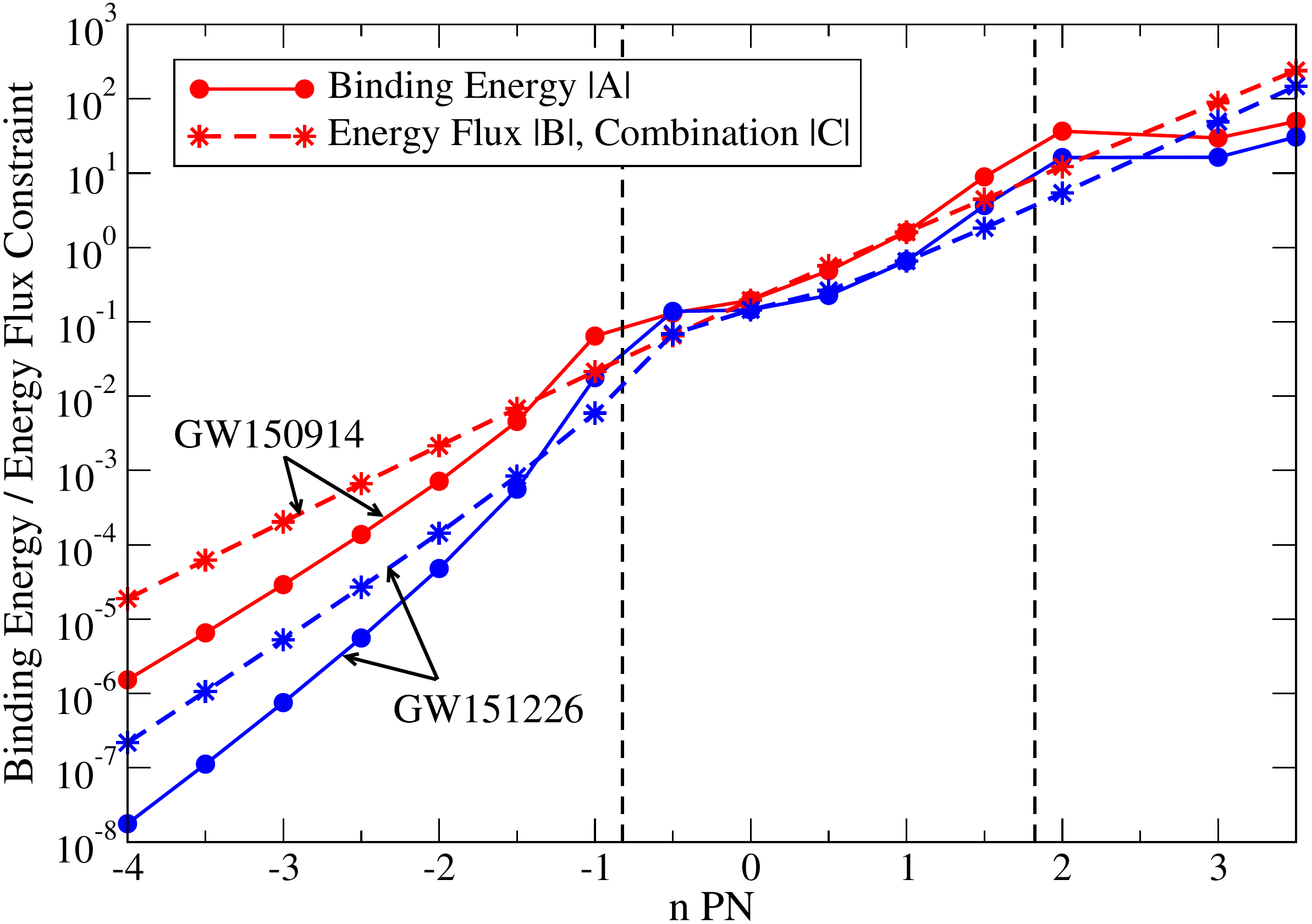} 
\caption{\label{fig:AB-constraint} (Color online) Upper bound on corrections to the binding energy $|A|$, energy flux $|B|$ [see Eq.~\eqref{eq:binding-energy-flux}] and a combination of these two $|C|$ [see Eq.~\eqref{eq:beta-ABC}] as a function of the PN order that they enter for GW150914 (red) and GW151226 (blue).
}
\end{center}
\end{figure}

What other generic features of GR in the generation phase can GW150914 and GW151226 constrain? To address this question, here we map the constraints on the ppE parameter $|\beta|$ in Fig.~\ref{fig:beta-cons} to those on generic corrections to the binding energy $E_{b}$ of a binary and the radiated energy flux $\dot E$. We follow the ppE treatment of~\cite{Chatziioannou:2012rf} and model such corrections as 
\be
\label{eq:binding-energy-flux}
E_{b} = E_{b,\GR} \left( 1 + A v^{2p}  \right)\,, \qquad \dot E = \dot E_\GR \left( 1 + B v^{2q}  \right)\,,
\ee
where $v = (\pi m f)^{1/3}$ corresponds to the relative orbital velocity, while $E_{b,\GR}$ and $\dot E_\GR$ denote the binding energy and energy flux in GR. Non-GR fractional corrections to $E_{b}$ and $\dot E$ have a magnitude $A$ and $B$ that enter first at $p$ and $q$ PN order respectively. Such corrections propagate to those in the gravitational waveform phase. When $p < q$, the dominant non-GR effect comes from the correction to the binding energy; we do not know of any theory where this is the case. When $p > q$, the dominant effect comes from the correction to the energy flux; examples of this include BD, EdGB and EA theory. When $p=q$, both corrections to $E_{b}$ and $\dot E$ are of comparable PN order, as is the case in dCS gravity. The mapping between these parameters and $\beta$ is given by~\cite{Chatziioannou:2012rf}\footnote{The $\beta$ used in~\cite{Chatziioannou:2012rf} is different from that in this paper by a factor of 2.} 
\be
\label{eq:beta-ABC}
\beta =
\begin{cases} 
      -\frac{5}{32} \frac{2 p^2 - 2 p - 3}{(4 - p) (5 - 2 p)} \eta^{-2 p/5} A & (p < q)\,, \\
      -\frac{15}{32} \frac{1}{(4 - q) (5 - 2 q)} \eta^{-2 q/5} B & (p > q)\,, \\
     -\frac{15}{32} \frac{1}{(4 - k) (5 - 2 k)} \eta^{-2 k/5} C & (p=k=q)\,,
   \end{cases}
\ee
with $C \equiv [(2 k^2 - 2 k - 3) A +3 B]/3$.

Figure~\ref{fig:AB-constraint} presents the upper bound on $|A|$, $|B|$ and $|C|$ obtained by mapping the bound on $|\beta|$ in Fig.~\ref{fig:beta-cons} via Eq.~\eqref{eq:beta-ABC}. This figure shows that the GW data bounds the magnitude of corrections to $E$ and $\dot E$ to much better than unity in the negative PN region. GW151226 places stronger constraints than GW150914, as expected from Fig.~\ref{fig:beta-cons}. Since the mapping for $p>q$ and $p=k=q$ has the same structure, the bound on $|B|$ and $|C|$ coincide. $\beta$ with $p < q$ vanishes when $p = (1 \pm \sqrt{7})/2 \sim -0.82$ and $1.8$, and hence, the constraint on $|A|$ becomes significantly weaker close to these two values of $p$, as shown by the vertical dashed lines. 

These generic constraints on the binding energy and energy flux can be used to place bounds on generic scalar field interactions in any theory. For example, if the BH components of a binary acquire scalar hair of $\ell$th multipole order (or $\ell$th scalar hair), the interaction of this scalar field will produce a correction to the binding energy at $2\ell$~PN order and a correction to the energy flux at $(3 \ell -1)$~PN order~\cite{Stein:2013wza}. The scalar field in EdGB gravity and in scalar-tensor theories gives rise to BH scalar hair of $\ell=0$ (monopole) order, which modifies the binding energy and energy flux at 0PN and $-1$PN order respectively, the latter being the well-known dipole radiation. On the other hand, in dCS gravity BHs acquire scalar hair of $\ell=1$ multipole order, and thus, the correction to the binding energy and energy flux enter both at 2PN order, with the latter being scalar quadrupolar radiation. 

Another generic feature that the GW events can constrain is the sudden activation or deactivation of dipole radiation at a given transition frequency $f_*$ during BH binary inspirals. Such an abrupt activation and deactivation is known to arise in certain modified theories in the presence of matter. An example is \emph{dynamical} scalarization in scalar-tensor theories~\cite{ST1,Palenzuela:2013hsa,Shibata:2013pra,Taniguchi:2014fqa,sennett,Sampson:2014qqa}, during which the scalar charge of a NS in a binary grows suddenly at a given threshold binding energy or frequency, abruptly turning dipole radiation on. A similar mechanism arises in scalar-tensor theories without dynamical scalarization but with a massive scalar field~\cite{Cardoso:2011xi,Yunes:2011aa,Alsing:2011er,Berti:2012bp}, during which  scalar dipole radiation activates at a transition frequency related to the mass of the scalar field. Scalar field deactivation occurs in scalar-tensor theories that allow for \emph{induced} scalarization~\cite{damour_esposito_farese,ST0,ST1,Palenzuela:2013hsa}, during which the scalar charge of one of the NSs in a binary induces a scalar charge in its binary companion, suppressing scalar dipolar radiation since this is proportional to the square of the difference in scalar charge~\cite{ST1,Palenzuela:2013hsa}. 

Whether a sudden activation or deactivation of dipole radiation is possible in vacuum spacetimes has not been investigated in the theoretical realm. One possibility is to consider EdGB gravity with a massive scalar field, in which case one would expect scalar dipole radiation to turn on at a threshold frequency related to the scalar mass. A mass for the EdGB (dilaton) scalar field could arise due to super-symmetry breaking~\cite{O'Neill:1993qz}, in which case the mass would be of the order of the super-symmetry breaking scale. If one wishes for super-symmetry to resolve the hierarchy problem, then this scale must be larger or comparable to $10^{12}$ eV, which then leads to a very massive dilaton ($\sim 10^{12}$eV), and thus, a very high threshold frequency ($\sim 10^{26}$Hz) that is well outside what can be probed with these GW observations. Another possibility is to consider scalar-tensor theories with non-trivial initial data~\cite{Healy:2011ef} for the scalar field, in which case the scalar field will evolve until it is absorbed by the BHs or it scatters to infinity, at which point dipole radiation will cease. The deactivation of dipole radiation in a vacuum spacetime could also be present if Kerr BHs can acquire scalar hair and then lose it during the inspiral, which could occur in GR~\cite{Herdeiro:2014goa,Herdeiro:2015gia,Herdeiro:2015tia,Herdeiro:2015waa} and in (complex-boson) scalar-tensor theories~\cite{Kleihaus:2015iea}.

Observations of GWs with aLIGO can constrain such sudden scalar field activation, as first studied in~\cite{Sampson:2013jpa}. This work showed through a Bayesian, model-hypothesis study that the simplest (one-parameter) ppE model is sensitive to a sudden turn on/off of dipole radiation if present in the data. Furthermore, a modification where this ppE model's phase term is multiplied by a Heaviside function with a new threshold frequency parameter is even more effective at detecting an abrupt activation or deactivation of dipole radiation anywhere in the aLIGO band.  
 
\begin{figure}[t]
\begin{center}
\includegraphics[width=\columnwidth,clip=true]{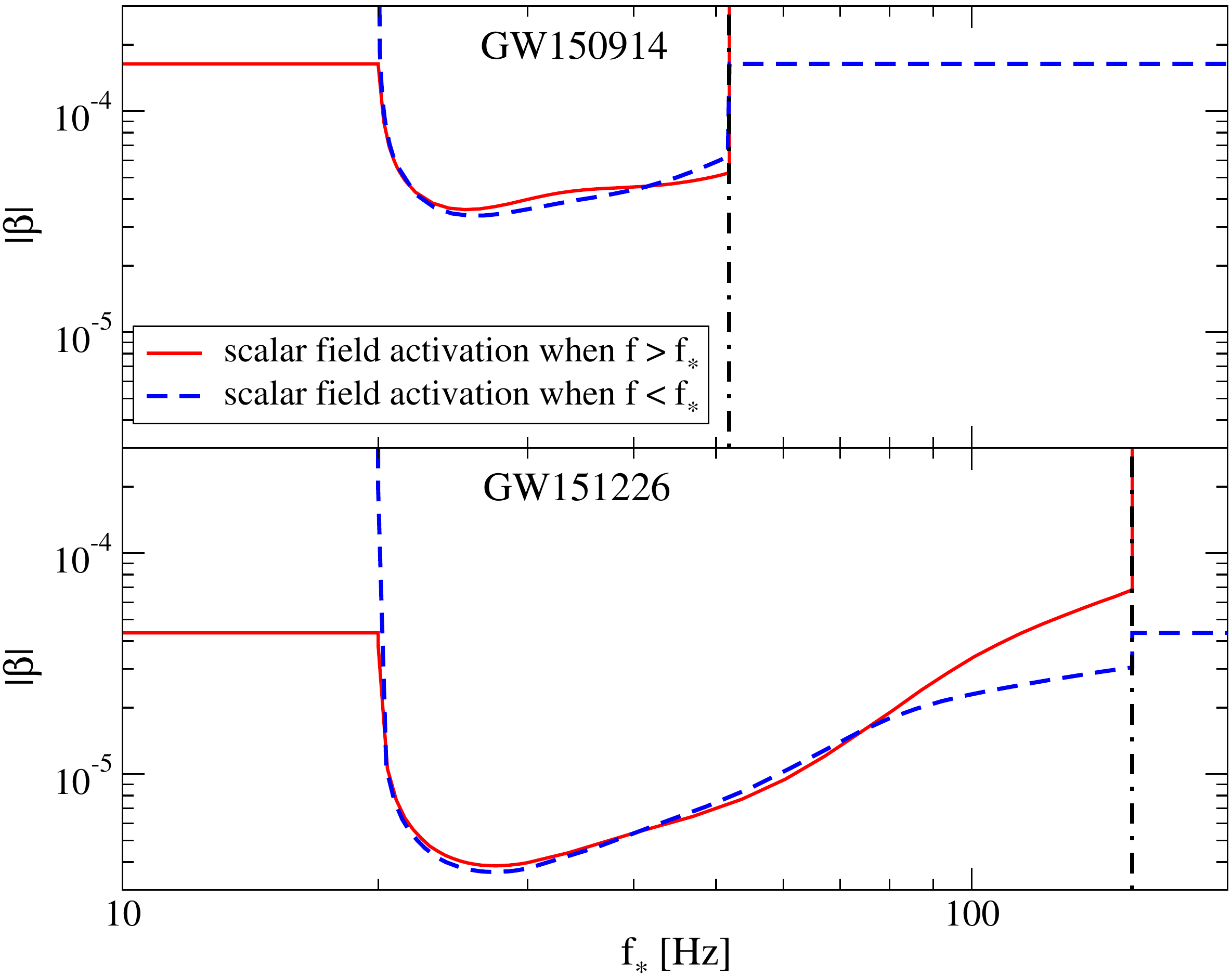} 
\caption{\label{fig:dyn-scalarization} (Color online) 90\%-confidence upper bound on the ppE parameter $|\beta|$ at $-1$PN as a function of the transition frequency $f_*$ for GW150914 (top) and GW151226 (bottom), for theories in which the scalar field only activates when $f > f_*$ (red solid) and $f < f_*$ (blue dashed). The vertical dotted-dashed line corresponds to the transition frequency $f_\mrm{Int}$ between the inspiral and intermediate phase in the IMRPhenom waveform. 
}
\end{center}
\end{figure}

GW150914 and GW151226 can thus place generic constraints on abrupt dipole-like changes to GW generation. The red solid curves in Fig.~\ref{fig:dyn-scalarization} present the upper bound on $|\beta|$ at $-1$PN versus $f_*$, assuming that the scalar field only activates when $f > f_*$. Since the correction is included in the inspiral phase only, one cannot place constraints in the region $f_* > f_\mrm{Int}$ with this method, where $f_* = f_\mrm{Int}$ is shown by the vertical dotted-dashed line. The constraint becomes stronger when the scalar field evolves during the observed inspiral phase $(20 \mrm{Hz} < f_* < f_\mrm{Int})$ compared to the case where the scalar field has already activated before the signal enters the aLIGO observation band $(f_* < 20\mrm{Hz})$. For example, the GW150914 constraints on $|\beta|$ with $f_*=40$Hz (the scalar field activates while the signal is in the observational frequency band of aLIGO) and $f_*=10$Hz (the scalar field is already on when the signal enters the band) are $|\beta|<4.5 \times 10^{-5}$ and $|\beta|<1.6 \times 10^{-4}$ respectively. This is because the former has a very distinct feature which helps break degeneracies between $\beta$ and other parameters. The constraint on $|\beta|$ does not go smoothly to infinity at $f_* = f_\mrm{Int}$. This is because when $f_* < f_\mrm{Int}$, the correction introduced in the inspiral phase also propagates to intermediate and merger-ringdown phases through the smooth matching condition of the phase at interfaces, while such corrections disappear completely from the template when $f_* > f_\mrm{Int}$  (see Appendix~\ref{app:BvsD} and~\cite{Mandel:2014tca} for a related discussion). Similar features can be seen for the case where the scalar field only activates when $f < f_*$ (blue dashed curves). A comparison between GW150914 (top) and GW151226 (bottom) shows that smaller mass systems allow one to probe scalarization effects in a wider frequency band with better accuracy. Lacking a concrete theory that predicts dynamical scalarization in the coalescence of black hole binaries, we cannot map the constraints on $\beta$ to bounds on fundamental constants of any known theory.  

\subsection{Implications on the Propagation of GWs}
\label{sec:mod-disp-rel}

We now study how strongly one can constrain the modified dispersion relation of the graviton using GW150914 and GW151226 (see also~\cite{Wu:2016igi,Kahya:2016prx} for possible constraints on the equivalence principle with GW150914 through a Shapiro time delay measurement). As described in Sec.~\ref{subsec:par-def-to-physics}, we include $\beta_\MD$ in Eq.~\eqref{zeta} in the IMRPhenom waveform in all phases (inspiral, merger and ringdown). We then carry out a Fisher analysis and derive upper bounds on $|\mathbb{A}|$ as a function of $\alpha$. Such an analysis corresponds to extending that in~\cite{mirshekari} by including also the merger and ringdown effects and using the specific parameters of the two GW events.
\begin{figure}[htb]
\begin{center}
\includegraphics[width=\columnwidth,clip=true]{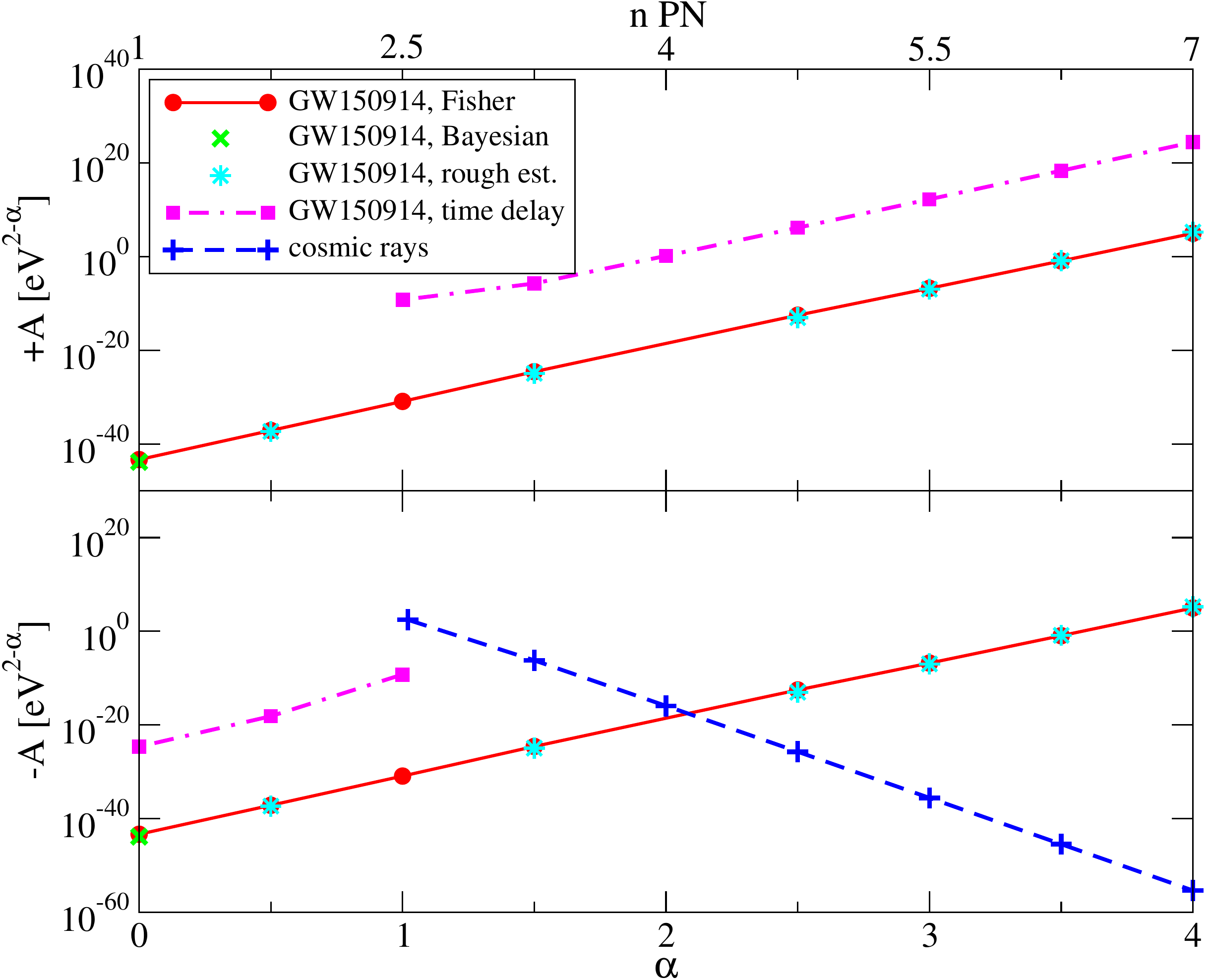} 
\caption{\label{fig:A-constraint} (Color online) Upper bound on the amplitude of the correction to the graviton dispersion relation $\mathbb{A}$ in Eq.~\eqref{dispersion} from GW150914 (the GW151226 bound is almost indistinguishable) as a function of $\alpha$ for $\mathbb{A}>0$ (top) and $\mathbb{A}<0$ (bottom) obtained from 90-\% confidence constraints on the ppE parameter $\beta$. The top axis shows the corresponding PN order at which the correction enters. The red circles are the Fisher estimates derived in this paper, while the green crosses are a mapping of the Bayesian bound in~\cite{TheLIGOScientific:2016src} on the graviton mass through $m_g = \sqrt{\mathbb{A}}$ at $\alpha=0$. Cyan stars are rough bounds on $\mathbb{A}$ in Eq.~\eqref{eq:rough-bound} based on the Bayesian bound at $\alpha=0$. The magenta squares correspond to a bound derived from the time of arrival of GW150914 at Hanford and Livingston~\cite{Blas:2016qmn}. Blue pluses present the bound on $\mathbb{A}$ from the absence of gravitational Cherenkov radiation in cosmic ray observations~\cite{Kiyota:2015dla}, assuming that cosmic ray particles of $p=10^{11}$GeV arrive from a distance of 100Mpc.
The GW150914 observation constrains $|\mathbb{A}|$, while cosmic ray observations only constrain the negative sector of $\mathbb{A}$. The former places a stronger bound on the $\mathbb{A} < 0$ region than the latter for $\alpha \lesssim 2$, while it places a unique bound on the $\mathbb{A} > 0$ region. 
}
\end{center}
\end{figure}

The upper bound on $|\mathbb{A}|$ from event GW150914 using a Fisher analysis is shown with red circles in Fig.~\ref{fig:A-constraint}. We do not show the bound at $\alpha = 2$ as $\beta_\MD$ is degenerate with $t_c$ in this case. For reference, we also show the bound on the superluminal propagation of GWs with magenta squares, derived in~\cite{Blas:2016qmn} from the difference in arrival times at Hanford and Livingston. When mapping the bound in~\cite{Blas:2016qmn} to that on $\mathbb{A}$ in Fig.~\ref{fig:A-constraint}, we assumed a GW frequency of $f = 100$Hz, corresponding to roughly the peak of the GW150914 signal. The new Fisher constraint is always stronger than the bound from the arrival time delay of GWs by roughly 20 orders of magnitude, except when $\alpha=2$ which cannot be constrained from the Fisher analysis presented here.

The GW151226 constraint on $\mathbb{A}$ is very similar to that from GW150914, but is \emph{weaker} by a factor of $\sim 5$ for large $\alpha$, which is the opposite of what happens in the generation mechanism case in Fig.~\ref{fig:beta-cons}. This is because at fixed frequency the velocities of the binary constituents are smaller for GW151226, which makes the ppE correction term with large $\alpha$ smaller, deteriorating the bound compared to GW150914. We also estimated the combined bound on $\mathbb{A}$ from both GW150914 and GW151226. We find that such a bound is stronger by $\sim 25\%$ compared to the bound from each event for smaller $\alpha$ where the GW150914 and GW151226 bounds are comparable, while the combined bound is dominated by the GW150914 bound for larger $\alpha$.

We can derive a simple, approximate expression for the GW150914 bound on $\mathbb{A}$. The Bayesian bound on $m_g$ in~\cite{TheLIGOScientific:2016src}, which corresponds to the $\mathbb{A} = m_{g}^{2}$ and $\alpha=0$ case for a simple dispersion relation, can be interpreted as a constraint on the propagation speed of GWs via Eq.~\eqref{eq:prop-speed} and $E = h f$, given by $|\delta_g| \equiv |1-v_g/c| \lesssim 4.5 \times 10^{-20}$ assuming $f = 100$ Hz. Reference~\cite{Ellis:2016rrr}\footnote{Reference~\cite{Ellis:2016rrr} actually used $f = \omega/2\pi= (100/2\pi)$ Hz and derived $|\mathbb{A}| \lesssim 10^{-5}$eV$^{-1}$. If one uses the more appropriate choice of $f = 100$Hz, one finds $|\mathbb{A}|  \lesssim 10^{-7}$eV$^{-1}$, which is consistent with the Fisher analysis in this paper to roughly $\sim 30\%$.} then obtained a rough bound on $|\mathbb{A}|$ at $\alpha=3$ from the constraint on $\delta_g$ above and Eq.~\eqref{eq:prop-speed}. Applying the same assumption to arbitrary $\alpha$, one finds the rough bound 
\be
\label{eq:rough-bound}
|\mathbb{A}| \lesssim \frac{1.5 \cdot 10^{-44} \; \mrm{eV}^{2-\alpha}}{1-\alpha} \left( \frac{10^{13}}{4.1} \right)^{\alpha} \left( \frac{f}{100\mrm{Hz}} \right)^{-\alpha}  \quad (\alpha \neq 1)\,.
\ee
which is shown with cyan stars in Fig.~\ref{fig:A-constraint}. Notice how accurate this order of magnitude estimate is relative to the more precise Fisher analysis carried out here. A similar analysis was performed in~\cite{Calcagni:2016zqv} to derive an order-of-magnitude GW bound on multifractional spacetime theories with $\alpha = 5/2$.

We now compare the GW bound on $\mathbb{A}$ to other existing bounds. In Fig.~\ref{fig:A-constraint}, we show the upper bound on $-\mathbb{A}$ from the absence of gravitational Cherenkov radiation in cosmic ray observations~\cite{Kiyota:2015dla}\footnote{The assumptions used in~\cite{Kiyota:2015dla} to derive constraints on $\mathbb{A}$ are valid only when $\alpha > 1$. One needs to re-derive the constraint without these assumptions to obtain constraints when $0 < \alpha < 1$.}. Such an observation can only constrain $\mathbb{A}<0$ sector, since otherwise there is no Cherenkov radiation. In this sector, the GW bound is stronger relative to the cosmic ray bound when $\alpha \lesssim 2$. On the other hand, in the positive $\mathbb{A}$ sector, the GW event places a unique constraint (one that is not possible with the Cherenkov argument). The GW bound, unfortunately, is very weak for high values of $\alpha$. For example, when $\alpha = 3$ or $4$, the bound on $\mathbb{A}$ normalized to the Planck energy $E_p$ becomes $|\mathbb{A} E_p| < \mathcal{O}(10^{20})$ and $|\mathbb{A} E_p^2| < \mathcal{O}(10^{60})$ respectively. Regarding table-top experiments, Blas and Lim~\cite{Blas:2014aca} derived the constraints $|\mathbb{A}| < 10^{8}$eV$^{-4}$ with $\alpha=6$. Using Eq.~\eqref{eq:rough-bound} with $\alpha=6$, one finds the GW bound of $|\mathbb{A}| < 6 \times 10^{29}$eV$^{-4}$, which is much weaker than the table-top bound.

Finally, we map constraints on $\mathbb{A}$ to example theories listed in Sec.~\ref{subsec:par-def-to-physics}. The results are summarized in the second half of Table~\ref{tab:summary2}, together with the current bounds obtained from e.g.~Solar System experiments and cosmic ray observations. We find the following theoretical implications of GW150914 and GW151226 on GW propagation:

\begin{itemize}[itemsep=0.1pt, topsep=5pt, partopsep=0pt]

\item \emph{Massive gravity}:
GW150914 (GW151226) constrains the mass of the graviton as $m_g < 2.2 \times 10^{-22}$eV ($< 2.3 \times 10^{-22}$eV) (see also~\cite{Bicudo:2016pps} for other constraints on the graviton screening mass from GW150914). These Fisher bounds are in good agreement with the bounds derived from a Bayesian analysis by the LVC~\cite{TheLIGOScientific:2016src}. The GW bounds are a few times stronger than the current Solar System constraint~\cite{talmadge} and more than two orders of magnitude stronger than the binary pulsar one~\cite{sutton}, although comparable to the superradiance bound of~\cite{Brito:2013wya}. On the other hand, they are weaker than the bound from galaxy cluster observations~\cite{goldhaber,Hare:1973px}, though the latter have larger systematic errors due to uncertainties in the dark matter distribution of the Universe. Such bounds can be applied to certain theories in which the graviton has a mass, such as Fierz-Pauli theory~\cite{Fierz:1939ix} and Lorentz-violating massive gravity~\cite{Rubakov:2004eb,Dubovsky:2004sg,Rubakov:2008nh}, but not to all such theories. In particular, these bounds cannot be applied to theories like bigravity~\cite{Hassan:2011zd} because, even though gravitons have a mass in this theory, they oscillate between physical and reference sectors, making the dispersion relation much more complicated~\cite{DeFelice:2013nba,Narikawa:2014fua}.

\item \emph{Multifractional spacetime}: The GW events place constraints on the characteristic energy scale $E_*$ for both timelike and spacelike fractal spacetimes. The former bound is unique while the latter bound is weaker than that from cosmic ray observations.

\item \emph{Double Special Relativity}: GW150914 and GW151226 constrain the characteristic length scale $\eta_\mrm{drst}$ for both positive and negative values. The former is unique while the latter is weaker than the cosmic ray bound.

\item \emph{Extra dimension theories}: GW150914 and GW151226 constrain the characteristic length squared $\alpha_\mrm{edt}$ for both positive and negative values. The former  is weaker than the cosmic ray bound while the latter is unique.

\item \emph{Gravitational SME}:
Table~\ref{tab:summary2} summarizes the GW150914 and GW151226 constraints on $\mathring{k}_{(I)}^{(4)}$, $\mathring{k}_{(V)}^{(5)}$ and $\mathring{k}_{(I)}^{(6)}$. For $\mathring{k}_{(V)}^{(5)}$, we present the bound obtained by Kostelecky and Mewes~\cite{Kostelecky:2016kfm} due to the apparent lack of birefringence, which modifies the real part of the dispersion relation and the propagation speeds of the plus and cross polarization modes\footnote{The absence of the birefringence in GW150914 can either mean that Lorentz violation effects are too small that the delay in the arrival time between the two tensor modes was not detected, or such effects are so large that the slower mode arrived when the detector was offline, or has not even arrived yet. Reference~\cite{Kostelecky:2016kfm} assumed the former, but due to the possibility of the latter, one can only exclude a certain finite range in the parameter space of the Lorentz violation coefficients in gravitational SME. For example, given that the O1 run lasted for 130 days, the data cannot rule out the parameter region that is above the threshold value corresponding to 9 orders of magnitude larger than the upper bound claimed in~\cite{Kostelecky:2016kfm}.}. Such a bound turned out to be slightly stronger than the Fisher bound on $\mathring{k}_{(V)}^{(5)}$. On the other hand, the bounds on $\mathring{k}_{(I)}^{(4)}$ and $\mathring{k}_{(I)}^{(6)}$ are new, and are complementary to cosmic ray bounds due to the absence of the gravitational Cherenkov radiation. In fact, the GW bounds are cleaner in the sense that they are bounds on the pure-gravity sector, whereas the latter is affected by the coupling between the matter and gravity sectors. Moreover, gravitational Cherenkov radiation may even be forbidden for certain ranges of the coefficients~\cite{Kostelecky:2016kfm}, and obviously the Cherenkov bound becomes invalid in this case. 

\item \emph{Ho\v rava-Lifshitz gravity}: GW150914 and GW151226 constrain a combination of the coupling parameters $\kappa_\mrm{hl}^4 \, \mu_{hl}^2$. Such bounds are unique and cannot be constrained from cosmic ray observations.

\end{itemize}

\section{Theoretical Implications for Exotic Spacetimes}
\label{sec:generictests}

The observation of the ringdown of GW150914 is consistent with the merger forming a single Kerr BH, and can be used to place stringent constraints on exotic compact objects alternatives for the remnant. Ref.~\cite{Chirenti:2016hzd} compared the measured ringdown frequency and damping time of GW150914 with the QNMs of a rotating gravastar, and found that the GW150914 remnant is unlikely to be such an object. Similarly, Ref.~\cite{Konoplya:2016pmh} compared the same frequency and damping time to the dominant QNM of a scalar field propagating in a parametrically deformed Kerr spacetime to place a constraint on the latter, though there is significant degeneracy here with the spin of the deformed Kerr object. Such tests, and the overall line of reasoning, however, is much more nuanced than it would at first seem for the following three reasons.  

First, such tests require that one chooses an exotic compact object or a specific Kerr deformation to compare GW150914 against, but there are many alternatives, most of which have severe theoretical problems from the start. The gravastars used in~\cite{Chirenti:2016hzd}, for example, are ``cut-and-paste'' spacetimes where an interior de Sitter metric is glued to an exterior BH metric through a boundary layer of exotic matter; to our knowledge, such constructions are not realized naturally in GR or in any known modified theories of gravity. Moreover, all horizonless compact objects with stable circular photon orbits, including gravastars, are likely to be unstable due to ergoregion instabilities if they are spinning rapidly~\cite{Cardoso:2007az,Pani:2010jz}. The deformed Kerr metric used in~\cite{Konoplya:2016pmh} has an identical quadrupole moment to Kerr, but with a ``shifted'' event horizon. Yet, no known theory predicts such a deformed metric, and thus it is unclear what new physics this metric encodes. Thus, whether one can use these exotic objects as an ``in-principle'' caveat to the evidence of BH existence with GW150914 is, at best, questionable.

Second, pure ringdown tests of the Kerr hypothesis -- that the exterior metric of compact objects is given by the Kerr metric -- do not address how the perturbed Kerr spacetime arose to begin with. For GW150914 the presumed Kerr remnant is clearly produced by the inspiral of two compact objects, each consistent with being Kerr BHs as well. Since the properties of this phase (e.g.~its duration, power spectrum, etc.)~must be consistent with the binary merger problem in GR, the aLIGO observation places stringent constraints on the \emph{dynamics} of the compact objects that merged. Thus, whether alternatives such as gravastars, wormholes, or parametrically deformed BHs should be taken seriously in light of the GW150914 data is further questionable in that they do not have a sound theoretical underpinning to describe their dynamics, and consequent GW emission, during a collision.

Third, the observation of the beginning of the ringdown (right after merger) is not necessarily sufficient to distinguish between exotic compact objects that possess similar light rings. Recently, Refs.~\cite{Barausse:2014tra,Barausse:2014pra,Cardoso:2014sna,Cardoso:2016rao} argued that the frequency and damping time of the GWs emitted during the beginning of the ringdown are related to the orbital frequency and the instability timescale of circular null geodesics, roughly associated, in turn, with the light ring of the spacetime (see also~\cite{Nakamura:2016gri,Nakano:2016sgf,Nakamura:2016yjl} on relations between BH QNMs and the BH light ring, ergoregion and horizon). This is the case even if the GWs emitted at late times, when the QNMs dominate, are drastically dissimilar for different exotic compact objects, as is the case for wormholes~\cite{Cardoso:2016rao}. GW150914 did not have a SNR large enough to measure the late-time, purely QNM-dominated phase of the signal if the amplitude of such QNMs is small, and so it cannot constrain this class of exotic BH alternatives (though note the current examples arguing for this possibility suffer from the problems discussed in points one and two above).

With these issues in mind, this section explores the theoretical implications that one can infer from event GW150914\footnote{We do not consider GW151226 as the post-merger SNR is much smaller than that of GW150914~\cite{Abbott:2016nmj,TheLIGOScientific:2016pea}.} on the nature of exotic spacetimes from the combined late-inspiral, merger and ringdown in three ways. First, we describe how the connection between the inspiral, merger and ringdown can provide information about the spacetime through the location of an effective ISCO, and how this could place constraints on modified gravity theories and on generic metric deformations. Unlike the prior study in~\cite{Konoplya:2016pmh} that focused on the ringdown phase, we use the entire late-inspiral, merger and ringdown phases of the GW150914 event. 

Second, we study what inferences one can draw from GW150914 on the nature of an exotic compact object remnant without appealing to any particular theory. We cast these inferences in the form of effective bulk and shear viscosities that would be required to explain the rapid damping time that aLIGO measured, assuming the bulk dynamics of the remnant can be characterized by viscous hydrodynamics (with appropriately exotic equation of state and transport properties). Even for non-material exotic alternatives this could still be a useful way to understand their dynamical behavior, akin to the membrane paradigm description of BHs in GR~\cite{membrane}. 

Third, we study how the GW150914 observation constrains the amplitude of a second oscillation mode, which can either be a higher-order, subleading BH QNM or a mode caused by additional matter oscillations of an exotic compact object. In contrast to other work~\cite{Chirenti:2016hzd}, our analysis is model independent and we do not assume specific properties of such exotic compact objects. We conclude by discussing how our result can be used to constrain actual QNMs of exotic compact objects with a light ring~\cite{Barausse:2014tra,Barausse:2014pra,Cardoso:2014sna,Cardoso:2016rao}.

\subsection{Implications on the ISCO Properties} 
\label{sec:ISCO} 

GW150914 is a so-called golden binary merger event, i.e.~one that allows the accurate extraction of the total mass lost during the merger~\cite{Hughes:2004vw}. Using such binaries, one can, for example, first estimate the final mass $M_f$ and the magnitude of the spin parameter vector $a_f$ (where $a_{f} = |\vec{a}_{f}| = |\vec{S}_{f}|/M_f$, with $\vec{S}_{f}$ the final spin angular momentum) of the remnant BH after merger from the inspiral part of the GWs using the phenomenological fit in~\cite{Healy:2014yta,Barausse:2009uz} \emph{and} assuming GR is correct. One can then compare this fitted prediction to the posterior distribution of the mass and spin parameter of the remnant BH extracted using only the ringdown (or post-inspiral) part of the waveform; the posterior then provides a set of best-fit parameters $\Delta M_f$ and $\Delta a_f$ for the deviation from GR, together with statistical uncertainties. The power of such a test was recently demonstrated using both Fisher~\cite{Nakano:2015uja} and Bayesian~\cite{Ghosh:2016qgn} methods. The latter method was first applied to the GW150914 data in~\cite{TheLIGOScientific:2016src}, demonstrating consistency with GR, albeit with a relatively large $90\%$ confidence contour about the GR value in the $(\Delta M_f/M_f,\Delta a_f/a_f)$ plane, due to the low (for this test) SNR of the ringdown part of the event.

Inspired by these results, here we pursue a different approach to probe the extreme gravity nature of the compact objects that produced event GW150914. In GR, Ref.~\cite{Buonanno:2007sv} proposed that the final spin angular momentum $\vec{S}_f$ of a BH after merger with $M_f \sim m$ (the difference between $M_f$ and $m$ is not significant in their analysis) is approximately given by the sum of the two individual spin angular momenta before merger and the orbital angular momentum of a ``test particle'' with mass $\mu \equiv m \eta$ ($m=m_1+m_2$) at $r_\ISCO$ orbiting around a Kerr BH with the following final spin:
\begin{align}
\vec{S}_{f} &= \vec{L}_\mrm{orb} (\mu, r_\ISCO,a_f) + \vec{S}_{1} + \vec{S}_{2}\,.
\label{eq:mapping-Kidder}
\end{align}
Indeed, this equation correctly reproduces numerical relativity simulations of the magnitude of the final BH spin after a merger of two non-spinning BHs within an error of $\sim 3\%$~\cite{Buonanno:2007sv}; one could use a more accurate fit that includes precession, such as that in~\cite{Barausse:2009uz}, but we leave such refinements to future work. Taking the projection of Eq.~\eqref{eq:mapping-Kidder} along the unit orbital angular momentum vector $\hat{L}=\vec{L}_\mrm{orb}/|\vec{L}_\mrm{orb}|$, and substituting in that the spin angular momentum of the $A$th BH with mass $m_{A}$ and (dimensional) spin vector $\vec{a}_{A}$ is $\vec{S}_{A} = m_{A} \vec{a}_{A}$, we can re-arrange this equation as
\begin{align}
\frac{L_{\mrm{orb}} (\mu, r_\ISCO,a_f)}{m^{2}} &= \frac{a_{\|f}}{m} - \frac{a_{\|s}}{m} - \delta_{m} \frac{a_{\|a}}{m}\,.
\label{eq:BKL}
\end{align}
Here $\vec{a}_s \equiv (\vec{a}_1 + \vec{a}_2)/2$ and $\vec{a}_a \equiv (\vec{a}_1 - \vec{a}_2)/2$
are the symmetric and antisymmetric combinations of the spin vector, $\delta_m = (m_1 - m_2)/m$, and the subscript ${ }_\|$ denotes projection of the corresponding vector quantity along $\hat{L}$.  We then see that a measurement of the individual component spins of the binary and of the final spin of the merged object can be used to infer the location of the ISCO through $L_{\mrm{orb}}$.

How accurately is the orbital angular momentum inferred from the GW150914 event? The individual masses and dimensionless spins associated with it were determined to be $(m_1, m_2) = (36.2^{+5.2}_{-3.8},29.1^{+3.7}_{-4.4})M_\odot$ and $(\chi_1,\chi_2) := (|\vec{a}_{1}|/m_{1},|\vec{a}_{2}|/m_{2}) = (0.32^{+0.47}_{-0.29},0.48^{+0.47}_{-0.43})$~\cite{TheLIGOScientific:2016wfe,TheLIGOScientific:2016pea}. The LVC also found that the so-called effective dimensionless spin $\chi_\mrm{eff}$, related to the projected symmetric spin combination by $a_{\|s} = m \chi_\mrm{eff} /2$, could be extracted to $\chi_\mrm{eff} = -0.06^{+0.14}_{-0.14}$ ~\cite{TheLIGOScientific:2016wfe,TheLIGOScientific:2016pea}. From the merger-ringdown phase, the final dimensionless spin was inferred from fitting formulas to numerical simulations to be $\chi_f := |\vec{a}_{f}|/M_f = 0.68^{+0.2}_{-0.58}$, where the error is extracted from the post-inspiral posterior distribution in the $\chi_f$--$M_f$ plane in the top panel of Fig.~3 in~\cite{TheLIGOScientific:2016src}. Since the median of such a posterior distribution is unclear from the figure, we simply adopt 0.68, which is the median value derived from a full inspiral-merger-ringdown analysis~\cite{TheLIGOScientific:2016wfe,TheLIGOScientific:2016pea}\footnote{In principle, one needs to rederive the median and error for the final spin measurement using a non-GR template. However, we are assuming here that the non-GR contribution to the radiated energy and angular momentum is negligible in the merger-ringdown phase. Thus, we use the error for the final spin extracted purely from the post-inspiral phase in GR and neglect possible non-GR contributions.}. To apply these aLIGO measurements to Eq.~\eqref{eq:BKL}, for simplicity, we assume $a_{\|f} = |\vec{a}_f|$, namely the final spin is aligned with the orbital angular momentum. For GW150914 we can also neglect contributions from the second and third terms in Eq.~\eqref{eq:BKL} since the former has been shown to be small from the measurement of $\chi_\mrm{eff}$, while the latter is suppressed by a factor of $\delta_m \sim 0.1$. Thus, using the error on $\chi_f$ from the merger-ringdown phase to estimate the error on $L_\mrm{orb}$ for GW150914 gives
\be
\label{eq:L-constr}
 \frac{L_\mrm{orb} (\mu, r_\ISCO,a_f)}{m^2} \approx 0.68^{+0.2}_{-0.58}\,.
\ee
This large error justifies us having ignored the intrinsic error of $\sim 3\%$ in Eq. ~\eqref{eq:BKL} coming from neglecting the angular momentum radiated after merger in GR.

The above calculation suggests that GW observations can constrain the predicted orbital angular momentum of a test particle at the ISCO, but with a strong caveat: such a constraint is only valid provided the mapping constructed in~\cite{Buonanno:2007sv} remains valid when non-GR physical mechanisms are active. In particular, the following two conditions need to be satisfied: (i) the non-GR contribution to the total angular momentum radiated during merger is negligible compared to the non-GR correction to the orbital angular momentum at ISCO and its location for a rotating BH, and (ii) the two-body dynamics can still be well described as a deformed effective one-body model in non-GR theories. Wether this is the case or not depends on the particular theory in question and should be studied through numerical simulations (that are currently not available) on a case-by-case basis.

Assuming the mapping in Eq.~\eqref{eq:mapping-Kidder} holds, we can then study the implications of Eq.~\eqref{eq:L-constr} on the hypothesis that the spacetime of BHs is that of the Kerr metric. This hypothesis is violated in a large class of modified gravity theories, where the  orbital angular momentum of a test particle at the equatorial ISCO can be written as
\begin{align}
L_{\mrm{orb}}(r_{\ISCO}) &= L_{\mrm{orb}}^{\Kerr}(r_{\ISCO}^{\Kerr}) 
\nn \\
&+ \zeta \left[ \left.\frac{\partial L_{\mrm{orb}}^{\Kerr}}{\partial r} \right|_{r_{\ISCO}^{\Kerr}} \!\!\! \delta r_{\ISCO} + \delta L_{\mrm{orb}}(r_{\ISCO}^{\Kerr}) \right] 
\nn \\
&+ {\cal{O}}(\zeta^{2})\,.
\end{align}
We have here expanded to linear order in the small deformation parameter $\zeta$, modeling the angular momentum of a test particle as $L_{\mrm{orb}} = L_{\mrm{orb}}^{\Kerr} + \zeta \; \delta L_{\mrm{orb}}$ and the location of the ISCO as $r_{\ISCO} = r_{\ISCO}^{\Kerr} + \zeta \; \delta r_{\ISCO}$. GW150914 then places a constraint on the combination
\begin{align}
\label{eq:gen-bound}
\frac{1}{m^2} \left( \left.\frac{\partial L_{\mrm{orb}}^{\Kerr}}{\partial r} \right|_{r_{\ISCO}^{\Kerr}} \!\!\! \delta r_{\ISCO} + \delta L_{\mrm{orb}}(r_{\ISCO}^{\Kerr}) \right) \lesssim {\cal{O}}(1)\,,
\end{align}
where on the right-hand side we have used that the error in Eq.~\eqref{eq:L-constr} is of order unity. This constrains a \emph{combination} of the modification to the angular momentum of a test-particle in a non-Kerr spacetime and a modification to the location of the ISCO. 

The constraint above can be refined further by specifying a parametrically deformed Kerr spacetime~\cite{mn,Collins:2004ex,glampedakis,vigelandhughes,vigeland,johannsen-metric,vigelandnico,Johannsen:2013rqa,Cardoso:2014rha,Rezzolla:2014mua,Johannsen:2015pca,Lin:2015oan,Konoplya:2016jvv}. To give a concrete example, let us consider the quasi-Kerr metric~\cite{glampedakis}, which is constructed to represent a generic deformation to the Kerr metric through a correction in its quadrupole moment
\be
Q = Q_K \left(1 + \frac{\zeta_\QK}{\chi^2} \right)\,,
\ee
where $\zeta_\QK$ is supposed to be a small (dimensionless) deformation parameter that controls the magnitude of the Kerr deviation. Such a spacetime describes a generic, asymptotically-flat and slowly-rotating vacuum spacetime in GR, including the exterior spacetime for slowly-rotating gravastars~\cite{Pani:2015tga}, provided the deformation away from Kerr is small. Using the orbital angular momentum and the shift in the location of the ISCO for a test-particle in an equatorial orbit of the quasi-Kerr metric in Eq.~\eqref{eq:L-constr} or~\eqref{eq:gen-bound}, one finds $-2.0 \lesssim \zeta_\QK \lesssim  5.3 \times 10^{-3}$, which is a constraint of order unity in the deformation parameter. 

A final refinement of this constraint is to consider specific modified gravity theories that violate the Kerr hypothesis, such as EdGB and dCS gravity~\cite{Ayzenberg:2016ynm,kent-CSBH}. In these two theories, the location of the ISCO of a BH with mass $M$ and dimensionless spin $\chi$ is modified from the GR prediction by~\cite{Ayzenberg:2014aka,kent-CSBH} 
\begin{align}
\delta r_\ISCO^\EDGB &= -\frac{16297}{9720} \zeta_\EDGB M \nn \\
& \times \left(1 + \frac{205982 \sqrt{6}}{440019} \chi - 
   \frac{1167369773}{9702418950} \chi^2 \right)\,, \\
\delta r_\ISCO^\dCS &= \frac{77 \sqrt{6}}{5184} \zeta_\dCS  M \chi \left(1 - \frac{9497219}{19559232} \chi \right)\,, 
\end{align}
while corrections to the orbital angular momentum can be found in Eq.~(68) of~\cite{Ayzenberg:2014aka} and Eq.~(99) of~\cite{kent-CSBH}, assuming slowly-rotating BHs to quadratic order in spin and working in the small-coupling approximation $\zeta \ll 1$ (where recall that $\zeta_\EDGB \geq 0$ and $\zeta_\dCS \geq 0$ by definition). Substituting these expressions in Eq.~\eqref{eq:L-constr} or~\eqref{eq:gen-bound} and truncating all expressions at quadratic-order in spin, we find the constraints $\zeta_\EDGB \lesssim 5.2$ and $\zeta_\dCS \lesssim 1.2 \times 10^3$. 

These bounds, however, are not compatible with the small-deformation approximation ($|\zeta_{\QK}| \ll 1$) and the small-coupling approximations ($\zeta_{\EDGB} \ll 1$ and $\zeta_{\dCS} \ll 1$) that were heavily used to derive the above expressions. Thus, we conclude that GW150914 cannot place meaningful constraints on mechanisms that modify the orbital angular momentum of a test particle at the ISCO if such a mechanism is built as a small deformation from Kerr. Either the mechanism must be known to all orders in the deformation parameter, such as in Lorentz violating theories of gravity~\cite{Barausse:2011pu,Barausse:2013nwa}, or we must wait for higher SNR GW observations that can constrain $a_{f}$ more accurately.

\subsection{Implications on the Effective Hydrodynamic Properties of Exotic Matter}  
\label{sec:exotic-hydro}

We now consider properties that exotic matter alternatives to BHs would need to have to be consistent with the signal seen by aLIGO.
We begin by treating such exotic compact objects within the framework of hydrodynamics by estimating the \emph{effective} viscosity that would be required to explain the observed damping time of $\tau = 4$ ms~\cite{TheLIGOScientific:2016src}, assuming large amplitude matter oscillations were produced by the merger. There are numerous ways to proceed here, and the specific numbers and physical properties will depend on the model. However, we emphasize that our treatment itself does not depend on the theory or nature of the exotic compact object; it is merely a way to characterize the properties of the exotic object using a mundane object whose properties we understand. 

For simplicity then, we consider our model to be a Newtonian, quasi-incompressible star with a density $\rho$ and radius $R$, perturbed by a spherical harmonic mode $Y_{\ell m}$, for which the following relationships have been derived~\cite{1987ApJ...314..234C}\footnote{Equation~\eqref{eq:shear} is valid for incompressible Newtonian stars, while Eq.~\eqref{eq:bulk} was derived for Newtonian stars with a non-relativistic Fermi gas equation of state (the prefactor of 5/3 corresponds to the adiabatic index of the fluid).}:
\begin{align}
\label{eq:shear}
\bar \eta &= \frac{1}{(\ell -1) (2 \ell +1)} \frac{\rho R^2}{\tau_{\bar\eta}}\,, \\ 
\label{eq:bulk}
\bar \zeta &= \left( \frac{5}{3} \right)^4 \frac{2 (2 \ell+3)}{\ell^3} \frac{\rho R^2}{\tau_{\bar \zeta}}\,,
\end{align}
where $\bar \eta$ and $\bar \zeta$ are the shear and bulk viscosity respectively, while $\tau_{\bar\eta}$ and $\tau_{\bar \zeta}$ are the damping time of oscillations associated with each type of viscosity. We restrict attention to the least damped mode, $\ell=2$, which is also the dominant GW generating mode that will be present  in the initial remnant following a two body, near equal mass collision. Eliminating $\rho$ from these expressions with
\be
\label{eq:rho}
\rho = \frac{m}{(4\pi/3) R^3}\,,
\ee
we obtain 
\begin{align}
\label{eq:eta-rem}
\bar \eta_\mrm{eff} &\sim 4 \times 10^{28} \frac{\mrm{g}}{\mrm{cm} \cdot \mrm{s}} \left( \frac{m}{65M_\odot} \right) \left( \frac{370\mrm{km}}{R} \right) \left( \frac{4\mrm{ms}}{\tau_{\bar \eta}} \right)\,, \\
\label{eq:zeta-rem}
\bar \zeta_\mrm{eff} &\sim 3 \times 10^{30} \frac{\mrm{g}}{\mrm{cm} \cdot \mrm{s}} \left( \frac{m}{65M_\odot} \right) \left( \frac{370\mrm{km}}{R} \right) \left( \frac{4\mrm{ms}}{\tau_{\bar \zeta}} \right)\,. 
\end{align}
We scaled these expressions by a fiducial damping time of $\tau = 4$ ms, a total mass of $m=65M_\odot$ and a radius of $R=370$ km (the orbital separation at the end of the inspiral $f=f_\Int$).
The above results should be interpreted as \emph{order-of-magnitude lower limits} to the effective viscosities of the matter comprising the remnant that aLIGO observed, \emph{assuming} the initial amplitude of the $\ell=2$ mode was such as to produce a GW signal close to the peak amplitude observed. For a typical material body collision, these viscosities would be an over-estimate, as part of the decrease in the amplitude after the merger is simply due to a decrease in the reduced quadrupole moment compared to that prior to contact. 

{
\newcommand{\minitab}[2][l]{\begin{tabular}{#1}#2\end{tabular}}
\renewcommand{\arraystretch}{1.2}
\begin{table}[tb]
\begin{centering}
\begin{tabular}{c|c|c|c|c|c}
\hline
\hline
\noalign{\smallskip}
& {\bf GW150914}  & BH & Boson star  & NS (n) & NS (B)\\
\hline
shear $\bar \eta$ & $\mathbf{4 \times 10^{28}}$  & $1 \times 10^{30}$ & $7\times 10^{26}$ & $2 \times 10^{14} $ &  $1 \times 10^{27}$\\
bulk $|\bar \zeta|$ & $\mathbf{3 \times 10^{30}}$ & $1 \times 10^{30}$ & $5\times 10^{28}$ & $6 \times 10^{28}$ &  ---\\ 
\noalign{\smallskip}
\hline
\hline
\end{tabular}
\end{centering}
\caption{%
Effective shear and bulk viscosities of compact objects in units of g~cm$^{-1}$~s$^{-1}$. The GW150914 viscosities are those of the remnant estimated in Eqs.~\eqref{eq:eta-rem} and~\eqref{eq:zeta-rem}. We assumed the BH and (solitonic) boson star mass of 65$M_\odot$ and the boson star radius of 1.5 times the Schwarzschild radius. ``NS (n)'' and ``NS (B)'' refer to NS effective viscosities due to neutron scattering and magnetic field damping respectively, with the stellar density, radius, temperature and magnetic field strength set to $10^{15}$g/cm$^3$, 12km, $10^{11}$K$\sim 10$MeV and $10^{15}$G respectively (typical values seen in simulations of magnetized NS mergers when a hypermassive remnant forms). See Appendix~\ref{app:viscosity} for more details.
}
\label{table:viscosity}
\end{table}
}

To put Eqs.~\eqref{eq:eta-rem} and~\eqref{eq:zeta-rem} in context, in Table~\ref{table:viscosity} we summarize effective viscosities of several known compact objects; the calculation of these numbers can be found in Appendix~\ref{app:viscosity}. Notice from Table~\ref{table:viscosity} that the magnitude of the viscosities of BHs are comparable to those of the remnant in Eqs.~\eqref{eq:eta-rem} and~\eqref{eq:zeta-rem}. On the other hand, for boson stars this is not the case, with effective viscosities that are 2 orders of magnitude smaller than those observed, assuming a solitonic configuration with mass of $65M_\odot$ and radius of $R=3M$. In fact, the frequency and damping time of the dominant QNM of such a boson star are $f \sim 160$Hz and $\tau \sim 320$ ms, which are also incompatible with the GW150914 event. Presumably, more compact boson stars would have QNMs with frequencies closer to the observed maximum frequency in GW150914, though the damping time would be more challenging to reduce to a level consistent with this observation. 

The viscosities of NSs with fiducial magnetic field strengths and temperatures (as measured in current binary NS merger simulations, see for e.g.~\cite{Kiuchi:2015sga}) are also in disagreement with those inferred from the remnant produced in the GW150914 event. If one can scale the relations for NS matter in Appendix~\ref{app:viscosity} to a 65$M_\odot$ NS-{\em like} object (which of course would require exotic fermionic matter as standard model NS equations of state cannot support masses above at most $\sim3M_\odot$), then an exotic remnant that at, or very shortly ($<4$ms) after merger already had a magnetic field strength of  $B_\mrm{rem} \gtrsim 4 \times 10^{16}$G and a temperature of $T_\mrm{rem} \gtrsim  150 \mrm{MeV} = 2 \times 10^{12}$K could be compatible with the signal of GW150914. 

\subsection{Implications on the Oscillation Modes of Exotic Objects} 
\label{subsec:osccill-modes} 

Can we infer the properties of exotic compact objects by constraining the amplitude of their oscillation modes with GW150914? If the binary constituents of GW150914 are Kerr BHs and the remnant is also a Kerr BH, the dominant QNM of the latter is the fundamental $\ell=m=2$ mode, with the next subleading modes being overtones of this and the $\ell = m = 3$ fundamental mode. The SNR, however, seems to be too low for the subdominant modes to be detectable~\cite{Berti:2007zu}. On the other hand, if the constituents and the remnant are exotic compact objects, they could produce matter oscillation modes that are longer lived than the $\ell=m=2$ mode of a Kerr BH~\cite{Abramowicz:1997qk}, and with amplitudes likely significantly larger than subleading modes of the Kerr-remnant case. In addition, Refs.~\cite{Barausse:2014tra,Barausse:2014pra,Cardoso:2014sna,Cardoso:2016rao} pointed out that when an exotic compact object with a light ring is perturbed by a test particle, the dominant GW mode will resemble that of a regular BH, followed by a set of exotic compact object QNMs, which will have a smaller amplitude than the primary $\ell=m=2$ mode but a longer damping time. The goal of this subsection is to place constraints on one additional mode (on top of the primary $\ell = m = 2$ mode) in terms of the new mode's oscillation frequency $\bar f_\RD$ and damping time $\bar \tau$ in a manner as agnostic as possible. 

We begin by explaining how we model the GWs emitted during QNM ringing for Kerr BHs in GR. The IMRPhenom waveform models the GW amplitude $A_\MR$ of the merger-ringdown phase as a product of a Lorentzian function and an exponential decay:
\be
\label{eq:amp-MR}
A_\MR = A_0 f^{-7/6} \frac{\gamma_1}{m}  \frac{\gamma_3  f_\mrm{damp}}{(f- f_\RD)^2+(\gamma_3 f_\mrm{damp})^2} e^{-\frac{\gamma_2 (f-f_\RD)}{\gamma_3 f_\mrm{damp}}}\,,
\ee
where $A_0$ is an overall amplitude factor that is common to the inspiral, intermediate and merger-ringdown phases, while the coefficients $\gamma_i$ are given by fits in terms of the symmetric mass ratio and spins~\cite{Khan:2015jqa}. The oscillation and the damping frequencies of the $\ell=m=2$ mode are $f_\RD$ and $f_\mrm{damp}\equiv 1/(2 \pi \tau)$, where $\tau$ is the mode's damping time. aLIGO measured $f_\RD$ and $\tau$ from the GW150914 event to be $f_\RD = 251\pm8$~Hz and $\tau = 4.0\pm0.3$~ms, using the full inspiral-merger-ringdown waveform in GR. Due to the exponential factor in Eq.~\eqref{eq:amp-MR}, the amplitude peaks at a frequency that is slightly \emph{lower} than $f_\RD$, i.e.~at $f_\MR = 222$~Hz, which also corresponds to the transition frequency between the intermediate and merger-ringdown phase.

Let us now explain how we model the GW amplitude of the exotic compact object's additional ringdown mode. For simplicity, we adopt the same model of Eq.~\eqref{eq:amp-MR}, but with the replacements $(\gamma_i, f_\RD, f_{\mrm{damp}}) \to (\bar \gamma_i, \bar f_\RD, \bar f_\mrm{damp})$; notice that this additional ringdown mode is very similar to the original ppE ringdown waveform of~\cite{PPE}, and is also similar to the extension used in~\cite{Sampson:2013jpa} to calculate projected constraints on the ringdown phase of future aLIGO observations.  The choice of $\bar \gamma_3$ does not affect our analysis as shifting this parameter is equivalent to redefining $\bar f_\mrm{damp}$; in fact, even in GR, $\gamma_3$ is completely degenerate with $f_\mrm{damp}$ in Eq.~\eqref{eq:amp-MR}. We thus set $\bar \gamma_3 = \gamma_3$ so that $\bar \tau$ can be directly compared to $\tau$. The choice of $\bar \gamma_2$ is more subtle; we set it to zero, $\bar \gamma_2 = 0$, to avoid having an artificial exponential enhancement in the amplitude when $f_\RD$ is much larger than $f$.

With these models at hand, we carry out a Fisher analysis to estimate a bound on $\bar \gamma_1$. We inject a GR GW signal ($\bar \gamma_1 = 0$) with the central values $f_\RD$ and $f_\mrm{damp}$ that aLIGO measured\footnote{This SNR is different from the post-inspiral SNR of $\sim 16$ found in~\cite{TheLIGOScientific:2016src}, as the latter was calculated from $f_\Int = 132$~Hz.}, which leads to an SNR of $\sim 7$. The parameters in our Fisher analysis are $(\gamma_1, \gamma_2, f_\RD, \tau, \bar \gamma_1)$. We also search for parameters of the primary oscillation mode, since we do not assume that the binary components are BHs, and hence we cannot use the fitting formulas presented in~\cite{Husa:2015iqa,Khan:2015jqa}. We only use the GW signal with $f \geq f_\MR$ and use the prior $\bar f_\RD \geq f_\MR$\footnote{A similar analysis could be used to place limits on lower frequency modes. The exponential term in Eq.~\eqref{eq:amp-MR} would then need to be modified to avoid issues when $f<f_{\MR}$. For brevity we do not do so here, focusing instead on the $f>f_{\MR}$ case as an illustrative example.}. Other parameters in the additional mode's amplitude, such as $\bar f_\RD$ and $\bar \tau$, cannot be measured when $\bar \gamma_1 = 0$ since derivatives of the amplitude with respect to such parameters vanish. Since here we are only constraining the amplitude of the additional mode, we only consider the amplitude in the waveform and set the waveform phase to be effectively independent of the above parameters when calculating the Fisher matrix.

\begin{figure}[htb]
\begin{center}
\includegraphics[width=\columnwidth,clip=true]{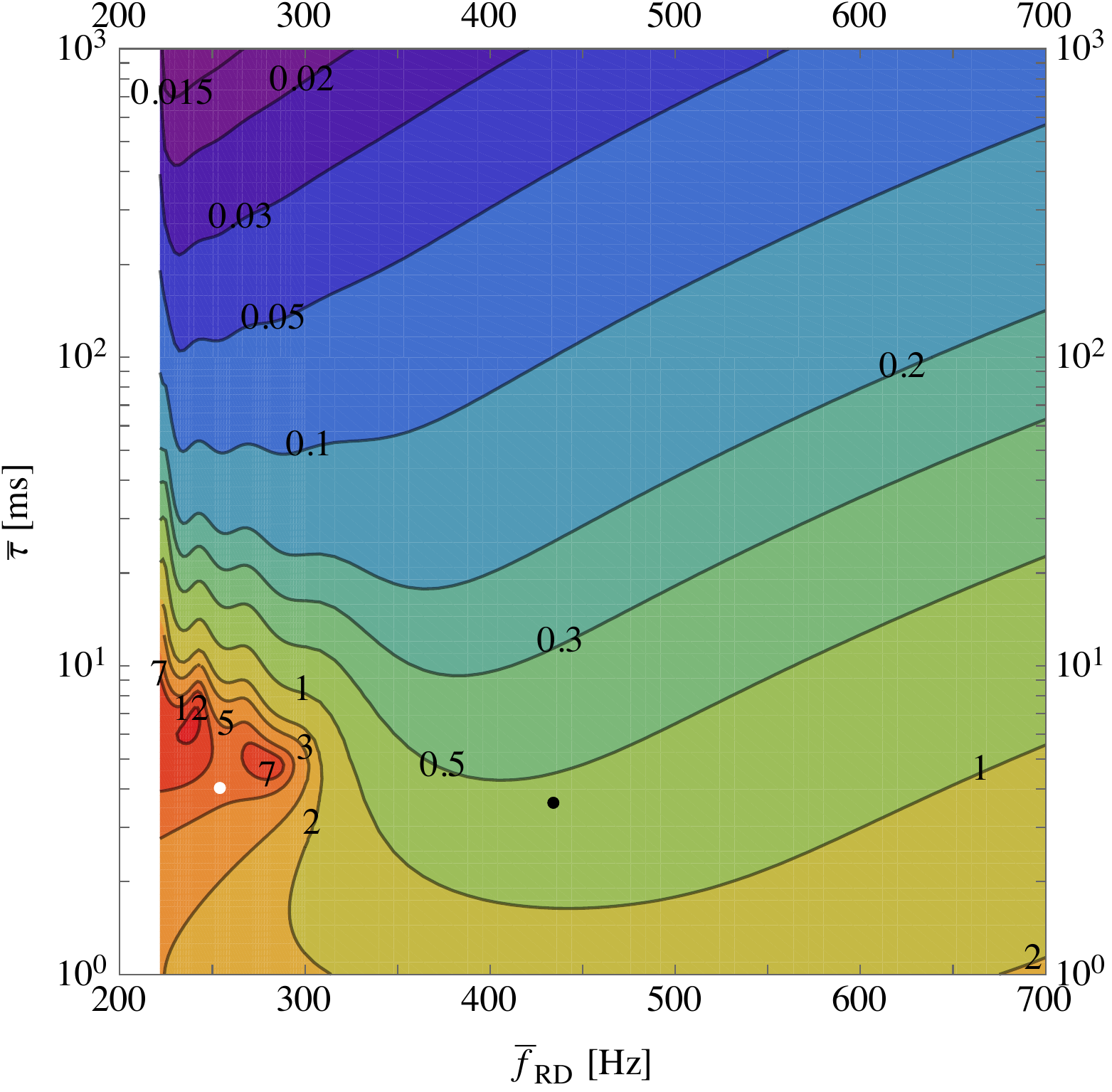} 
\caption{\label{fig:QNM} (Color online) 90\% confidence upper bound on the amplitude of an additional ringdown mode $\bar \gamma_1$ relative to the amplitude $\gamma_1$ of the primary $\ell=m=2$ mode as a function of the former's ringdown frequency $\bar f_\RD$ and damping time $\bar \tau$. $\bar \gamma_{1}$ can be constrained to be less than $\sim 10\%$ of the primary mode's amplitude if the damping time is larger than $100$ms, which is typical for boson star QNMs. The constraint becomes relatively weak around the white dot corresponding to the frequency and damping time of the primary $\ell=m=2$ mode, due to degeneracies between the primary mode and the additional mode. The black dot represents the subleading $\ell = m =3$ mode of a BH~\cite{2006PhRvD..73f4030B}, whose amplitude is smaller than $10\%$~\cite{Berti:2007zu}.
}
\end{center}
\end{figure}

Figure~\ref{fig:QNM} presents the upper bound on $\bar \gamma_1/\gamma_1$ as a function of $\bar f_\RD$ and $\bar \tau$. The constraint on the additional mode's amplitude becomes stronger when the oscillation frequency is close to $f_\MR$ and when $\bar \tau$ is large. In particular, the constraint is better than $10\%$ when $\bar{\tau} \gtrsim 100$ ms. This is as expected since the SNR of the additional mode becomes larger in this case with a fixed $\bar \gamma_1 \neq 0$. On the other hand, when $(\bar f_\RD,\bar\tau)$ are close to the frequency and damping time of the primary $\ell=m=2$ mode (white dot in the figure), the parameters become degenerate, which weakens the constraint. One can alternatively see Fig.~\ref{fig:QNM} as showing upper bounds on $\tau$ for a given $\bar \gamma_1$ and $\bar f_\RD$, which can be mapped to lower bounds on the viscosity via Eqs.~\eqref{eq:eta-rem} and~\eqref{eq:zeta-rem}. For example, when we assume $\bar \gamma_1 / \gamma_1 \leq 0.1$ and $\bar f_\RD = f_\RD$, one finds $\tau \lesssim 50$ms, which maps to $\bar \eta \gtrsim 3 \times 10^{27}$g/cm/s and $\bar \zeta \gtrsim 2 \times 10^{29}$g/cm/s.

Let us now discuss the implications of Fig.~\ref{fig:QNM} on the properties of the compact object remnant, assuming that the remnant of GW150914 was a BH. In such a case, the remnant would have emitted subdominant modes, such as the $\ell = m = 3$ one, whose amplitude is typically smaller than $10\%$ of the dominant mode~\cite{Berti:2007zu}. The frequency and damping time of such a mode can be derived from the fit in~\cite{2006PhRvD..73f4030B} to be $f_\RD = 433$~Hz and $\tau =3.6$~ms for a 62.3$M_\odot$ remnant BH spinning at $\chi = 0.68$, which is shown by the black dot in Fig.~\ref{fig:QNM}. The figure shows that it would be difficult for aLIGO to detect such a signal, as one can only distinguish the dominant and subdominant modes if the latter's amplitude is larger than at least $\sim 60\%$ of the primary one. Such a finding is consistent with~\cite{Berti:2007zu}, which found that the \emph{ringdown} SNR needs to be $\gtrsim 100$ to distinguish the first two leading BH QNMs produced by the merger of two BHs with the mass ratio of $\sim1.2$.

Let us now discuss the implications of Fig.~\ref{fig:QNM} on the properties of exotic compact object remnants. Reference~\cite{Cardoso:2016rao} showed that exotic compact objects with a light ring can produce GWs whose dominant modes are similar to those of a Kerr BH at early times, but differ at late times through subleading modes that correspond to exotic QNMs. This conclusion was arrived at by studying how a test particle falls into an exotic object, which is quite different from the merger of comparable mass objects. Nonetheless, assuming this conclusion remains true for comparable-mass mergers, Fig.~\ref{fig:QNM} then implies that aLIGO can bound the amplitudes of additional, slower damped QNMs of an exotic object. For a damping time 5 times longer than the primary $\ell=m=2$ mode, the secondary mode amplitude must be less than $\sim 50\%$ that of the primary mode, with the limits strengthening as the damping time of the secondary mode increases. Regarding boson star mergers, as we discussed in Sec.~\ref{sec:exotic-hydro}, one may expect the damping time to be $\mathcal{O}(100\mrm{ms})$ (see also Appendix~\ref{app:viscosity}). The amplitude of such a boson star QNM can then be constrained from the GW150914 measurement to be less than 10\% of the primary mode's amplitude.

Let us end this subsection by discussing how a possible time delay between the primary and secondary oscillation modes may affect the constraint on the amplitude of the latter in Fig.~\ref{fig:QNM}. Since the above analysis assumes a damping sinusoidal waveform in the time domain for both the primary and secondary oscillation modes, we effectively assumed that the two modes were excited approximately at the same time. On the other hand, Ref.~\cite{Cardoso:2016rao} showed that QNMs of exotic compact objects with a light ring are typically excited after the primary mode excitation (at merger). If the time delay between the excitation of the two modes is relatively large, one can treat them independently with the secondary one modeled by a Gaussian sinusoidal (or a sine-Gaussian) waveform. One can then easily estimate the upper bound on the amplitude of such a secondary mode relative to the primary one as a function of $\bar f_\RD$ and $\bar \tau$, requiring that the SNR of the secondary mode be smaller than a threshold SNR of $\sim 5$. Such a bound leads to results similar to those presented in Fig.~\ref{fig:QNM}, with the only exception around the white dot in the figure, where parameter degeneracies would become negligible in such a new analysis. To give an example on the bound comparison, we found that the upper bound on the secondary mode relative amplitude with $\bar f_\RD = f_\RD$ and $\bar \tau = 5 \tau$ is 0.15, while that in Fig.~\ref{fig:QNM} is 0.3. Given the similarity between the two analyses, we expect Fig.~\ref{fig:QNM} to be a valid order of magnitude estimate, even if one allows for a finite time delay between the two modes, with the bound around the white dot becoming stronger as the time delay becomes larger. 

\section{Theoretical Implications of an Electromagnetic Counterpart to GW150914}
\label{sec:spec-imps}

\label{sec:counterpart}

The Fermi collaboration announced that the GBM in the Fermi spacecraft detected a gamma-ray signal that was coincident with event GW150914~\cite{Connaughton:2016umz} (see also~\cite{Bagoly:2016}). This signal lasted for roughly 1 second and it started $0.4$ seconds after GW150914. With a false alarm probability of roughly $0.002$, this is not a high-$\sigma$ signal. Moreover, the signal was detected only in the GBM offline search~\cite{Abbott:2016gcq} (not as a GBM trigger) and not in any other instrument (like the Fermi Large Area Telescope~\cite{Fermi-LAT:2016qqr}, INTEGRAL~\cite{Savchenko:2016kiv} or Swift~\cite{Evans:2016mta}) or by any other particle detector (like neutrino detectors~\cite{Adrian-Martinez:2016xgn}). The properties of the signal make it look like a weak short GRB, but if so, it is unclear how it was generated; typically, short GRBs are expected to be produced by the merger of binary NSs or a mixed BH/NS system, and not by a binary BH merger. Some astrophysical scenarios have been proposed for the generation of such a short GRB, which include emission from a circumbinary accretion disk with possible future afterglows~\cite{Loeb:2016fzn,Li:2016iww} (see also~\cite{Woosley:2016nnw}). 

In broad terms, electromagnetic counterparts can be classified in two groups: precursor-emission signals or prompt/delayed-emission signals. In the first scenario, the electromagnetic counterpart is produced during the inspiral phase, for example due to interactions of the binary with a circumbinary accretion disk (see e.g.~\cite{Bode:2009mt,Bogdanovic:2010he}). In this case, assuming GR is correct, the electromagnetic signal can arrive before or with the peak of the GW strain. In the second scenario, the electromagnetic signal is produced after the compact objects have merged, for example due to the production of a short GRB (see e.g.~\cite{Zhang:2003uk,Capozziello:2010sm}). In this case, the electromagnetic counterpart arrives a certain time after the peak of the detected GW. The Fermi GBM observation would fit in this second scenario (as a prompt/delayed-emission signal), if GWs travel at or slower than the speed of light. If GWs travel at superluminal speeds, however, then the Fermi GBM signal could have been emitted before the GW signal, with the latter arriving first due to its faster propagation speed.   

Let us study then what theoretical implications one can infer \emph{if} the GBM signal were indeed interpreted as an electromagnetic counterpart to GW150914\footnote{Recent studies have argued that the GMB signal may not even be of astrophysical origin, let alone be a counterpart to GW150914~\cite{Xiong:2016ssy,Greiner:2016dsk}.}. The most obvious implication is a model-independent test of the speed of gravity by simply comparing the times of arrival of the two signals~\cite{Nishizawa:2014zna}. In the prompt/delayed-emission scenario, if GWs travel at the speed of light, the difference in the arrival times can, at most, be due to the intrinsic time delay in the emission of photons after the GW emission has ended. For NS mergers in the standard short GRB scenario, this time delay can be anywhere between O($1$) seconds and O($100$) seconds, with variations dependent on the particular details of the astrophysical model. The fractional difference between the speed of light and the speed of GWs, $\delta_{g} = 1 - v_{g}/c$, can then be constrained to~\cite{Nishizawa:2014zna}
\be
|\delta_{g}| < \frac{c \Delta \tau_{\rm int}}{D_{L}}\,,
\ee
where $D_{L}$ is the luminosity distance. 
\begin{figure}[htb]
\begin{center}
\includegraphics[width=\columnwidth,clip=true]{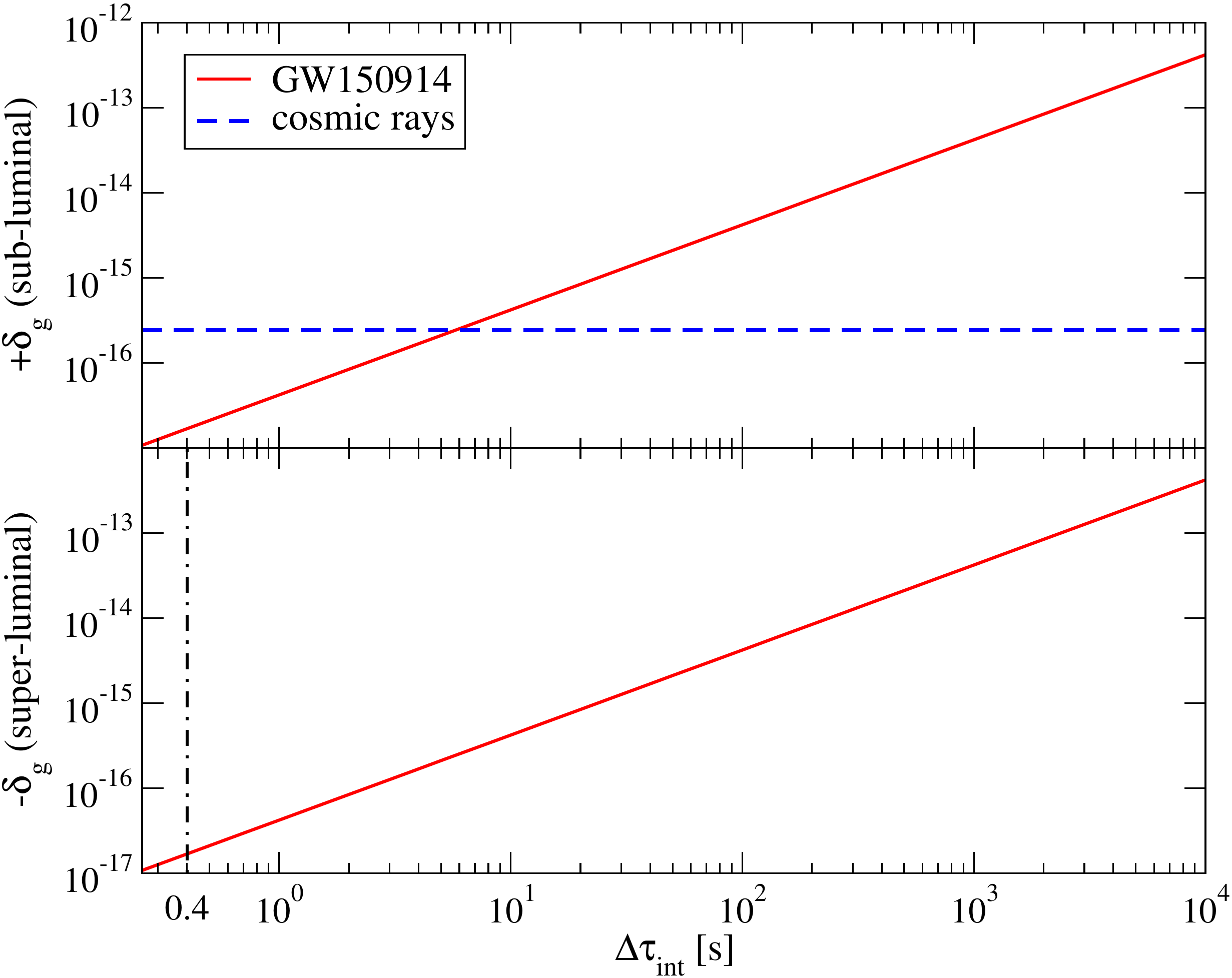} 
\caption{\label{fig:deltag} (Color online) GW150914 constraints on the fractional deviation in the propagation speed of GWs away from the speed of light for the $\delta_g > 0$ (top) and $\delta_g < 0$ (bottom) region, assuming that the event Fermi observed was associated with GW150914. We also show the cosmic ray constraints from the absence of the gravitational Cherenkov radiation in Eq.~\eqref{eq:GCR} in the top panel. The vertical dotted-dashed line in the bottom panel corresponds to $\Delta \tau_\mrm{int} = 0.4$s assumed in~\cite{Li:2016iww,Ellis:2016rrr}. We do not show the region with $\Delta \tau_\mrm{int} < 0.256$s, since then the binning of the Fermi observation in the time dominates the error budget over $\Delta \tau_\mrm{int}$.
}
\end{center}
\end{figure}

Let us now investigate how strongly the GW150914 event constrains $|\delta_g|$ if the Fermi event was a prompt/delayed counterpart to GW150914. Using that the GW inferred distance $D_{L} = 420^{+150}_{-180}$ Mpc, one can place a conservative bound on the speed of gravity, but only as a function of the unknown $\Delta \tau_\mrm{int}$. The top (bottom) panel of Fig.~\ref{fig:deltag} shows constraints in the subluminal (superluminal) region, with the region above the curves excluded. Since the Fermi GBM time binning is $0.256$s~\cite{Connaughton:2016umz}, we do not show constraints on $|\delta_g|$ with $\Delta \tau_\mrm{int} < 0.256$s. One could also map constraints on $|\delta_g|$ to constraints on $\mathbb{A}$ via Eq.~\eqref{eq:prop-speed}, assuming e.g.~a GW frequency of $f \sim 100$Hz. However, such constraints are weaker than those in Fig.~\ref{fig:A-constraint} by more than two orders of magnitude, except in the $\alpha = 2$ case, which cannot be constrained strongly from GW observations alone. The point of this figure is to show that the constraint depends sensitively on the unknown intrinsic time delay parameter, without which a constraint cannot actually be placed.  

Let us discuss in detail constraints on the subluminal propagation of GWs. The top panel of Fig.~\ref{fig:deltag} shows the hypothetical Fermi/GW150914 constraint, together with the bound from cosmic ray observations due to the absence of gravitational Cherenkov radiation~\cite{Moore:2001bv}:
\be
\label{eq:GCR}
\delta _g \leq 2.45 \times 10^{-16} \left( \frac{E}{10^{11}\mrm{GeV}} \right)^{-3/2} \left( \frac{D_L}{1\mrm{Mpc}} \right)^{-1/2}\,,
\ee
where we assumed, as in Fig.~\ref{fig:A-constraint}, that cosmic ray particles with an energy $E = 10^{11}$ GeV have traveled $D_L = 1$ Mpc to reach Earth. The GW/Fermi coincident constraint is more stringent than the cosmic ray bound, provided that $\Delta \tau_{\mrm{int}} < 60$ s. Given that there are no agreed-upon models for the electromagnetic emission detected by Fermi, it is not clear whether such an intrinsic time delay is reasonable.   

Let us now discuss in detail constraints on the superluminal propagation of GWs. The bottom panel of Fig.~\ref{fig:deltag} shows the coincident Fermi/GW150914 constraint, which again depends on the intrinsic time delay $\Delta \tau_{\mrm{int}}$. If one assumes the Fermi event was a prompt/delayed scenario, the most conservative bounds on negative $\delta_g$ is obtained when $\Delta \tau_{\mrm{int}} = 0.4$ s, i.e.~when we set the delay to be exactly the observed arrival time delay between the GW observation and the GBM observation (shown with a dotted-dashed line), which gives $\delta_g \gtrsim -10^{-17}$~\cite{Li:2016iww,Ellis:2016rrr}\footnote{The deviation in the propagation speed of high-energy photons $v_p$ with energy $E_p$ from that of low-energy photons $c$ is constrained by $|v_p/c-1| \leq 4.8 \times 10^{-22} (E_p/1\mrm{MeV})$ or $2.4 \times 10^{-28} (E_p/1\mrm{MeV})^2$~\cite{HESS:2011aa}, which is much smaller than $10^{-17}$, and hence can be neglected.}. On the other hand, Collett \text{et al.}~\cite{Collett:2016dey} assumed that GWs and gamma rays are emitted simultaneously at merger, which corresponds to $\Delta \tau_{\mrm{int}} = 0$ and leads to $\delta_g = -1.0^{+0.8}_{-1.9} \times 10^{-17}$, where the errors are propagated from the Fermi timing bins and from errors in the luminosity distance measurement. The error bar does not contain $\delta_g=0$ in this case, which means that GWs must propagate superluminally under the assumption that gravitons and photons were emitted simultaneously. 

Assuming that our understanding of the astrophysical emission mechanisms improve in the future and constraints can be placed from a coincident Fermi/GW observation, let us investigate the theoretical implications of the resultant model-independent constraint on the speed of gravity. The most obvious implication is a severe constraint on gravitational Lorentz violation~\cite{Hansen:2014ewa}. EA theory~\cite{Jacobson:2000xp,Eling:2004dk} breaks gravitational Lorentz invariance by introducing a vector field that couples to the metric tensor; this theory is the most generic modification to Einstein's theory that contains a (unit timelike) vector field and (at most) quadratic combinations of its first derivative. Khronometric theory~\cite{Blas:2009qj,Blas:2009yd} breaks gravitational Lorentz invariance by introducing a globally preferred frame selected by a scalar field (the ``khronon''); this theory arises as the low-energy limit of the ultraviolet complete and power-counting renormalizable Ho\v rava-Lifshitz theory~\cite{Horava:2009uw}. In these theories, the speed of GWs is corrected through a fractional modification of the form~\cite{Foster:2006az,Blas:2011zd} 
\begin{align}
\delta_{g}^{\EA} &= 1 - \left(1 - c_{+}\right)^{-1/2}\,,
\\
\delta_{g}^{\KG} &= 1 - \left(1 - \beta_{\KG}\right)^{-1/2}\,,
\end{align}
where $c_{+}$ is a combination of coupling constants in EA theory, while $\beta_{\KG}$ is a coupling constant in khronometric gravity. 

We can now easily map constraints on $\delta_g$ to constraints on Lorentz-violation mechanisms. A bound on $\delta_{g}$ of order $10^{-17}$ implies a constraint on gravitational Lorentz violation at the same level:
\be
\label{eq:c+-constraint}
c_{+} \lesssim 10^{-17}\,,
\qquad
\beta_{\KG} \lesssim 10^{-17}\,.
\ee
These constraints are 15 orders of magnitude more stringent than any other constraint on gravitational Lorentz violation. The EA and khronometric modification to GW propagation cannot be constrained from the bound on $\mathbb{A}$ with the Fisher analysis of Fig.~\ref{fig:A-constraint} and only GW observations. This is because such a modification corresponds to the $\alpha=2$ case in Eq.~\eqref{dispersion}, and thus, it is degenerate with the time of coalescence in the waveform phase, as explained in Sec.~\ref{sec:mod-disp-rel}. On the other hand, one can apply the GW150914 (GW151226) bound on $\delta_g$ from the GW arrival time delay between Hanford and Livingston detectors~\cite{Blas:2016qmn}, which yields $(c_+,\beta_{\KG}) \lesssim 0.7$ ($\lesssim 19$), as summarized in Table~\ref{tab:summary2}. Although such a bound is weaker than the putative constraints in Eq.~\eqref{eq:c+-constraint}, obviously the former is more robust, given the uncertainties associated with the Fermi GBM event. A simultaneous measurement of GWs and gamma rays also allows us to place constraints on a more generic Lorentz-violating framework, the gravitational SME framework, in particular on non-dispersive and non-birefringent coefficients like $\mathring{k}_{(V)}^{(4)}$, as discussed in Sec.~\ref{subsec:par-def-to-physics}.

Another theoretical implication that could be derived from a coincident GW/electromagnetic observation is a severe constraint on gravitational parity violation~\cite{Yunes:2010yf,Alexander:2007kv,Yunes:2008bu}. If gravity breaks parity, then, generically, left- and right-polarized GWs will obey different propagation equations, with the amplitude of one mode suppressed and the other enhanced. To constrain this effect, a network of GW detectors~\cite{Aasi:2013wya} would then need to separate the two GW polarization amplitudes. Due to parameter degeneracies, however, such a test also requires that a coincident short GRB observation (a) constrain the inclination angle (the angle between the orbital angular momentum of the binary and the line of sight) and (b) provide a distance measurement through galaxy identification. Event GW150914 is particularly well-suited for this test, as it was observed nearly face-on and thus the signal arrives almost entirely circularly polarized. The aLIGO detectors, however, are essentially co-aligned, so the two GW polarizations could not be separated in event GW150914~\cite{TheLIGOScientific:2016src}. Moreover, the GBM signal was not bright enough to allow for galaxy identification and a measurement of distance. Therefore, a test of gravitational parity invariance cannot be carried out, even if one associated the GBM signal with a counterpart to GW150914; for this, one will have to wait for a network of GW detectors~\cite{Aasi:2013wya} to allow for the extraction of polarizations, as well as a coincident GRB signal that is sufficiently localized to allow for galaxy identification.  

\section{Conclusion}  
\label{sec:conclusions}

We have studied the theoretical physics implications of GW150914 and GW151226. The LVC has demonstrated that these events are entirely consistent with binary BH mergers in GR via constraints on deviations from the GR PN coefficients describing the inspiral, and for GW150914 that subtraction of the best-fit GR template from the data gives a residual consistent with noise. Our analysis has shown that more than simply verifying consistency with GR, the information contained in the GW events allows one to place limits on many physical phenomena that various modified gravity theories predict and could have been operational in a BH binary merger. 

Even though some of the constraints on these physical mechanisms are not as stringent as current bounds with binary pulsars, low mass X-ray binaries, Solar System experiments, table-top experiments on Earth or cosmological observations, constraints with GW150914 and GW151226 are of a completely different nature: they come directly from the extreme gravity environment of merging BHs. Moreover, we can anticipate that these bounds will steadily become stronger in the near future as (i) more GW observations are made (through stacking of multiple signals\footnote{For example, if aLIGO detects $N$ binary BH merger events, one can anticipate to statistically improve the upper bound on $|\beta|$ presented here by roughly a factor of $\sim\sqrt{N}$, an excellent prospect for the future considering that the expected number of highly  significant events at the end of the O3 run is above 35~\cite{Abbott:2016nhf, TheLIGOScientific:2016pea}. Though of course the exact enhancement factor depends on the distribution of sources in SNR and parameter space~\cite{Berti:2011jz}.}), (ii) different sources of GWs are observed (e.g. binary NS inspirals will constrain low-frequency mechanisms better than binary BH mergers), (iii) higher SNR events are observed (since the allowed magnitude of deviations quantified by the ppE parameter $\beta$ scales inversely with SNR), and (iv) multi-band GW observations of a heavier BH binaries (such as GW150914) with ground- and space-based interferometers may be possible~\cite{Sesana:2016ljz,Barausse:2016eii,Vitale:2016rfr}. 

Events GW150914 and GW151226 are fantastic probes of theoretical physics that have important implications for certain aspects of extreme gravity, but unfortunately not all. Future detections of GWs from NS binaries will allow us to probe different aspects of extreme gravity. The prime examples of this are theoretical models where gravity is described by a metric tensor with evolution equations that differ from the Einstein equations only through a modified ``right-hand side'' that depends on the matter stress-energy tensor~\cite{Pani:2013qfa}. Examples of such theoretical models are Eddington-inspired Born-Infeld (EiBI) gravity~\cite{Banados:2010ix} and Palatini $f(R)$ theories (see e.g. Sec.~9 of~\cite{DeFelice:2010aj} and references therein). Other examples include the activation of certain scalar or pseudo-scalar fields in the strong-gravity regime sourced by dense matter, such as Brans-Dicke theory~\cite{Fierz:1956zz,Jordan:1959eg,Brans:1961sx,zaglauer,Yunes:2011aa,Mirshekari:2013vb}, the scalar-tensor theory extension of Damour and Esposito-Far\`ese~\cite{damour_esposito_farese,ST0}, and $f(R)$ models as they can be mapped to scalar tensor theories~\cite{Sotiriou:2008rp}.

Let us end by stressing that the true potential of heavier BH mergers like GW150914 to test GR and exotic compact objects is limited by the lack of knowledge of how GWs behave during the merger phase in GR alternatives. This event has given us a remarkable glimpse into this regime of extreme gravity, which could in principle place very stringent constraints on modified gravity theories,   were their dynamics known in this regime. In our ppE analysis we only included non-GR corrections to the inspiral phase of coalescence. If one were to include modifications to the merger-ringdown phase, the bounds on various theoretical GW generating mechanisms presented in the top part of Table~\ref{tab:summary2} would become stronger. GW150914 therefore calls for a more concerted effort by the gravity and high-energy communities to explore the full non-linear regime of merging compact object binaries.

\acknowledgments

We thank 
Neil Cornish, Vasileios Paschalidis and Norbert Wex
for useful discussions and comments, and Katie Chamberlain for a code comparison.
We also thank Kenta Kiuchi for sharing with us gravitational waveforms from strongly-magnetized hypermassive NSs.
We further thank Enrico Barausse, Emanuele Berti, Diego Blas, Vitor Cardoso, Paolo Pani and Takahiro Tanaka for carefully reading the manuscript and giving us important comments.
N.Y.~acknowledges support from the NSF CAREER Grant PHY-1250636. 
K.Y.~acknowledges support from JSPS Postdoctoral Fellowships for Research Abroad. 
F.P.~acknowledges support from the NSF grant PHY-1305682 and the Simons Foundation.

\appendix

\section{Constraining GR Modifications with PhenomB and PhenomD Waveforms}
\label{app:BvsD}

\begin{figure*}[htb]
\begin{center}
\includegraphics[width=\columnwidth,clip=true]{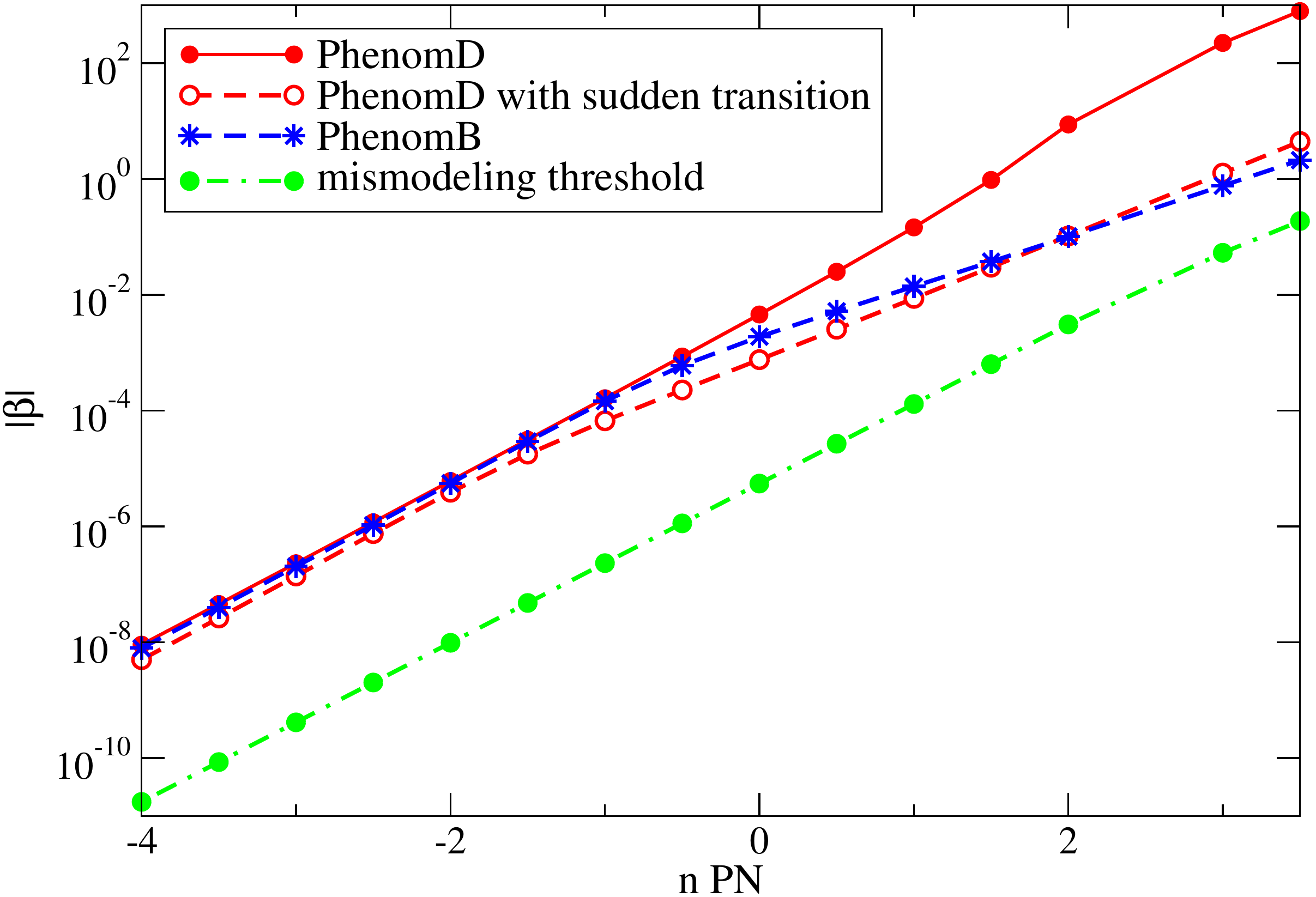} 
\includegraphics[width=\columnwidth,clip=true]{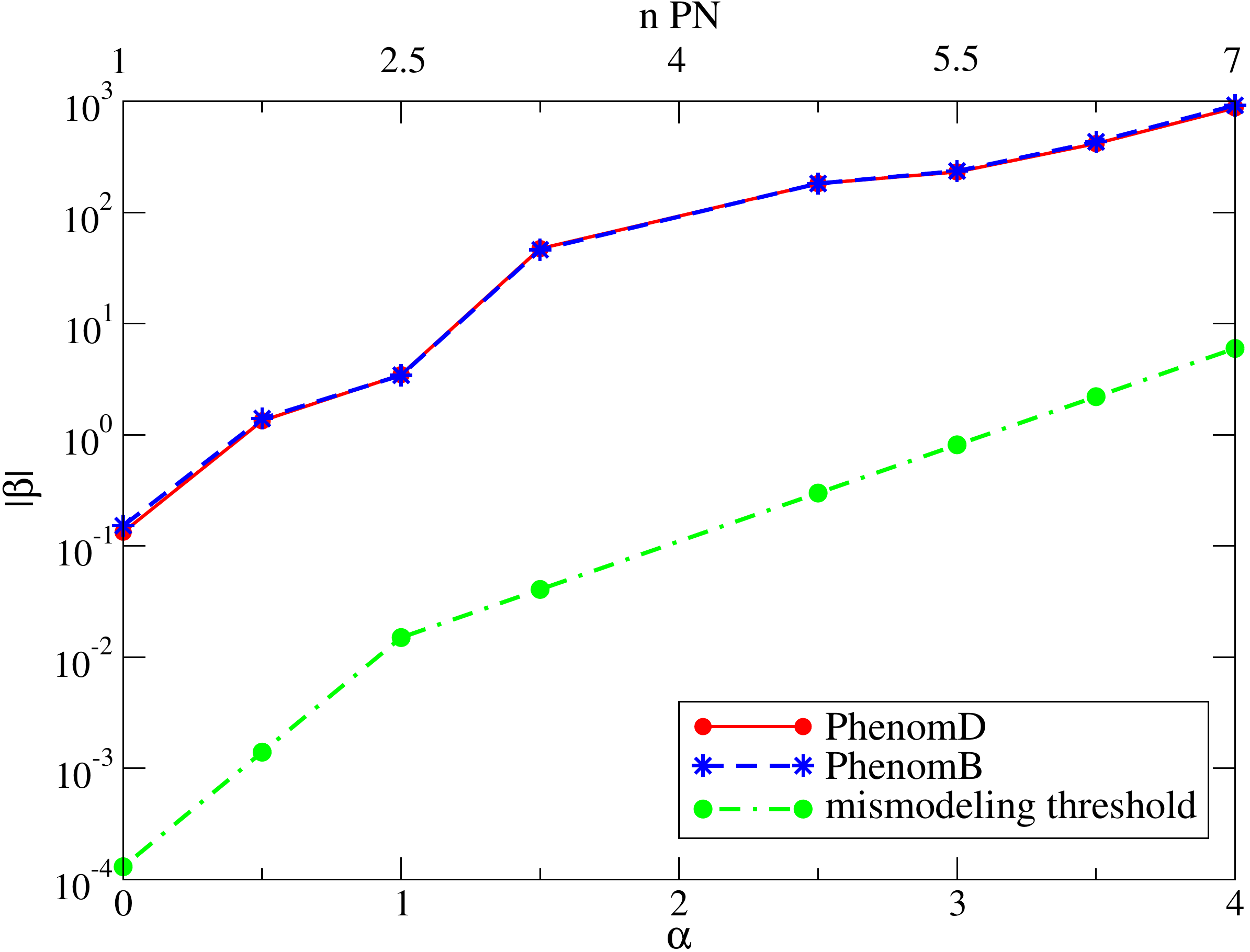} 
\caption{\label{fig:PhenomB-vs-D} (Color online) Comparison of 90\%-confidence GW150914 constraints on $|\beta|$ with PhenomD and PhenomB waveforms for modified generation (left) and propagation (right) effects on GWs. We also show the PhenomD result with a sudden transition of the non-GR effect at $f=f_\mrm{Int}$ like PhenomB, which corresponds to the sudden deactivation of the scalar field with a transition at $f_*=f_\mrm{Int}$ in Sec.~\ref{sec:gen-constr-in-generation} (but for arbitrary $b$). Green dotted-dashed curves present a rough estimate of the impact of mismodeling error in the PhenomD waveform on constraints on $|\beta|$, which serve as the threshold on future $|\beta|$ constraints.
}
\end{center}
\end{figure*}

In this appendix, we compare constraints on the ppE parameter $\beta$ using PhenomB~\cite{Ajith:2009bn} and PhenomD~\cite{Husa:2015iqa,Khan:2015jqa} waveforms. Although we only used the latter for parameter estimation, it is still interesting to see whether the constraints we found are affected by the GR waveform model employed (especially given that GW150914 is in a regime where some of the approximations used to create these models become questionable). Through this comparison, we provide a rough estimate of the impact of mismodeling error on the constraints on $\beta$. As we will show, the impact of mismodeling error is minimal and unimportant on the bounds reported in this paper.

Let us then begin by reviewing the similarities and differences between these two waveform models. Both waveforms were obtained by first constructing hybrid PN and numerical relativity waveforms in the time domain, and then Fourier transforming them into the frequency domain. The PhenomB waveform was calibrated over the mass ratio range $1 \leq q \leq 10$ and the spin range $-0.85 \leq \chi_A \leq 0.85$. Each part of the PhenomB waveform (inspiral, merger and ringdown) was then fitted by polynomials of the form of Eq.~\eqref{eq:waveform}, with $\Phi_i = \Phi_{\EI}$ in Eq.~\eqref{eq:early-insp-IMR}. On the other hand, the PhenomD waveform was calibrated over a larger sector of parameter space: $1 \leq q \leq 18$ and $-0.95 \leq \chi_A \leq 0.95$. Moreover, the PhenomD waveform amplitude and phase in each segment (described in Sec.~\ref{sec:IMRPhenom-GR}) are matched together to ensure continuity and differentiability at the interfaces.

These PhenomB and PhenomD waveforms in GR can be extended to capture non-GR effects by adding a ppE correction term in Eq.~\eqref{eq:ppEphase} in the waveform phase. We include such a term only in the inspiral phase when investigating modified GW generation mechanisms. However, such a correction propagates to the intermediate and merger-ringdown phases in PhenomD due to the continuity and differentiability requirements at interface frequencies. This does not occur in the modified PhenomB model, and thus, the model is discontinuous at the interface between the inspiral and merger phases when $\beta \neq 0$. When studying modified GW propagation effects, we include the ppE correction term in all phases, which renders both the modified PhenomB and PhenomD models continuous and differentiable at the interfaces.

Figure~\ref{fig:PhenomB-vs-D} compares the upper bound on $|\beta|$ from GW150914 as a function of the leading PN order of the ppE correction using the ppE-modified PhenomB and PhenomD models. The left and right panels show the bound on modified generation and propagation mechanisms. The two models give almost identical bounds at any PN order in the propagation case, but only when $n \leq -1$ in the generation case, with larger differences arising at positive PN order. This is because the generation-modified PhenomB model has a correction only in the inspiral phase, with the correction shutting off suddenly at the inspiral-merger interface, while the propagation-modified model is always continuous and differentiable as the PhenomD model is. This non-smooth feature of the generation-modified PhenomB model makes such a correction unique, allowing $\beta$ to be less degenerate with other parameters relative to the PhenomD case (see also the related work of~\cite{Mandel:2014tca}). In order to check this, we constructed a modified PhenomD waveform by adding Eq.~\eqref{eq:early-insp-IMR} in the inspiral phase but not imposing continuity and differentiability at the interface frequency in the non-GR part of the phase. Thus, such a waveform suddenly changes to the GR model in the post-inspiral phases, which is similar to what occurs in the scalar field deactivation waveform with a transition frequency of $f_* = f_\Int$ (see Sec.~\ref{sec:gen-constr-in-generation}) but an arbitrary $b$ instead of $b = -7/3$. As we expect, the bound on $|\beta|$ shown by the red dashed curve in the left panel of Fig.~\ref{fig:PhenomB-vs-D} is similar to the bound obtained from the PhenomB model even for positive PN corrections. 

The main conclusion this result allows us to draw is that if one wishes to obtain an accurate constraint on a possible deviation from GR using GW data, one should use the ppE-modified PhenomD model (or a model as accurate or more accurate than this one), as this guarantees smoothness at the transition frequencies. Indeed, one does not expect that GR deviations will lead to non-smooth GWs; even when there is a sudden activation or de-activation of scalar dipole radiation, this ought to occur smoothly (even though such a transition has sometimes been modeled with a Heaviside function in the literature~\cite{Alsing:2011er,Cardoso:2011xi,Berti:2012bp,Sampson:2013jpa}). 

We conclude this appendix with a rough estimate of the impact of mismodeling error of the GR part of the waveform on $|\beta|$ constraints. Although both PhenomB and PhenomD waveforms are approximations to numerical relativity waveforms, the fact that both of these give similar bounds on $|\beta|$ when the ppE modifications are introduced in the same way suggests that the effect of GR waveform mismodeling does not strongly affect the $\beta$ bounds. In order to quantify this statement, consider the following. The difference between the waveform phase in the PhenomD model and the numerical relativity waveform for an equal-mass, non-spinning BH binary (i.e.~the mismodeling error of the GR phase) is $\sim 0.015$ rad at most at any frequency~\cite{Khan:2015jqa}. This suggests that the peak of the posterior distribution for $\beta$ may shift away from $\beta = 0$, producing systematic errors. Setting $|\beta| (\pi \mathcal{M} f)^{b/3} \lesssim 0.015$ and \emph{maximizing} $\beta$ over the frequency\footnote{The frequency range of GW150914 for maximizing $\beta$ is chosen to be $f \in (20,52)$Hz for the generation mechanism constraint (as the ppE modification is only introduced in the inspiral phase) and $f \in (20,300)$Hz for the propagation mechanism one.}, we can find the maximum systematic error on $\beta$ assuming that the mismodeling error is completely absorbed by the ppE phase. This, in turn, determines the \emph{minimum} value of $|\beta|$, the \emph{mismodeling threshold}, that can be constrained without contamination from GR mismodeling error, i.e.~if the SNR were large enough to allow for a constraint on $\beta$ that is smaller than this minimum value, such a constraint would be limited by mismodeling error in the GR part of the waveform. The mismodeling threshold for GW150914\footnote{The mismodeling threshold for GW151226 is almost identical to that of GW150914. Although the statistical error on $\beta$ is much smaller for GW151226 especially on negative PN modifications, the mismodeling error is still smaller than such a statistical error by a factor of $\sim 5$ even at -4PN order.} is shown with green dotted-dashed curves in Fig.~\ref{fig:PhenomB-vs-D}. The constraints on $\beta$ would have to be much tighter (the blue or red lines would have to be much lower, by a factor of 100--5000) for GR mismodeling to have an effect. Given that GR mismodeling is independent of the SNR, while the constraints on $\beta$ scale linearly with SNR, we conclude that GR mismodeling would become important for SNR $\gtrsim 2400$. 

This rough estimate does not account for how GR mismodeling error affects other parameters, and in turn, the mismodeling threshold. Namely, inferring systematic errors from the dephasing alone and comparing them against statistical errors is not a robust approach to estimate the former properly. A much better approach is to maximize the overlap between the signal and template waveforms over all parameters, which could be achieved through a Bayesian analysis of the GW signal with the ppE-modified IMRPhenomD templates and zero noise realization (as specific noise realizations shift the peak of the posterior distribution~\cite{Sampson:2013lpa}), which is beyond the scope of this paper.

\begin{figure*}[htb]
\begin{center}
\includegraphics[width=\columnwidth,clip=true]{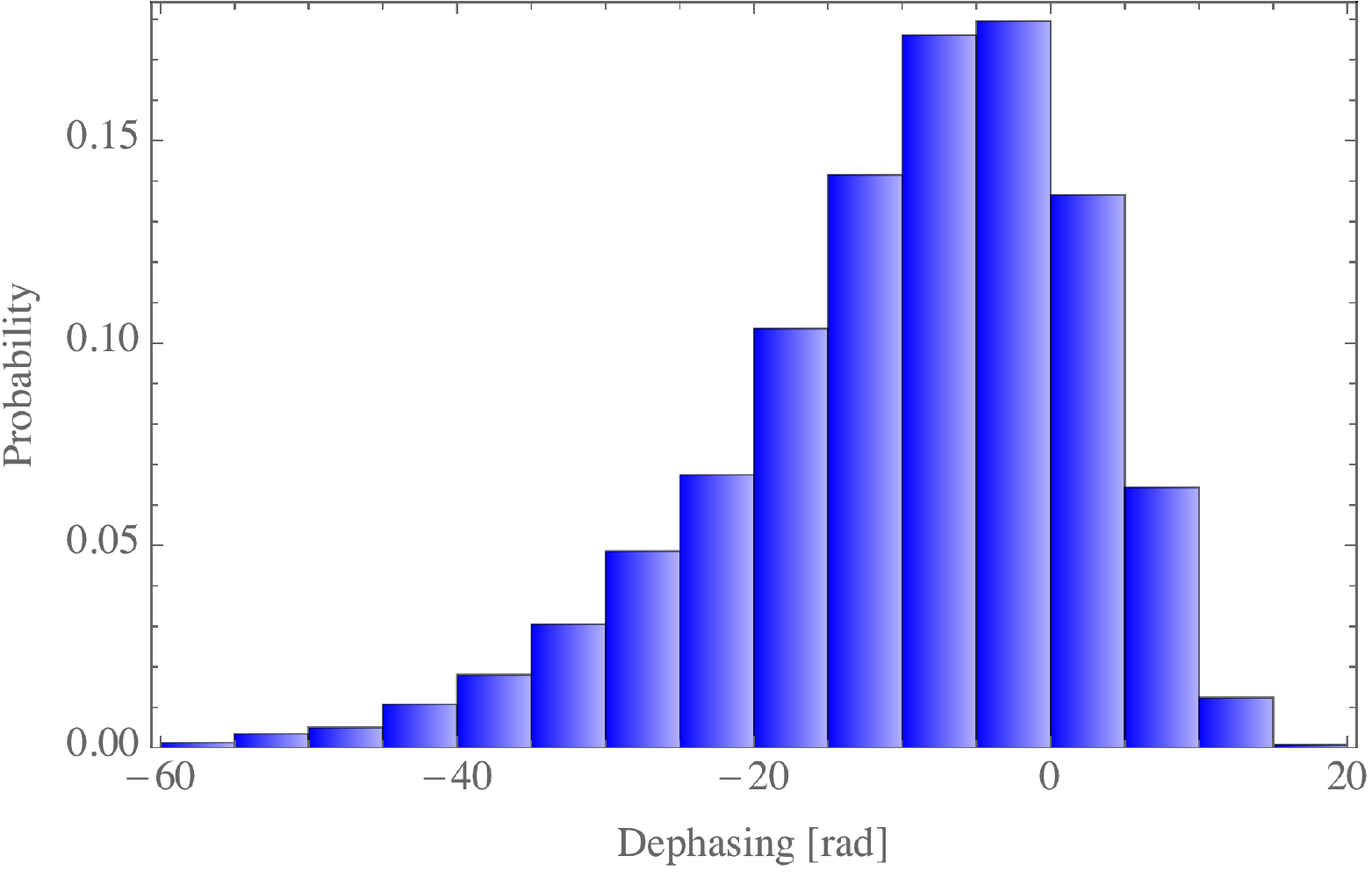} 
\includegraphics[width=8.cm,clip=true]{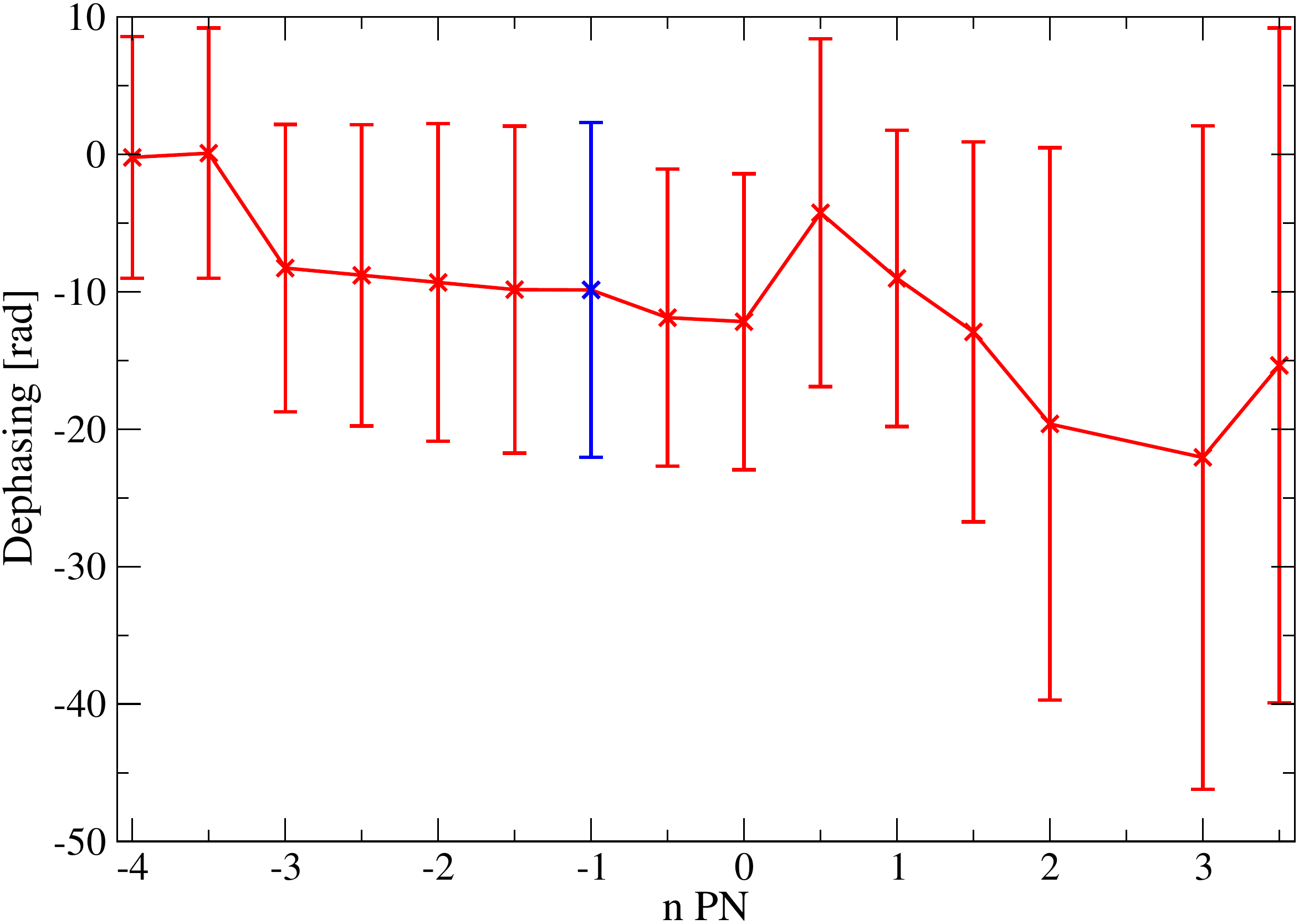} 
\caption{\label{fig:dephasing} (Color online) (Left) Probability distribution of the PhenomD waveform dephasing between the injected parameters and parameters within statistical errors at $f=50$Hz with a ppE modification at -1PN order. The asymmetry in the distribution arises from the condition $\eta \leq 0.25$.
(Right) The mean of the probability distribution of the dephasing as a function of the PN order of the ppE modification, together with 1$\sigma$ error bars. The mean and 1$\sigma$ error of the left panel (at -1PN order) are shown in blue. The absolute dephasing within 1$\sigma$ errors of the distribution can be as large as 10--50 rads, which is much larger than the maximum mismodeling dephasing of 0.015 rad. This suggests that the latter is negligible in constraining $\beta$.
}
\end{center}
\end{figure*}

Having said this, one can take an alternative approach to map the statistical errors to the dephasing and compare the latter to the maximum mismodeling dephasing of 0.015 rad within the PhenomD model. Such an approach allows us to circumvent the problem of estimating systematic errors properly, and yet provides us with more trustworthy results than the above mismodeling threshold argument. To achieve this, we carried out the following Monte Carlo simulations. First, we draw a point in the parameter space, based on the Fisher matrix derived with injected parameters (see the beginning of Sec.~\ref{sec:parametrizedtests}) consistent with the LVC measurement for GW150914; such a Fisher matrix defines a multivariate Gaussian probability distribution within the parameter space that defines a proposal function\footnote{A new point $\bm \theta$ in the parameter space is chosen via
\be
\bm \theta = \bm \theta_\mathrm{(inj)} + \sum_{A=1}^D \frac{\alpha_A}{\sqrt{\lambda_A}} \bm{V}_A\,,
\ee
where $\bm \theta_\mathrm{(inj)}$ are injected parameters with dimension $D$, $\lambda_A$ and $\bm{V}_A$ are eigenvalues and unit eigenvectors of the Fisher matrix, and $\alpha_A$ are random numbers drawn from a Gaussian distribution with zero mean and variance of $1/D$.}. 
Next, we calculate the ppE-modified PhenomD waveform phase at this new point in parameter space. With that in hand, we evaluate the dephasing $\delta \Psi$ between that non-GR model $\Psi^{\rm{mod}} (\bm \theta;f)$ (at the new point in parameter space)  and the GR model $\Psi^{\GR} (\bm \theta_\mathrm{(inj)};f)$ (at the the injected parameters)  evaluated at a fixed frequency $f^{*}$, i.e.
\be
\delta \Psi \equiv \Psi^{\rm{mod}} (f^{*};\bm \theta) - \Psi^{\GR} (f^{*};\bm \theta_\mathrm{(inj)})\,.
\ee
We then repeat this calculation over $10^4$ times to construct a normalized histogram that defines a probability distribution function for the dephasing. This probability distribution is a dephasing measure of the statistical uncertainties. 

The left panel of Fig.~\ref{fig:dephasing} presents such a distribution at $f^{*}=50$Hz for the -1PN ppE modification. We evaluate the dephasing at this frequency because we only include ppE corrections in the inspiral part of the waveform, and this phase ends at 52Hz for the GW150914 event. The asymmetry in the distribution mainly comes from the requirement that $\eta \leq 0.25$. The mean and standard deviation of this distribution are -9.9 rads and 12 rads, which are shown in the right panel of Fig.~\ref{fig:dephasing}, together with those for ppE modifications at other PN orders. The absolute dephasing within 1$\sigma$ errors of the distribution can be as large as 10--50 rads. Such dephasings at one sample frequency of 50Hz is already much larger than the maximum mismodeling dephasing (0.015 rad) of the PhenomD waveform (and the dephasing becomes even larger if we were to maximize it over frequency). This suggests that such mismodeling errors are negligible compared to statistical parameter uncertainties, a finding that is consistent with the mismodeling threshold argument of Fig.~\ref{fig:PhenomB-vs-D}.

\section{Effect of Higher PN Order Corrections in the ppE Formalism}  
\label{app:effect-of-hi-PN}

In this appendix, we study whether the quantitative inferences derived from the GW150914 event by using an inspiral-only analysis with a leading-PN order deformation (\emph{a la} simple ppE) is affected by our ignorance of higher PN order terms induced in the late inspiral and merger. Indeed, we do not possess predictions for the GWs emitted during the entire inspiral-merger-ringdown coalescence that includes modifications to Einstein's theory, such as the activation of a scalar field in a vacuum spacetime, or from the presence of large extra dimensions. This appendix will demonstrate that it is not {\em a priori} necessary to have knowledge beyond the leading-order PN modification to derive some level of meaningful inferences from GW observations, i.e.~knowledge of the higher-order terms and merger phase may strengthen the constraints derived from the analysis presented here (since e.g.~one would be able to integrate the signal to higher frequencies), but it does not invalidate our analysis.

Consider the question of how much are bounds on modified gravity affected by the inclusion of higher PN order modified gravity terms in the inspiral phase. For a specific calculation, let us consider a $-1$PN deformation from GR, for example as induced by the activation of a scalar monopole charge in BD theory. The waveform in such a theory is known to $2.5$PN order relative to the leading $-1$PN order term in the test-particle limit for non-spinning BHs. To this order then, the inspiral Fourier phase is given by 
\begin{align}
\Phi_{\I}^{\BD}(f) &= \Phi_{\I}^{\GR}(f) + \beta_{\BD} \left(\pi {\cal{M}} f\right)^{b_{\BD}} 
\nn \\
& \left[1 + \sum_{i=2}^{5} \delta\phi_{i}^{\BD}(\eta) \left(\pi {\cal{M}} f\right)^{i/3} \right]\,,
\label{eq:mod-phase}
\end{align}
where $b_{\BD} = -7/3$ and~\cite{Yunes:2011aa}
\begin{align}
\label{eq:BD-EMRI-2}
\delta \phi_{2}^{\BD} &= -\frac{7}{2} \eta^{-2/5}\,,
\\
\delta \phi_{3}^{\BD} &= 5 \pi \eta^{-3/5}\,,
\\
\delta \phi_{4}^{\BD} &= -\frac{350}{9} \eta^{-4/5}\,,
\\
\label{eq:BD-EMRI-5}
\delta \phi_{5}^{\BD} &= \frac{84}{5} \pi \eta^{-1}\,.
\end{align} 
Although $\beta_{\BD} = 0$ for BH binaries even if the BD parameter is finite, one can still estimate the bound on $\beta_{\BD}$ with GW150914 to see if such a measurement is consistent with the BH no-hair theorem in BD theory. 

The subsequent coefficients in BD theory shown above present the familiar structure of the PN series: alternating signs, absence of a relative $0.5$PN order modification in the Fourier phase, and growing coefficients as the PN order increases. Of course, this neglects mass-ratio corrections, as the modifications were calculated in the test-particle limit; however, in GR the PN series in the test-particle limit presents more (asymptotic) divergent features than in the comparable-mass limit, i.e.~the coefficients of the series grow more rapidly with PN order in the test-particle limit. Thus, by using the $\delta \phi_{i}$ above in the test-particle limit we are exaggerating the effect of higher PN terms in the waveform, which will suffice to make conservative statements. 

\begin{figure}[t]
\begin{center}
\includegraphics[width=\columnwidth,clip=true]{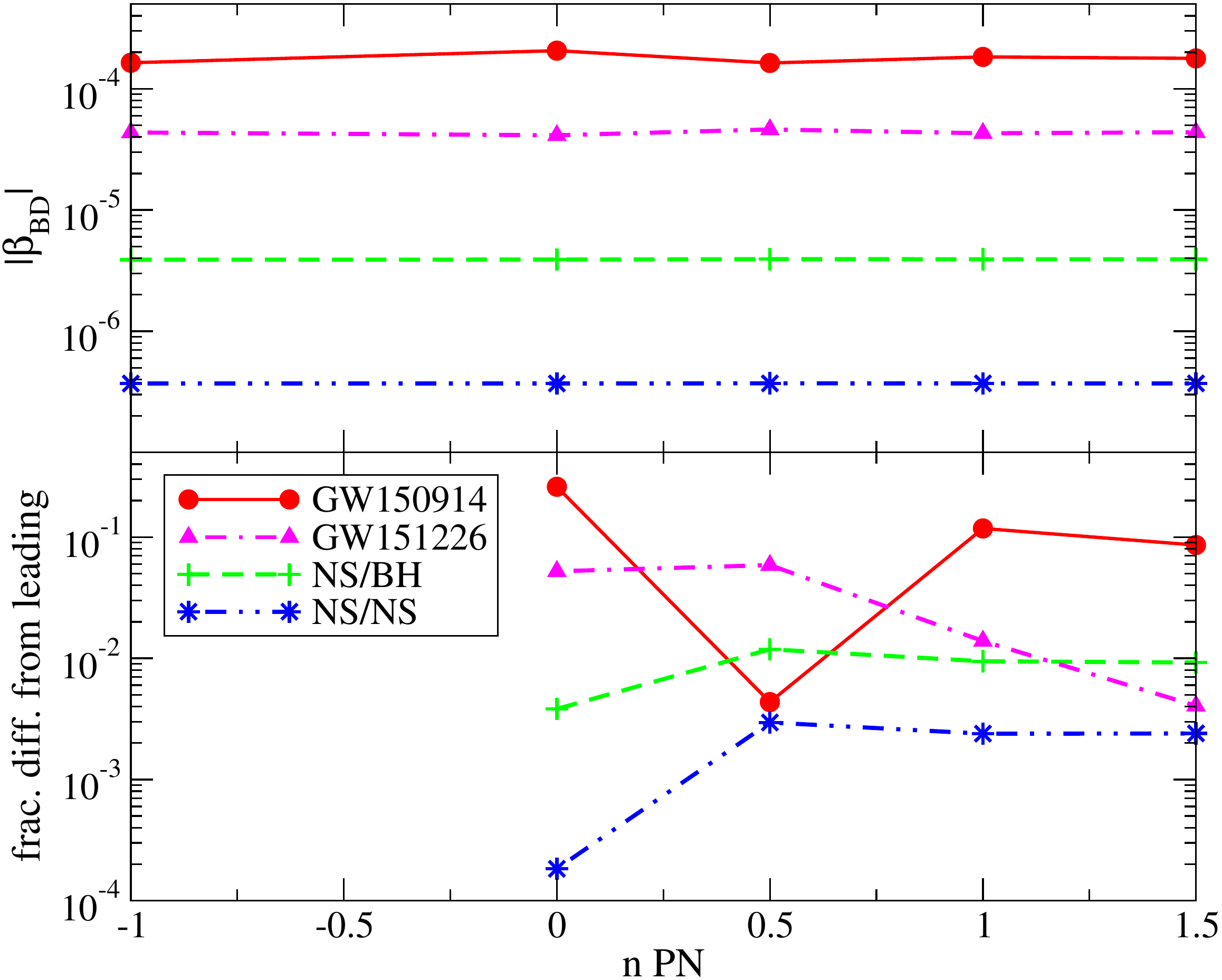} 
\caption{\label{fig:BD-EMRI} (Color online) (Top) Fisher estimates of the accuracy one can constrain the ppE parameter $\beta_{\BD}$ at $-1$PN with GW150914 and GW151226, as a function of the highest PN order included in the modified waveform phase. The bound at $-1$PN order is the same as that in Fig.~\ref{fig:beta-cons}. For reference, we also show the bound from a NS/BH and NS/NS binary with SNR=24, and with masses $(10,1.4)M_\odot$ and $(1.5,1.3)M_\odot$ respectively. (Bottom) The absolute fractional difference of the bound on $|\beta_{\BD}|$ as a function of PN order. Including higher order corrections only affects the bound obtained with only the leading PN order phase correction by at most $\mathcal{O}(10\%)$ for GW150914. The fractional difference for GW151226, NS/BH and NS/NS binaries is even smaller. 
}
\end{center}
\end{figure}

With this in mind, we carried out five different Fisher analysis:~one analysis used only the leading order ($-1$PN) phase modification, while the others included higher PN corrections. Figure~\ref{fig:BD-EMRI} shows Fisher estimates of the accuracy to which $\beta_{\BD}$ can be constrained from GW150914 and GW151226 as a function of the highest PN order included in the modified phase. For example, $n=0$ corresponds to Fisher studies where the correction to the waveform phase includes the leading-order ($-1$PN) piece and its first PN-order correction (0PN). Including higher-order PN terms barely modifies the strength of the constraint one can place on $\beta_{\BD}$. The difference shown in the bottom panel shows nice convergence as one increases the order of higher PN corrections included. We conclude that, if the event has already been shown to be consistent with GR and one is trying to constrain deviations from Einstein's theory, then including only the leading-order PN term in the analysis suffices. 

To show that the behavior described in the previous paragraph is not specific to BH binaries, we also present in Fig.~\ref{fig:BD-EMRI} how higher PN corrections in BD affect the measurement accuracy of $\beta_\BD$ with a NS/BH and NS/NS binary, with masses $(10,1.4)M_\odot$ and $(1.5,1.3)M_\odot$ respectively. To compare with the GW150914 result, we set the SNR to 24. We still use the IMRPhenom waveform and neglect any finite size effects in NSs, which would first enter at 5PN order~\cite{flanagan-hinderer-love} and thus be weakly correlated with $\beta_\BD$. We also neglect conservative corrections to the waveform phase, which are not included in Eqs.~\eqref{eq:BD-EMRI-2}--\eqref{eq:BD-EMRI-5} and are absent for BH binaries in the test-particle limit~\cite{Yunes:2011aa}. Figure~\ref{fig:BD-EMRI} shows that the higher PN order corrections in NS/BH and NS/NS binaries are even less important than in BH/BH binaries. This is because at a fixed frequency, the orbital velocity of the binary constituents is smaller for the former, which makes the higher order PN effects less important.

\begin{figure}[t]
\begin{center}
\includegraphics[width=\columnwidth,clip=true]{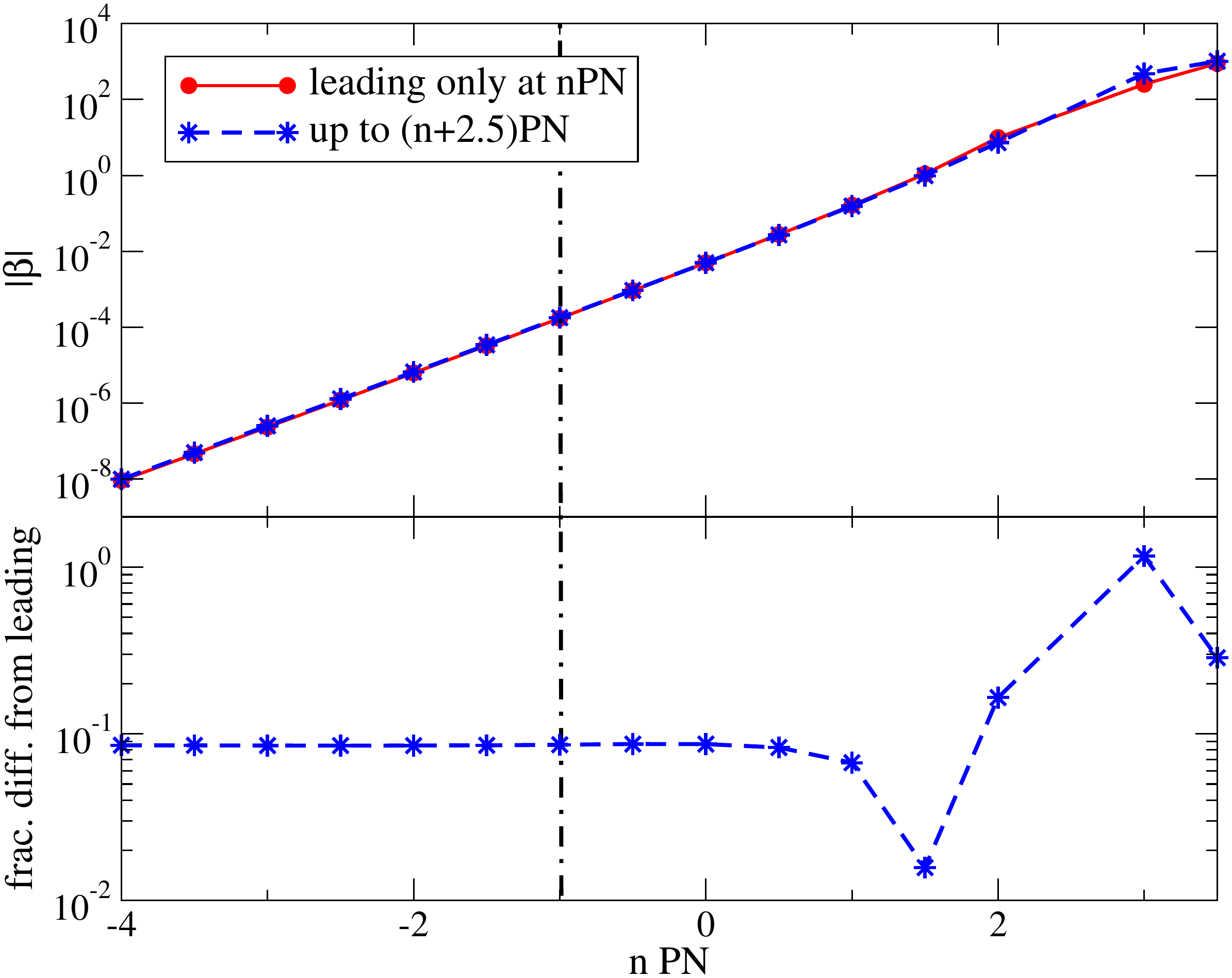} 
\caption{\label{fig:higher-PN} (Color online) (Top) Upper bound on the leading ppE parameter $|\beta|$ for GW150914 with only the leading PN correction at $n$PN order added (red solid) and with relative corrections the same as BD theory added up to $(n+2.5)$PN  order (blue dashed). The vertical dotted-dashed line shows $n=-1$, which corresponds to the BD case in Fig.~\ref{fig:BD-EMRI}.  (Bottom) The fractional difference between the two curves in the top panel. Such a difference generally becomes larger for high PN terms but is smaller than $\sim 20\%$ in most cases. 
}
\end{center}
\end{figure}

How do higher PN order corrections affect constraints on ppE parameters in theories other than BD? To a limited extent, we can address this question by repeating the calculation explained above, but varying the ppE exponent. The top panel of Fig.~\ref{fig:higher-PN} compares constraints on the ppE parameter $\beta$ from GW150914 (for which higher PN terms have a larger contribution than for GW151226) obtained with only the leading PN order correction (red solid) and with corrections up to $2.5$PN order higher than the leading order term (blue dashed). To model the latter, we adopt the same relative corrections from the leading order term as in the BD case. The bottom panel shows the fractional difference of the two constraint curves. The difference remains around $10\%$ for all $n \leq 1$. Although this difference generally grows with $n$, it typically remains smaller than $\sim 20\%$ in most cases. When $n=3$, a partial degeneracy with the phase of coalescence at $2.5$PN order deteriorates the bound. 

Let us now move away from the ppE framework and discuss how higher PN order corrections may affect constraints on $\beta$ in more general theories. Consider a non-GR theory with a single coupling constant that admits a PN expansion, i.e.~one in which the solution to the field equations admits a perturbative solution in $v \ll 1$. The coefficients in the PN expansion are functions of the system parameters (like the masses and spins) and the coupling constant of the non-GR theory (that controls the magnitude of the GR deformation).  One can classify a non-GR theory by the behavior of these coefficients into one of the following three classes:
\begin{enumerate}
\item[(i)] There are no values of the system parameters for which the coefficient of the leading PN order correction is suppressed. 
\item[(ii)] There is a set of values of the system parameters for which the coefficient of the leading PN order correction is moderately suppressed. 
\item[(iii)] There is a set of values of the system parameters for which the coefficient of the leading PN order correction is strongly suppressed and may vanish exactly. 
\end{enumerate}
For theories in class (i), the leading PN order correction always dominates any higher PN order corrections and the constraints on $\beta$ derived in this paper are valid. This is the case for BD theory, as already discussed previously. 
For theories in class (ii), there may be a small subset of systems for which the leading PN order correction becomes comparable to the next-to-leading order correction in a given velocity range. Therefore, one can further split such theories into the following two subclasses:
\begin{enumerate}
\item[(ii-1)] There are no values of the system parameters for which the coefficient of the leading PN order correction cancels with the next-to-leading order correction.
\item[(ii-2)] There is a set of system parameters for which the above cancellation occurs, forcing the next-to-next-to-leading order correction to be dominant.
\end{enumerate}
For theories in case (ii-1), the bound on $\beta$ obtained from the leading PN correction is still valid as an order of magnitude estimate. No known theory falls into class (ii-2), but if one existed, constraints on $\beta$ derived by including only the leading PN order correction could be too strong; for correct estimates one would have to map the constraint on $\beta$ to the particular coupling constants of the theory for a value of $b$ that corresponds to  the next-to-next-to leading order term. Class (ii-2), however, would likely require fine-tuning of the system, i.e.~the masses and spins would have to be just right so that the cancellation occurs. Even if this fine-tuning did happen in a given (as-of-yet unknown) theory, it is incredibly unlikely that it would happen for both events (GW150914 \emph{and} GW151226) simultaneously. 
 
For theories in class (iii), there is a small subset of systems for which the leading PN order correction becomes subdominant relative to the next-to-leading order correction in a given velocity range. If this is the case, constraints derived from the leading PN order correction would be too weak, i.e.~they would be \emph{conservative} because if one had included the next-to-leading PN order term the constraints would have been stronger. This is probably the case for EdGB gravity, as discussed previously. 
 
Since the second case requires a fine-tuned system and a theory that has not yet been proposed or studied, we conclude that in all \emph{known} modified gravity cases the bound on $\beta$ presented in this paper is a solid conservative estimate for theories with a single coupling parameter that admits a PN expansion. 

What about modified theories that either possess more than one coupling parameter or do not admit a PN expansion?  EA theory is an example of a model with more than one coupling parameter. All coupling parameters are likely to enter the GW phase, with different combinations entering at different PN orders. In such a case, including more than the leading PN order term in the waveform phase, as was done in e.g.~\cite{Sampson:2013lpa,TheLIGOScientific:2016src}, is critical to break degeneracies between the coupling parameters and constrain them individually. Certain scalar-tensor theories (i.e.~those that admit dynamical scalarization) are examples of models that do not necessarily admit a PN expansion. In such a case, including more than a single ppE parameter, as shown in~\cite{Sampson:2014qqa}, is critical to properly constrain the modified GR effect. We have discussed the latter in more detail in Sec.~\ref{sec:gen-constr-in-generation}.

Whether one should include higher PN order terms in modified waveforms depends sensitively on whether the event in question has been shown to be consistent with GR. Let us imagine that a new GW observation is made. The first step should then be to determine whether this observation is consistent with GR or whether anomalies are present in the data. The verification of consistency can be made through the residual SNR argument suggested in~\cite{cornish-PPE,Sampson:2013jpa,Vallisneri:2013rc} and performed for GW150914 by the LVC~\cite{TheLIGOScientific:2016src}. The search for anomalies could be done through a parameterized model, as the ppE framework\footnote{Absence of a residual from the best-fit GR template does not necessarily imply the data is also consistent with the absence of anomalies, as in some cases a parameter bias in the GR template could ``fit'' the anomaly, producing \emph{stealth bias}~\cite{cornish-PPE,Vallisneri:2013rc,Sampson:2013jpa,Vitale:2013bma}.}. Reference~\cite{Sampson:2013lpa} has shown that using a single parametric deformation in the waveform phase is ideal to {\em detect} anomalies; the inclusion of simultaneous multiple deformations dilutes the power of such an analysis. However, it was further shown that if an anomaly is present, a single parametric deformation will not be able to pinpoint exactly what type of modification to GR is present in the data. It is only in such cases, i.e. when the data points to the presence of a statistically significant anomaly, that a higher PN order parametric deformation may be necessary to properly characterize it. 

\section{Noise Spectrum Fit}
\label{app:noise-fit}

We construct a fit for the Hanford noise spectrum data~\cite{TheLIGOScientific:2016zmo,noise-data} through the polynomial
\ba
\label{eq:fit}
\sqrt{S_n(f)} &=& \sqrt{S_0} \exp\left( a_0 + a_1 x + a_2 x^2 + a_3 x^3 + a_4 x^4  \right. \nn \\
& + &\left.  a_5 x^5+ a_6 x^6   \right)\,,
\ea
where $S_0 = 0.8464/\mrm{Hz}$, $x \equiv \ln[f/(1\mrm{Hz})]$ and the coefficients $a_i$ are given by Table~\ref{tab:fit}. We assumed that the Livingston noise spectrum is identical to the Hanford one, for simplicity\footnote{We checked that the fractional difference between the Hanford and Livingston detectors on the upper bound on $\beta$ in Fig.~\ref{fig:beta-cons} is always smaller than 20\%.}. None of the conclusions derived in this paper are affected by that assumption.
 
{
\newcommand{\minitab}[2][l]{\begin{tabular}{#1}#2\end{tabular}}
\renewcommand{\arraystretch}{1.2}
\begin{table}[hb]
\begin{centering}
\begin{tabular}{c|c}
\hline
\hline
\noalign{\smallskip}
$a_0$ & $47.8466 \pm 5.38$ \\
$a_1$ & $-92.1896 \pm 6.41$ \\
$a_2$ & $35.9273 \pm 3.07$\\
$a_3$ & $-7.61447 \pm 0.759$ \\
$a_4$ & $0.916742 \pm 0.103$ \\
$a_5$ & $-0.0588089 \pm 0.00721$\\
$a_6$ & $0.00156345 \pm 0.000206$ \\
\noalign{\smallskip}
\hline
\hline
\end{tabular}
\end{centering}
\caption{Fitting coefficients and their standard deviation for the fitting function of Eq.~\eqref{eq:fit}, which approximates the aLIGO noise spectrum during O1, and in particular, around the time of the GW150914 observation.
}
\label{tab:fit}
\end{table}
}

\begin{figure}[htb]
\begin{center}
\includegraphics[width=8.5cm,clip=true]{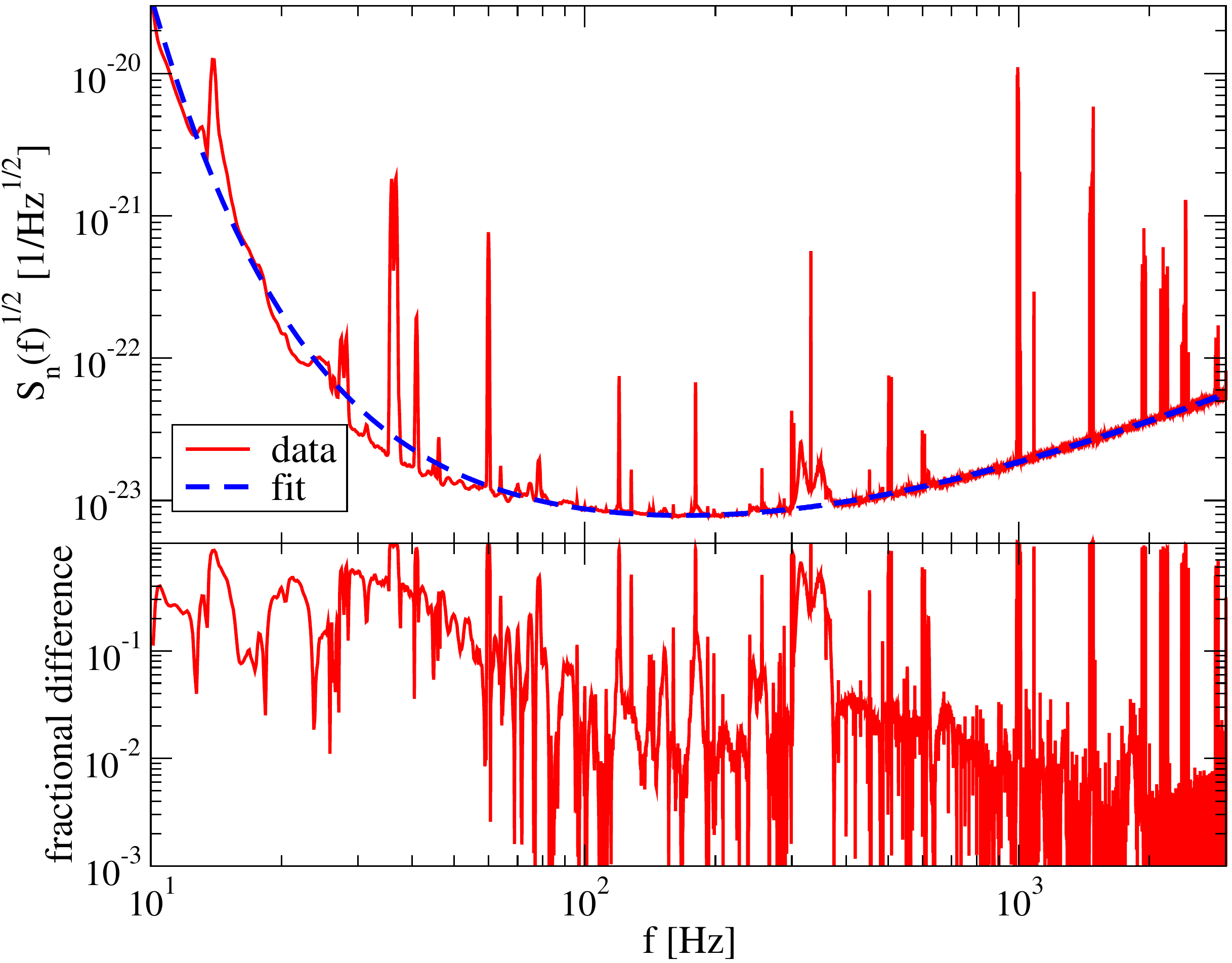} 
\caption{\label{fig:noise} (Color online) (Top) Square-root of the noise spectral density as a function of frequency (in Hz) for the Hanford detector during the O1 (red solid) and using the fitted function of Eq.~\eqref{eq:fit} (blue dashed). (Bottom) The relative fractional difference between the data and the fit. The data contains spikes that are absent from the fit, but we have checked that these spikes do not significantly affect the constraints quoted in this paper, if the SNR is fixed.  
}
\end{center}
\end{figure}

Although the fit provided above is good ($r^{2} = 0.99995$), it is by no means perfect. To see this graphically, the top panel of Fig.~\ref{fig:noise} shows the actual Hanford noise spectral density during the O1 run, together with the fit as a function of frequency in Hz. The data contains many spikes, which the fit smoothes over. The bottom panel shows the fractional difference between the fit and the data. On average, the fit accurately describes the data to $\mathcal{O}(1\%)$ accuracy in the $f > 10^2$ Hz region, while the fractional difference becomes $\mathcal{O}(10\%)$ in the $f <60 $ Hz region. We could have constructed a more accurate fit to the data, but we found that this was not necessary. 

One may wonder whether the spikes in Fig.~\ref{fig:noise} affect the constraints derived in this paper. The answer is no. The spikes do affect the SNR that would be measured at Hanford and at Livingston, but we have here chosen the waveform amplitude such that the SNR with the fitted noise curve is exactly what aLIGO measured. With the SNR properly adjusted, we have checked that the difference between the fit and the data only affects the constraints on $|\beta|$ at $-1$PN order by $3\%$--$6\%$ at most relative to what we quote in this paper.   

\section{Constraining EdGB Gravity with GW150914}
\label{app:EdGB}

In this appendix, we study whether the GW150914 observation by aLIGO allows us to place constraints on EdGB gravity from the absence of scalar dipolar radiation. As shown in Fig.~\ref{fig:beta-cons}, such an observation places a bound on the ppE parameter $|\beta|$ at $-1$PN of $|\beta| \leq 1.7 \times 10^{-4}$. One can map this constraint to that on the coupling constant $\alpha_\EDGB$ using Eq.~\eqref{eq:beta-EdGB}. Assuming the injected values of spins ($\chi_1 = 0 = \chi_2$) and masses that we used to derive the bound on $|\beta|$ via a Fisher analysis, one finds $\sqrt{|\alpha_\EDGB|} \leq 22$km. However, to be as conservative as possible, one needs to study how such a bound depends on the injected values of binary parameters such as individual spins $\chi_A$, as the latter are only weakly constrained ($|\chi_1| \leq 0.79$ and $|\chi_2| \leq 0.95$~\cite{TheLIGOScientific:2016pea,TheLIGOScientific:2016qqj}), even when the effective spin parameter is better constrained $\chi_\mrm{eff} = - 0.06 ^{+0.14} _{-0.14}$~\cite{TheLIGOScientific:2016wfe,TheLIGOScientific:2016pea}. 

\begin{figure}[htb]
\begin{center}
\includegraphics[width=7.cm,clip=true]{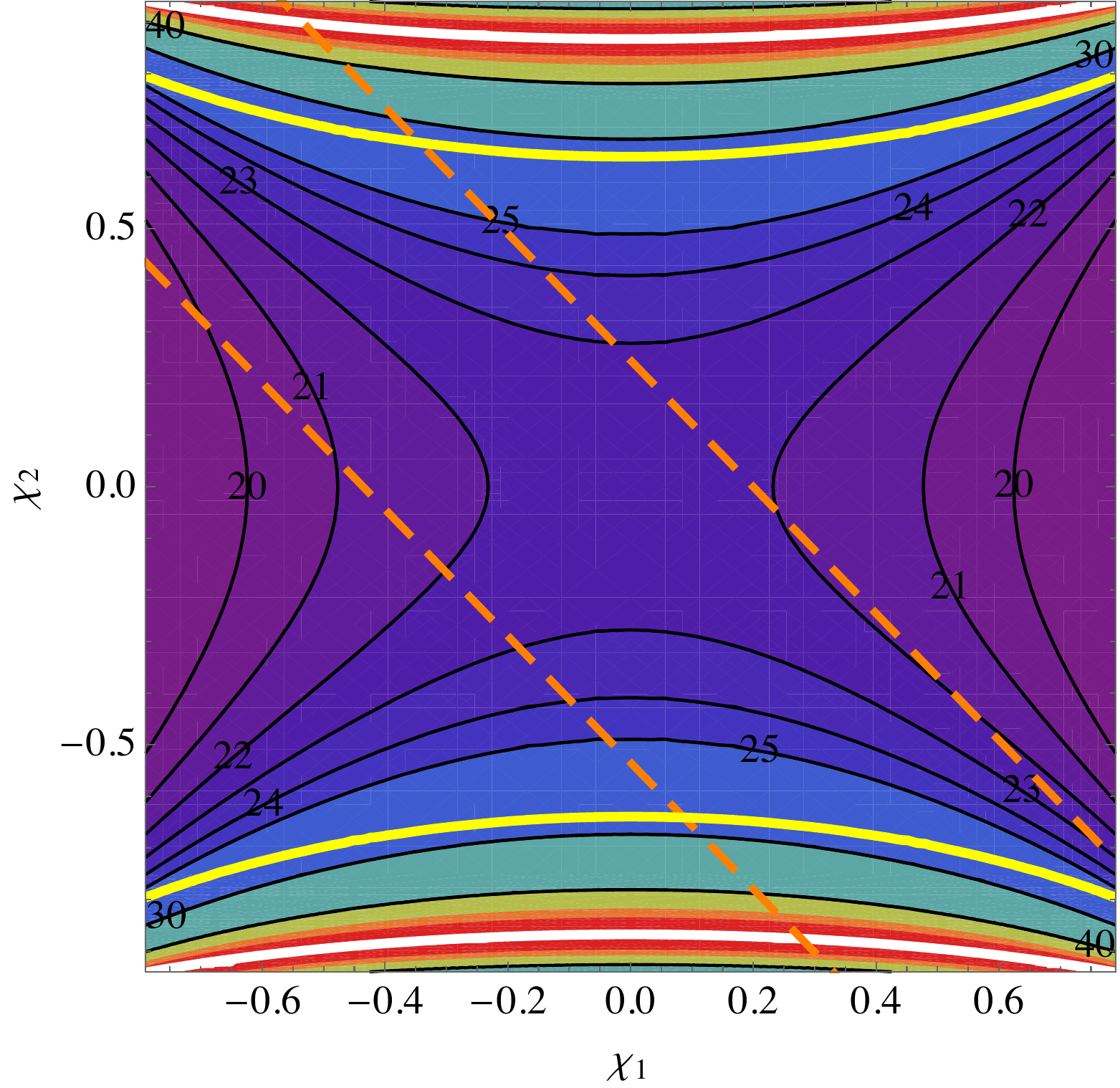} \qquad
\caption{\label{fig:EdGB} (Color online) GW150914 upper bound on $\sqrt{|\alpha_\EDGB|}$ in km derived by mapping the constraint on $\beta$ at $-1$PN order with each ($\chi_1$, $\chi_2$) allowed from the aLIGO measurement. Orange dashed lines show the allowed region from the effective spin $\chi_\mrm{eff}$ measurement, while white curves show spins that shut off dipole radiation completely. However, note that the small coupling approximation used to derive these bounds is only valid within the region between the yellow curves. 
}
\end{center}
\end{figure}

Figure~\ref{fig:EdGB} presents the upper bound on $\sqrt{|\alpha_\EDGB|}$ in kms obtained by mapping the bound on $|\beta|$ but varying $\chi_1$ and $\chi_2$. The region of $\chi_{1}$ and $\chi_{2}$ spanned by the posterior distribution is within the orange dashed lines. The bound weakens rapidly as one approaches spins that shut off dipole radiation completely (shown by white curves). Because of the width of the $\chi_{1}$ and $\chi_{2}$ posterior distribution, this immediately shows that one cannot place a bound on $\sqrt{|\alpha_\EDGB|}$ with the GW150914 event. 

Figure~\ref{fig:EdGB} also shows that the bound on $\sqrt{|\alpha_\EDGB|}$ weakens significantly as $\chi_{1}$ and $\chi_{2}$ approach unity. These bounds, however, were obtained within the small coupling approximation, which requires $16 \pi \alpha_\EDGB^2/r_{\hor}^4 \ll 1$ with $r_\hor$ corresponding to the horizon size of the smaller BH. This approximation, thus, is valid only for sufficiently small values of $\sqrt{|\alpha_\EDGB|}$, i.e.~those within the region enclosed by horizontal yellow curves. Therefore, if GW150914 was produced by BHs with large spins, and if aLIGO had been able to measure these spins accurately, one would still not be able to place bounds on $\sqrt{|\alpha_\EDGB|}$, because these would be outside the regime of validity of the approximation used to derive such bounds. 

\section{Effective Viscosities of Compact Objects}
\label{app:viscosity}

In this appendix, we derive the effective viscosities of compact objects that are summarized in Table~\ref{table:viscosity}.
Let us first compute the viscosities of non-rotating BHs~\cite{Hawking:1972hy,Hartle:1973zz} with mass $M$. The membrane paradigm~\cite{membrane,Glampedakis:2013jya} allows us to estimate the kinematic viscosity $\bar{\nu}_{\BH} \sim M$, which is related to the shear viscosity by $\bar \eta_{\BH} = \rho_{\BH} \bar \nu_{\BH}$ and to the bulk viscosity by $\bar \zeta_\BH = - \bar \eta_\BH$~\cite{membrane}. Estimating the BH density as $\rho_\BH \sim M/[(4\pi/3) R_s^3]$ with $R_s = 2M$ corresponding to the Schwarzschild radius, one finds
\be
\bar{\eta}_{\BH} = - \bar{\zeta}_{\BH} \sim 1.3 \times 10^{30}  \frac{\mrm{g}}{\mrm{cm} \cdot \mrm{s}} \left( \frac{m}{65M_\odot} \right)^{-1}\,.
\ee
The fact that the sign of the bulk viscosity is negative is a well-known peculiarity of the effective fluid description of event horizons\footnote{The effective fluid description of dynamical horizons or future outer trapping horizons of BHs give a positive bulk viscosity with the same magnitude as that computed with event horizons~\cite{Gourgoulhon:2008pu}.}. Naive application of the Newtonian stellar fluid model in Eq.~\eqref{eq:zeta-rem} would then suggest BHs have unstable radial modes, which of course is not the case. It is unclear exactly how to interpret negative bulk viscosity in the case of BHs, and this illustrates that not all exotic compact objects may have dynamics that fit comfortably in an effective hydrodynamic framework.

Let us now compute the viscosities of non-rotating, unmagnetized NSs. The shear viscosity due to the scattering of neutrons is given by~\cite{1979ApJ...230..847F,1987ApJ...314..234C,1990ApJ...363..603C}
\be
\bar{\eta}_{\NS}^{(n)} \sim 2 \times 10^{14} \rho_{15}^{9/4}  T_{11}^{-2}   \frac{\mrm{g}}{\mrm{cm} \cdot \mrm{s}}\,,
\ee
while the bulk viscosity is given by~\cite{1989PhRvD..39.3804S,1990ApJ...363..603C}
\be
\label{eq:bulk-vis}
\bar{\zeta}_{\NS}^{(n)} \sim 6 \times 10^{28} \rho_{15}^{2}  T_{11}^{6}  \omega_{10}^{-2} \left( \frac{e^{\nu}}{0.1} \right) \frac{\mrm{g}}{\mrm{cm} \cdot \mrm{s}}\,,
\ee
where $\omega_{10}/2\pi$ corresponds to the oscillation mode frequency divided by $(10/2\pi)$kHz, $ \rho_{15}$ is the NS density divided by $10^{15}$g/cm$^3$, $e^{\nu}$ is the $(t,t)$ component of the NS metric, $T_{11}$ is the NS temperature divided by $10^{11}$K$\sim 10$MeV, with the latter corresponding to the typical temperature of hypermassive NSs formed after NS binary mergers.   

Let us then proceed to compute the shear viscosity of non-rotating but magnetized NSs. For a strongly-magnetized NS, this can be estimated by comparing the Alfv\'en timescale to the viscous timescale, given by Eqs.~(41) and~(42) in~\cite{Shapiro:2000zh} respectively: 
\be
\label{eq:magnetic-viscosity}
\bar{\eta}_{\NS}^{(B)} \sim 1.3 \times 10^{27} B_{15} R_{12} \sqrt{\rho_{15}} \frac{\mrm{g}}{\mrm{cm} \cdot \mrm{s}}\,, 
\ee
where $R_{12}$ is the NS radius divided by 12km, while $B_{15}$ is the magnetic field strength divided by $10^{15}$G. 

Let us finally compute the viscosities associated with boson stars. There are numerous models for boson stars~\cite{Liebling:2012fv}, though typically bosonic matter has very low effective viscosity, with the leading order dissipation of self-gravitating configurations coming from GW emission~\cite{Berti:2006qt,Macedo:2013jja} (this is similar to ideal fluid NSs). For example, from the calculation of the QNMs of a so-called solitonic boson star~\cite{Friedberg:1986tq} with radius $R\sim 3M$ presented in~\cite{Macedo:2013jja}, the damping time of the $\ell=2$ polar mode is $\tau\sim 10^3 M$. For a $65M_\odot$ boson star, the damping time is then $\tau\sim 320$ms, which leads to effective shear and bulk viscosities of $\bar{\eta}_{\BS} \sim 7\times 10^{26}$g~cm$^{-1}$~s$^{-1}$ and $\bar{\zeta}_{\BS} \sim 5\times 10^{28}$g~cm$^{-1}$~s$^{-1}$ via Eqs.~\eqref{eq:eta-rem} and \eqref{eq:zeta-rem} respectively.

\newpage
\bibliography{master}
\end{document}